\documentclass{aa}

\usepackage{graphicx}
\usepackage{graphbox}
\usepackage{txfonts}
\usepackage{amsmath}
\usepackage{stfloats}
\usepackage{caption}
\usepackage{subcaption}
\usepackage{placeins}
\usepackage{float}
\usepackage{color}
\usepackage{xcolor}
\usepackage{scalerel}
\usepackage{tabularx}
\usepackage{multicol}
\usepackage[colorlinks=true,
            linkcolor=blue,
            filecolor=magenta,
            urlcolor=blue,
            citecolor=blue]{hyperref}
\usepackage{natbib}
\newcommand\vfrac[2]{\ThisStyle{%
  \setbox0=\hbox{$\SavedStyle#1#2$}%
  \setbox2=\hbox{$\SavedStyle X$}%
  \ifdim\ht0>\ht2\setlength{\ht0}{\ht2}\fi%
  #1\mathord{\stretchto{\raisebox{2.3\LMpt}{$\SavedStyle/$}}{\ht0}}#2}}

\newlength{\bibitemsep}
\newlength{\bibparskip}
\let\oldthebibliography\thebibliography
\renewcommand\thebibliography[1]{%
  \oldthebibliography{#1}%
  \setlength{\parskip}{\bibitemsep}%
  \setlength{\itemsep}{\bibparskip}%
}

\newcommand\clearrow{\global\let\rowmac\relax}
\clearrow

\begin{document}

   \title{Metal-THINGS: gas metallicity gradients in nearby galaxies}

   \author{G. Valé\inst{1,2}
          \and
          M. A. Lara-López\inst{1,2}
          \and
          M. Valerdi\inst{3,4}
          \and
          I. Zinchenko\inst{5,6}
          \and         
          S. P. O'Sullivan\inst{1,2}
          \and
          L. S. Pilyugin\inst{6,7}
          \and
          J. Cepa\inst{8,9}
          \and \\
          V. Casasola\inst{10}
          \and
          M. E. De Rossi\inst{11, 12}
          \and
          S. Dib\inst{13}
          \and
          J. Fritz\inst{14}
          \and
          J. Gallego\inst{1,2}
          \and
          L. E. Garduño\inst{1}
          \and
          O. López-Cruz\inst{3}
          \and \\
          V. Tailor\inst{10, 15}
          \and
          J. Zaragoza-Cardiel\inst{16}
          }

   \institute{
    Departamento de Física de la Tierra y Astrofísica, Fac. de C.C. Físicas, Universidad Complutense de Madrid, E-28040 Madrid, Spain
    \and
    Instituto de Física de Partículas y del Cosmos, IPARCOS, Fac. C.C. Físicas, Universidad Complutense de Madrid, E-28040 Madrid, Spain
    \and
    Instituto Nacional de Astrofísica, Óptica y Electrónica (INAOE), Luis E. Erro No. 1, Sta. Ma. Tonantzintla, Puebla, C.P. 72840, México
    \and
    Consejo Nacional de Humanidades, Ciencias y Tecnologías, Av. Insurgentes Sur 1582, 03940, Ciudad de México, México
    \and
    Faculty of Physics, Ludwig-Maximilians-Universität, Scheinerstr. 1, 81679 Munich, Germany
    \and
    Main Astronomical Observatory, National Academy of Sciences of Ukraine, 27 Akademika Zabolotnoho St, 03680 Kiev, Ukraine
    \and
    Institute of Theoretical Physics and Astronomy, Vilnius University, Sauletekio av. 3, 10257 Vilnius, Lithuania
    \and
    Instituto de Astrofísica de Canarias, 38205 La Laguna, Tenerife, Spain
    \and
    Departamento de Astrofísica, Universidad de La Laguna (ULL), 38205 La Laguna, Tenerife, Spain
    \and
    INAF – Istituto di Radioastronomia, Via Piero Gobetti 101, 40129, Bologna, Italy
    \and
    Universidad de Buenos Aires, Facultad de Ciencias Exactas y Naturales y Ciclo Básico Común. Buenos Aires, Argentina
    \and
    CONICET-Universidad de Buenos Aires, Instituto de Astronomía y Física del Espacio (IAFE). Buenos Aires, Argentina
    \and
    Max-Planck-Institut f\"{u}r Astronomie, K\"{o}nigstuhl 17, 69117, Heidelberg, Germany
    \and
    Instituto de Radioastronomía y Astrofísica, Universidad Nacional Autónoma de México, Morelia, Michoacán 58089, Mexico
    \and
    Dipartimento di Fisica e Astronomia, Alma Mater Studiorum Università di Bologna,via Piero Gobetti 93/2, I-40129, Bologna, Italy
    \and
    Centro de Estudios de Física del Cosmos de Aragón (CEFCA), Plaza San Juan 1, 44001 Teruel, Spain
    }

   \date{Received XXXX XX, XXXX; accepted XXXX XX, XXXX}

  \abstract
   {}
   {This paper explores and analyses the gas metallicity gradients in a sample of 25 nearby galaxies using new Integral Field Spectroscopy observations from the Metal-THINGS survey, for a total of 102 individual pointings. We derive and analyse the resolved diffuse ionised gas content, Baldwin, Phillips \& Terlevich diagrams and gas metallicities for our entire sample, at spatial resolutions of 40–300 pc. Gas metallicity gradients are studied as a function of the galaxy's stellar mass, $\rm H$ \tiny{I }\small gas fraction, diffuse ionised gas content, and using different parametric length scales for normalisation.}
   {The metallicity gradients are analysed using Bayesian statistics based on data from the Metal-THINGS survey. Bayesian MCMC models are developed to explore how metallicity gradients vary with a galaxy's mass and how they correlate with properties such as the stellar mass or the atomic gas fraction. Additionally, we compare and contrast our results with those from other works that use the same metallicity calibration.}
   {For our sample, we find that the metallicity typically decreases with galactic radius, consistent with inside-out galaxy growth. We find a trend dependent on the stellar mass, with a break at $\log(M_{\rm star}/M_\odot) \simeq 9.5$, and another between the metallicity gradients and the atomic gas fraction ($f_{\rm g, \, H \, I}$) of a galaxy at $f_{\rm g, \, H \,I} \simeq 0.75$, indicating relatively shallower gradients for lower gas fractions. These results are consistent with previous studies for galaxies in comparable stellar mass regimes and morphologies. We find that normalisation using NUV-band effective radii are preferable for galaxies with a higher atomic gas content and lower stellar masses, while r-band radii are better suited for those with lower atomic gas fractions and more massive ones.}
   {Our results highlight a strong connection between gas content, stellar mass, and metallicity gradients. The breaks at $\log(M_\star/M_\odot) \simeq 9.5$ and $f_{\rm g, \, H \,I} \simeq 0.75$ mark shifts in chemical enrichment behaviour, with low-mass galaxies showing greater sensitivity to gas processes. Overall, this points to gas accretion and removal as key drivers of chemical evolution in low-mass systems.}


   \maketitle
%

\section{Introduction}
The interstellar medium (ISM) is a complex environment that plays a crucial role in shaping the dynamics and evolution of galaxies. The metal content and its distribution in galaxies result from the complex interplay of various processes that affect the ISM, including: (i) gas infall, which dilutes the ISM but can also trigger star formation \citep[e.g.,][]{intro_1}; (ii) feedback from active galactic nuclei (AGN), star formation and stellar evolution, driving outflows that redistribute or expel enriched gas \citep[e.g.,][]{intro_2}; (iii) metal injection into the ISM through stellar evolution \citep[e.g.,][]{intro_3}; (iv) turbulence leading to metal-mixing \citep[e.g.,][]{intro_4}; and (v) galaxy interactions and mergers, which can trigger enhanced star formation rates up to starbursts, AGN activity, or tidal perturbations \citep[e.g.,][]{Casasola_2004, Mo_Van_den_Bosch_White_2010}. Many of these processes are highly interconnected, complicating the overall picture. This way, the ISM acts as the critical link connecting processes on stellar scales to those on galactic scales \citep{Paron_2018}.

One of the clearest observational signatures of these processes is the presence of metallicity gradients in disc galaxies. 
Typically, the metal content decreases with increasing galactocentric radius (generating the so-called ``negative gradients''), a pattern that has been observed for decades \citep{Searle_1971, Lequeux_1979, Shaver_1983, Vila-Costas_1992} and that have been the subject of several observational studies \cite[e.g.][]{Afflerbach_1997, Deharveng_2000, Bresolin_2009a, Bresolin_2009b}. Large amounts of high quality data from integral field spectroscopy (IFS) now available, such as the Sydney–AAO Multi-object Integral field spectrograph (SAMI) Galaxy Survey \citep{SAMI_paper} and Mapping Nearby Galaxies at APO (MaNGA) \citep{MANGA_paper}, have enabled the exploration of the dependencies of the metallicity gradient (hereafter $\rm \nabla_{O/H}$) on a variety of galactic properties \citep[e.g.][]{SanchezMeguiano_2016, PerezMontero_2016, Belfiore_2017}.

However, most of these studies focus on intermediate and massive galaxies. In contrast, low-mass galaxies remain relatively unexplored in this context mostly due to observational challenges such as their low surface brightness. As a result, the behaviour of metallicity gradients in this mass regime is still poorly constrained. Investigating them can provide crucial insights into the role of feedback, gas accretion and different dynamical processes in shaping their chemical evolution.

Among the few studies targeting resolved metallicity gradients in low-mass galaxies are \citet{low_mass_grad_paper_1} (down to ${\log(M_{\textrm{star}}/M_\odot)} = 8.33$) and \citet{low_mass_grad_paper_2} (down to ${\log(M_{\textrm{star}}/M_\odot)} = 7.12$). 
From the large IFU surveys, the only one that has included low-mass galaxies in its sample is the SAMI survey. One notable study is \citet{Poetrodjojo_2021}, which uses SAMI data to estimate gas-phase metallicities and metallicity gradients in such galaxies. However, due to its limited spatial resolution, the SAMI survey is constrained to a lower stellar mass limit of approximately $\log(M_{\textrm{star}}/M_\odot) \simeq 8.5$.

Different observations of nearby galaxies allow us to probe their ISM and compare them with predictions from both theoretical models and simulations. For instance, strong negative gas-phase metallicity gradients in spiral galaxies indicate that the central regions are populated by more evolved, and thus more metal-rich, stellar populations, while metal-poor and less-evolved stars are located in their exterior regions \citep[e.g.][]{Zaritzsky_1994}. This is consistent with the inside-out formation scenario of spirals, where spiral discs form from gas accretion, as seen in infall models of galaxy formation \citep[e.g.][]{Matteucci_1989, BoissierPrantzos_1999, Perez_2013}.

Other cases involve galaxies undergoing mergers or interactions with other galaxies. These interactions disturb the gas and can create turbulent motions that cause loss of angular momentum and radially inwards flows of gas. While galaxy-galaxy interactions can initially flatten the metallicity gradient, the induced central star-formation can re-establish a negative gradient \citep[e.g.][]{Gibson_2013, Stanghellini_2014}, as metals are injected into the ISM by supernovae and stellar winds \citep[e.g.][]{LaraLopez_2022}.

For galaxies which are not part of interacting systems, a flat metallicity gradient has been proposed to arise from metal-mixing caused by radial gas flows \citep[][among others]{Lacey_1985, GoetzKoeppen_1992, Portinari_2000, Ferguson_2001, Schonrich_2009, Bilitewski_2012, Spitoni_2013}. Subsequent outflows of enriched gas driven by feedback from AGN, star-formation and stellar evolution can flatten the metallicity gradient at the outer regions \citep[e.g.][]{LaraLopez_2022}. This flattened metallicity gradient, or truncation in the outer parts, may also be linked to the presence of bars \citep[e.g.][]{Bresolin_2009a, Marino_2012, Rosales-Ortega_2011}.

\begin{table}[t]
    \centering
    \renewcommand{\arraystretch}{1.2}
    \begin{tabular}{>{\rowmac}c>{\rowmac}c>{\rowmac}c<{\clearrow}}
    \hline
    \hline
    1 & 2 & 3 \\
    \begin{tabular}[c]{@{}c@{}} Galaxy \\ \textbf{} \end{tabular} &
    \begin{tabular}[c]{@{}c@{}} Number of  \\ pointings \end{tabular} &
    \begin{tabular}[c]{@{}c@{}} Date of \\  observation \end{tabular} \\ \hline
    
    DDO 53          &      1        &    Jan. 19                         \\
    DDO 154         &      1        &    March 22                        \\
    Holmberg I      &      2        &    April 20                        \\
    M81 DwB         &      1        &    March 25                        \\
    Holmberg II     &      3        &    March 24; March 25              \\
    NGC 2366        &      4        &    Jan. 19; Dec. 19                \\
    NGC 4214        &      4        &    May 19; Dec 19; June 21         \\
    NGC 1569        &      2        &    Jan. 18; Oct. 20                \\
    IC 2574         &      2        &    March 24                        \\
    NGC 4449        &      4        &    June 21; March 22               \\
    NGC 2976        &      3        &    May 22; April 23                \\
    NGC 3077        &      1        &    April 22                        \\
    NGC 2403        &      10       &    Jan. 18; March 24; Dec. 19      \\
    NGC 925         &      14       &    2017-2019                       \\
    NGC 3198        &      1        &    March 25                        \\
    NGC 4826        &      2        &    March 25                        \\
    NGC 4736        &      4        &    April 23; March 25              \\
    NGC 3184        &      5        &    April 23; March 25              \\
    NGC 5457        &      6        &    June 21; March 22; April 23     \\
    NGC 6946        &      12       &    2017-2021                       \\
    NGC 5055        &      3        &    March 24                        \\
    NGC 5194        &      12       &    2018-2023                       \\
    NGC 3521        &      3        &    March 22                        \\
    NGC 2841        &      2        &    March 25                        \\
    NGC 7331        &      6        &    Sept 17; Oct. 18                \\
    \hline
    
    \end{tabular}
\caption{Summary of observations}
\label{tab:observation_table}
\vspace{-0.3cm}
\end{table}

Other studies also suggest that the morphology of galaxies and the shape of the radial metallicity profile may be related by certain processes that affect both, as different physical processes which can alter the gradients take place. In the case of spirals, included in our sample, three main aspects can be considered: how tightly wound the arms are, the presence or absence of a bar or bulge, and their relative size to that of the disc \citep{Sanchez-Menguiano_2016}.

In this context, it is of great interest to study the relation between metallicity gradients and the atomic gas or the dust content of galaxies, which is critical for understanding galactic evolution. While metallicity provides insight into the location of metals within galaxies, the distributions of atomic gas and dust trace the raw material for future star formation. Studying how these components correlate can help constrain models of chemical enrichment, feedback mechanisms, and the interplay between different components of the ISM \citep[e.g., ][]{met_gas_paper_1, met_gas_paper_2}.

\begin{table*}[t]
\vspace{0.25cm}
\centering
\renewcommand{\arraystretch}{1.2}
\begin{tabular}{>{\rowmac}c>{\rowmac}c>{\rowmac}c>{\rowmac}c>{\rowmac}c>{\rowmac}c>{\rowmac}c>{\rowmac}c>{\rowmac}c>{\rowmac}c>{\rowmac}c>{\rowmac}c<{\clearrow}}
\hline
\hline
1 & 2 & 3 & 4 & 5 & 6 & 7 & 8 & 9 & 10 & 11 \\
\begin{tabular}[c]{@{}c@{}} Galaxy \\ \textbf{} \end{tabular} &
\begin{tabular}[c]{@{}c@{}} Morph. \\ \textbf{} \end{tabular} &
\begin{tabular}[c]{@{}c@{}}RA (2000) \\ hh mm ss\end{tabular} &
\begin{tabular}[c]{@{}c@{}}DEC (2000) \\ dd mm ss\end{tabular} &
\begin{tabular}[c]{@{}c@{}}$\phi$ \\ $\rm ^o$\end{tabular} &
\begin{tabular}[c]{@{}c@{}}$i$\\ $\rm ^o$\end{tabular} &
\begin{tabular}[c]{@{}c@{}}$D$ \\ {[}Mpc{]}\end{tabular}  &
\begin{tabular}[c]{@{}c@{}}SFR \\ {[}$M_\odot \; \mathrm{yr}^{-1}${]}\end{tabular}  &
\begin{tabular}[c]{@{}c@{}}$\log M_{\rm star}$\\ {[}$\log M_{\odot}${]}\end{tabular} &
\begin{tabular}[c]{@{}c@{}}$\log M_{\rm H \, I}$\\ {[}$\log M_{\odot}${]}\end{tabular} &
\begin{tabular}[c]{@{}c@{}}$\log M_{\rm H_2}$\\ {[}$\log M_{\odot}${]}\end{tabular} \\ \hline


DDO 53      & Im          & 08 34 07.2 & +66 10 54   & 132 & 31 & 3.6  & 0.008 & 6.9   & 7.77   & $-$ \\
DDO 154     & IB(s)m      & 12 54 05.9 & +27 09 10   & 230 & 66 & 4.3  & 0.004 & 7.07  & 8.7    & 6.8 \\
Holmberg I  & IAB(s)m     & 09 40 32.2 & +71 10 56   & 50  & 12 & 3.8  & 0.006 & 7.4   & 8.3    & 7.2 \\
M81 DwB     & Im          & 10 05 30.6 & +70 21 52   & 321 & 44 & 5.3  & 0.005 & 7.8   & 7.4    & $-$ \\
Holmberg II & Im          & 08 19 05.0 & +70 43 12   & 177 & 41 & 3.4  & 0.07  & 8.3   & 8.9    & 7.6 \\
NGC 2366    & IB(s)m      & 07 28 53.4 & +69 12 51   & 40  & 64 & 3.4  & 0.13  & 8.41  & 8.81   & $-$ \\
NGC 4214    & IAB(s)m     & 12 15 39.2 & +36 19 37   & 65  & 44 & 2.9  & 0.05  & 8.7   & 8.7    & 7.0 \\
NGC 1569    & IBm         & 04 30 49.0 & +64 50 53   & 112 & 63 & 2.0  & 0.06  & 8.61  & 7.875  & $-$ \\
IC 2574     & SAB(s)m     & 10 28 27.7 & +68 24 59   & 56  & 53 & 4.0  & 0.12  & 8.72  & 9.3    & 7.9 \\
NGC 4449    & IBm         & 12 28 11.9 & +44 05 40   & 230 & 60 & 4.2  & 0.5   & 9.03  & 9.2    & 6.9 \\
NGC 2976    & SAc pec     & 09 47 15.3 & +71 10 56   & 335 & 65 & 3.6  & 0.10  & 9.09  & 8.3    & 7.8 \\
NGC 3077    & I0 pec      & 10 03 19.1 & +68 44 02   & 45  & 46 & 3.8  & 0.09  & 9.17  & 9.1    & 6.5 \\
NGC 2403    & SAB(s)cd    & 07 36 51.1 & +65 36 03   & 124 & 63 & 3.2  & 0.85  & 9.57  & 9.5    & 7.3 \\
NGC 925     & SAB(s)d     & 02 27 16.5 & +33 34 44   & 287 & 66 & 9.2  & 1.09  & 9.75  & 9.66   & 8.4 \\
NGC 3198    & SB(rs)c     & 10 19 55.0 & +45 32 50   & 215 & 72 & 13.8 & 0.85  & 10.05 & 10.1   & 8.8 \\
NGC 4826    & (R)SA(rs)ab & 12 56 43.6 & +21 41 00   & 121 & 65 & 7.5  & 0.82  & 10.2  & 8.15   & $-$ \\
NGC 4736    & (R)SA(r)ab  & 12 50 53.0 & +41 07 13   & 296 & 41 & 4.7  & 0.43  & 10.33 & 8.7    & 8.6 \\
NGC 3184    & SAB(rs)c    & 10 18 17.0 & +41 25 28   & 179 & 16 & 11.1 & 1.43  & 10.37 & 9.6    & 9.2 \\
NGC 5457    & SAB(rs)cd   & 14 03 12.6 & +54 20 57   & 39  & 18 & 7.4  & 2.49  & 10.39 & 10.151 & $-$ \\
NGC 6946    & SAB(rs)cd   & 20 34 52.2 & +60 09 14   & 243 & 33 & 5.9  & 4.76  & 10.5  & 9.8    & 9.6 \\
NGC 5055    & SA(rs)bc    & 13 15 49.2 & +42 01 45   & 102 & 59 & 10.1 & 2.42  & 10.72 & 10.1   & 9.7 \\
NGC 5194    & SA(s)bc pec & 13 29 52.7 & +47 11 43   & 172 & 42 & 8.0  & 6.05  & 10.73 & 9.5    & 9.4 \\
NGC 3521    & SAB(rs)bc   & 11 05 48.6 & $-$00 02 09 & 340 & 73 & 10.7 & 3.34  & 10.83 & 9.2    & 9.0 \\
NGC 2841    & SA(r)b      & 09 22 02.6 & +50 58 35   & 153 & 74 & 14.1 & 0.20  & 10.93 & 10.1   & 8.5 \\
NGC 7331    & SA(s)b      & 22 37 04.1 & +34 24 57   & 168 & 76 & 14.7 & 4.20  & 11.0  & 10.1   & 9.7 \\ 
\hline

\end{tabular}
\caption{Physical information of the observed galaxies. The parameters given in each column are as follows: 1) galaxy designation, 2) morphological type, 3) and 4) coordinates in J2000.0, 5) position angle, 6) inclination, 7) distances in Mpc, 8) star formation rates, 9) stellar mass, 10) atomic gas mass, and 11) molecular gas mass. Data for columns 2) to 8) can be found in \citet{Walter_2008_THINGS} and 9) are primarily from \citet{leroy_2013, leroy_2019}, among other sources, see Sect. 2 for details. For columns 10) and 11), the H{\small I} and H$_2$ gas masses are mostly taken from \citet{Leroy_2008}, see Sect. 2 for details. The SFRs for NGC 2366 and NGC 4449 are from \citet{NGC2366_sfr}, and \citet{ngc4449_sfr}, respectively.}

\label{tab:gal_data}
\end{table*}

To further study metallicity gradients, galactic chemical evolution models have been used historically to try to understand processes governing their shape. These simulations are capable of predicting metallicity gradients in simulated galaxies that match findings on observational data and support current models for their evolution. One of these, IllustrisTNG \citep{Illustris_paper}, is a set of state-of-the-art cosmological, magnetohydrodynamical simulations designed to simulate the formation and evolution of galaxies in a cosmological setting \citep[e.g.][]{Hemler_2021, Illustris_met3, Illustris_met2}.

Using simulations and observational data, there have been numerous studies on how the stellar mass is linked to the shape of the metallicity gradient \citep[e.g.][]{Sharda_2021}. These works suggest that massive galaxies tend to form earlier and in denser environments than their less massive counterparts. In contrast, less massive galaxies experience more prolonged gas accretion, as different physical processes occur within them \citep{Artemi_2023}.

An important result emerging from these findings is the correlation between stellar mass and gas-phase metallicity in galaxies, commonly referred as the mass-metallicity relation (MZR or MMR), which was firmly established by \citet{Tremonti_2004} using data for approximately $53,000$ galaxies in the Sloan Digital Sky Survey (SDSS). Subsequent studies \citep[e.g.][]{Ellison_2007, Dib_2011, Lara-Lopez_2013} have revealed an additional correlation between star formation rate (SFR) and metallicity, where galaxies with higher SFRs tend to exhibit lower metallicities.

The MMR can be explained by galactic outflows of metal-rich gas being more effective in low-mass galaxies, whereas galaxies with larger potential wells are able to retain most ejecta inside the halo, which can later be recycled \citep[e.g.][]{Tremonti_2004, Ma_2016}. 

One of the main objectives of Metal-THINGS is to combine the rich multiwavelength information available accross different spectral ranges to obtain a more complete understanding of the physical processes that govern galaxy evolution. This approach allows for the simultaneous analysis of stellar populations, gas content, dust and star formation. Currently there are only a few extragalactic surveys that provide a comprehensive multiwavelength coverage, \citep[i.e. the PHANGS survey,][]{Leroy_2021_phangs, phangs_spatial_res}, most of them focusing on more massive systems. This results in a more critical gap in the understanding on the processes taking place in low-mass galaxies, which remain underrepresented. Metal-THINGS tries to address this limitation by targeting galaxies in a wide range of stellar masses with spatially resolved data accross multiple wavelengths.

In this study, we characterize and analyse the metallicity gradients of the Metal-THINGS galaxy sample. Despite  extensive efforts, a comprehensive understanding of how metallicity gradients depend on certain galactic properties remains incomplete. We further investigate the role of the HI gas mass in shaping the overall metallicity gradients across the sample.

This paper is organized as follows: in Section 2 we describe the observations made for the Metal-THINGS survey along with the reduction process. In Section 3, we explain the methodology used for the extinction correction, the identification of diffuse ionized gas, ``Baldwin, Phillips \& Terlevich'' \citep[hereafter BPT,]{Baldwin_1981} diagnostic diagrams and the estimation of the gas-phase metallicities and their gradients in our sample. A discussion is presented in Section 4, and finally, our summary and conclusions are given in Section 5.

\begin{table*}[t]
\centering
\renewcommand{\arraystretch}{1.2}
\begin{tabular}{>{\rowmac}c>{\rowmac}c>{\rowmac}c>{\rowmac}c|>{\rowmac}c>{\rowmac}c>{\rowmac}c>{\rowmac}c<{\clearrow}}
\hline
\hline
1 & 2 & 3 & 4 & 5 & 6 & 7 & 8 \\
\begin{tabular}[c]{@{}c@{}} Galaxy \\ \textbf{} \end{tabular} &
\begin{tabular}[c]{@{}c@{}}$f_0$\\ {[}$\rm erg \, s^{-1} \, cm^{-2}${]}\end{tabular} &
\begin{tabular}[c]{@{}c@{}}$\beta$\\ \textbf{} \end{tabular} &
\begin{tabular}[c]{@{}c@{}}\% DIG \\ Model\end{tabular} &
\begin{tabular}[c]{@{}c@{}} \% \\ SF \end{tabular} &
\begin{tabular}[c]{@{}c@{}} \% \\ composite \end{tabular} &
\begin{tabular}[c]{@{}c@{}} \% \\ AGN \end{tabular} &
\begin{tabular}[c]{@{}c@{}} Number of \\ fibres in BPT \end{tabular} \\
\hline

DDO 53          & 8.03$\pm$0.05  & 1.267$\pm$0.006   & 28.03    & 100.0       & 0.0       & 0.0   & 60         \\
DDO 154         & 5.5$\pm$0.8    & 2.2$\pm$0.2       & 50.90    & 100.0       & 0.0       & 0.0   & 14         \\
Holmberg I      & 1.22$\pm$0.09  & 0.85$\pm$0.07     & 30.23    & 90.1        & 8.8       & 1.1   & 91         \\
M81 DwB         & 2.38$\pm$0.05  & 1.40$\pm$0.03     & 26.87    & 100.0       & 0.0       & 0.0   & 37         \\
Holmberg II     & 1.55$\pm$0.06  & 0.70$\pm$0.03     & 17.61    & 98.32       & 1.34      & 0.34  & 597        \\
NGC 2366        & 2.91$\pm$0.09  & 0.66$\pm$0.02     & 17.18    & 99.14       & 0.37      & 0.49  & 817        \\
NGC 4214        & 9.24$\pm$0.42  & 0.52$\pm$0.02     & 57.65    & 99.44       & 0.40      & 0.16  & 1261       \\
NGC 1569        & 15.9$\pm$1.0   & 0.38$\pm$0.01     & 31.98    & 99.33       & 0.40      & 0.27  & 752        \\
IC 2574         & 6.8$\pm$0.4    & 0.94$\pm$0.04     & 36.44    & 100.00      & 0.00      & 0.00  & 333        \\
NGC 4449        & 3.73$\pm$0.03  & 0.422$\pm$0.001   & 21.54    & 97.39       & 2.50      & 0.11  & 2680       \\
NGC 2976        & 1.69$\pm$0.07  & 0.46$\pm$0.02     & 25.89    & 96.66       & 3.34      & 0.00  & 1313       \\
NGC 3077        & 2.66$\pm$0.09  & 0.75$\pm$0.03     & 26.73    & 92.88       & 6.96      & 0.16  & 632        \\
NGC 2403        & 6.42$\pm$0.02  & 0.555$\pm$0.001   & 31.88    & 97.94       & 2.00      & 0.06  & 6584       \\
NGC 925         & 2.15$\pm$0.03  & 1.10$\pm$0.02     & 19.06    & 98.44       & 1.51      & 0.05  & 1919       \\
NGC 3198        & 4.1$\pm$1.0    & 0.78$\pm$0.011    & 6.16     & 77.02       & 22.65     & 0.32  & 618        \\
NGC 4826        & 6.0$\pm$1.0    & 0.27$\pm$0.03     & 53.13    & 16.63       & 41.01     & 42.36 & 517        \\
NGC 4736        & 0.97$\pm$0.12  & 0.26$\pm$0.02     & 41.80    & 21.13       & 63.40     & 15.47 & 2049       \\
NGC 3184        & 1.30$\pm$0.03  & 0.97$\pm$0.03     & 11.40    & 70.02       & 29.14     & 0.84  & 2028       \\
NGC 5457        & 2.62$\pm$0.03  & 0.541$\pm$0.007   & 25.04    & 76.01       & 23.30     & 0.69  & 3785       \\
NGC 6946        & 2.12$\pm$0.08  & 0.409$\pm$0.007   & 9.55     & 87.01       & 11.76     & 1.23  & 3095       \\
NGC 5055        & 6.5$\pm$0.2    & 0.86$\pm$0.02     & 12.23    & 72.85       & 26.57     & 0.58  & 2074       \\
NGC 5194        & 1.25$\pm$0.03  & 0.421$\pm$0.006   & 9.55     & 40.98       & 46.26     & 12.76 & 7157       \\
NGC 3521        & 14.5$\pm$0.2   & 0.574$\pm$0.004   & 17.20    & 53.26       & 46.00     & 0.74  & 1763       \\
NGC 2841        & 2.13$\pm$0.11  & 0.56$\pm$0.03     & 22.93    & 26.87       & 54.64     & 18.49 & 1098       \\
NGC 7331        & 0.91$\pm$0.28  & 0.22$\pm$0.06     & 12.33    & 66.52       & 31.38     & 2.10  & 1765       \\
\hline

\end{tabular}
\caption{Parameters that characterized the DIG models and BPT diagnostic diagrams for each galaxy. The information given in each column is as follows: 2) indicates the $\rm H\alpha$ flux threshold for 100\% DIG dominance; 3) indicates the exponent, $\beta$, of each model; 4) shows the percentage of DIG fibres with regard to all considered fibres for each galaxy; 5,6,7) show the percentage of star-forming, composite and AGN fibres in the sample, respectively; 8) denotes the total number of fibres used for the BPT classification.}
\label{tab:DIG_ident}
\end{table*}

\section{Observations and data reduction}
   
The observations used in this paper are part of the Metal-THINGS survey \citep{LaraLopez2021}, devised as a follow-up on the THINGS survey \citep{Walter_2008_THINGS}, with galaxies that cover a wide range of stellar masses and present different properties and morphologies. 

\begin{figure*}[p]
\centering
        \includegraphics[scale=0.54]{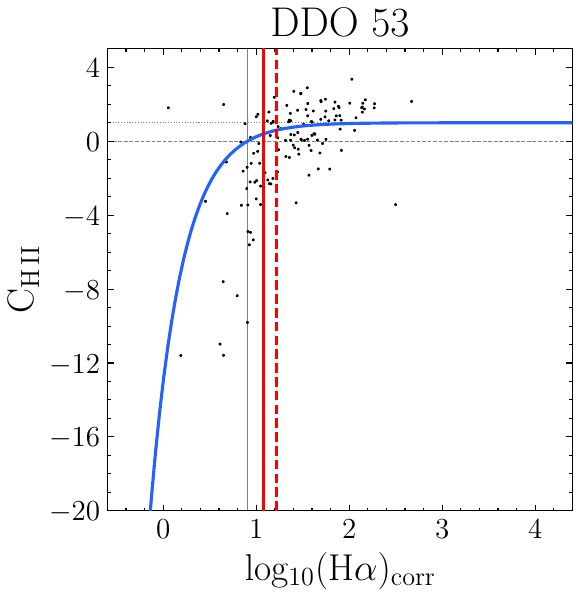}
        \hspace{0.35cm}
        \includegraphics[scale=0.54]{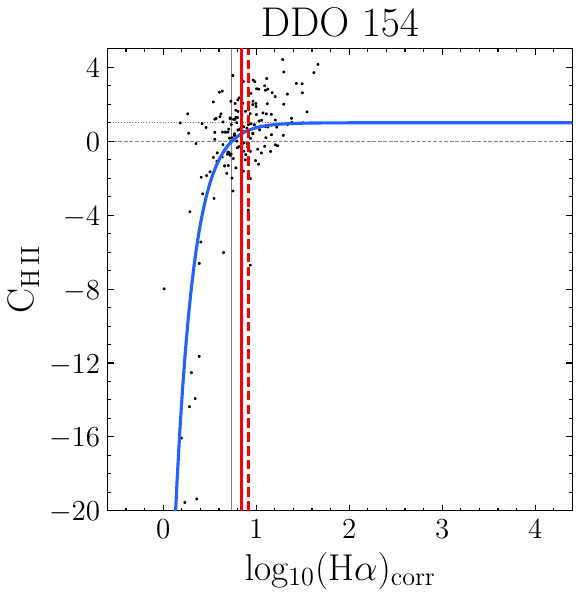}
        \hspace{0.35cm}
        \includegraphics[scale=0.54]{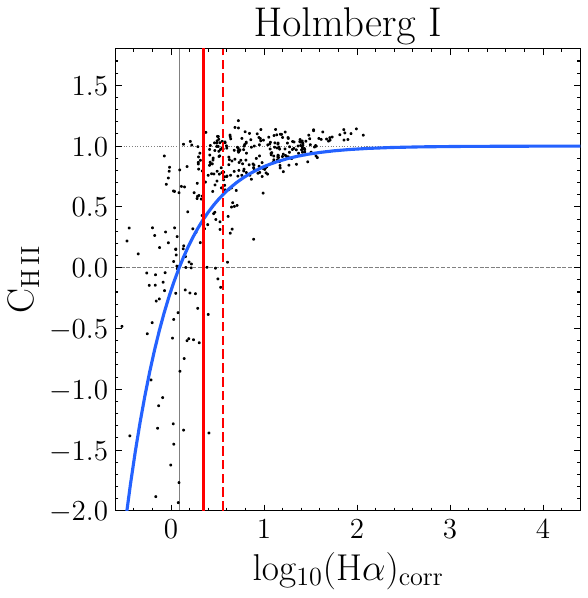}

        \vspace{0.15cm}

        \includegraphics[scale=0.54]{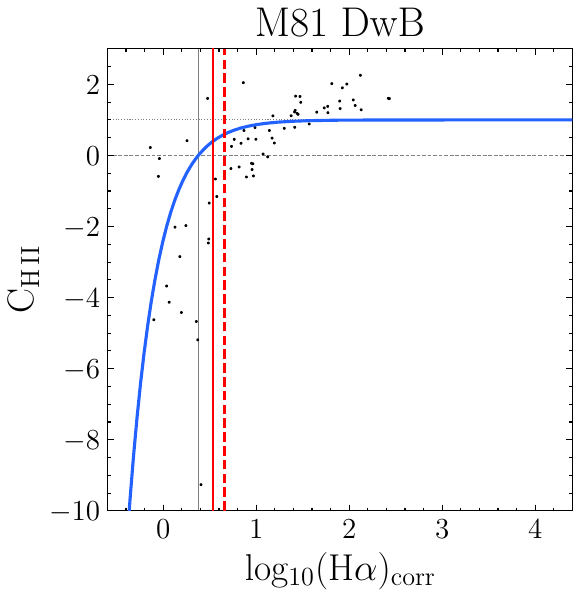}
        \hspace{0.35cm}
        \includegraphics[scale=0.54]{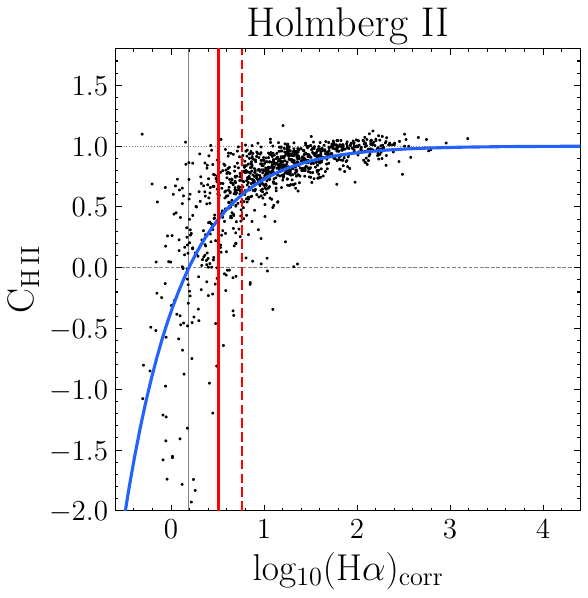}
        \hspace{0.35cm}
        \includegraphics[scale=0.54]{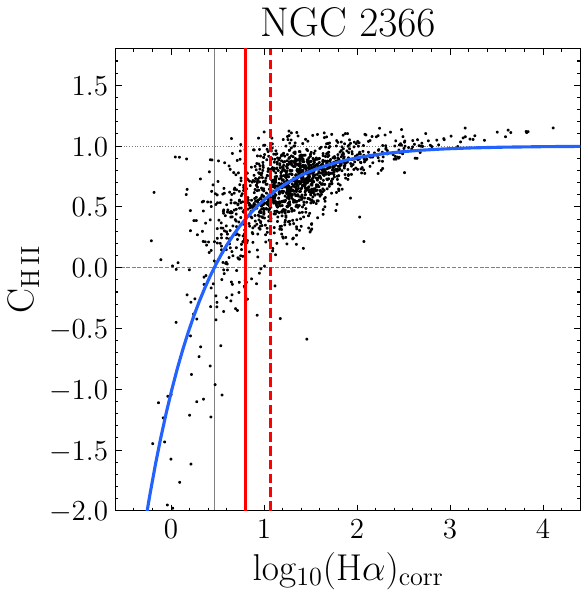}

        \vspace{0.35cm}

        \includegraphics[scale=0.54]{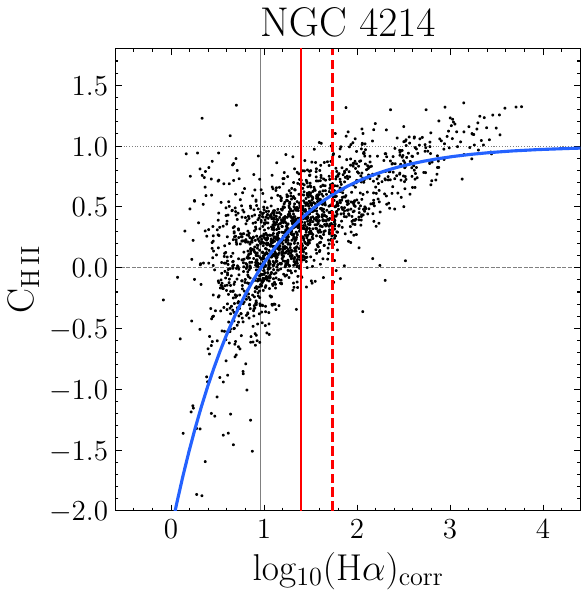}        
        \hspace{0.35cm}
        \includegraphics[scale=0.54]{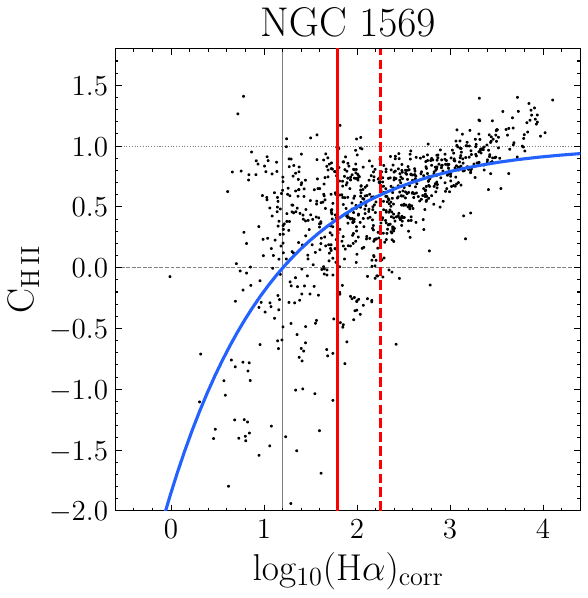}
        \hspace{0.35cm}
        \includegraphics[scale=0.54]{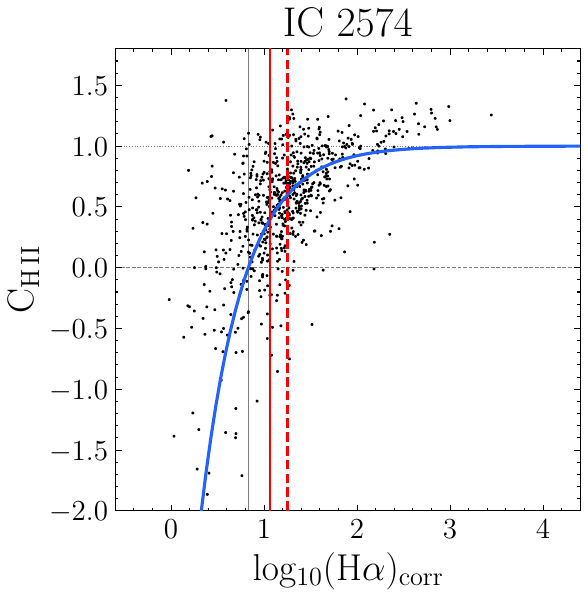}

        \vspace{0.15cm}

        \includegraphics[scale=0.54]{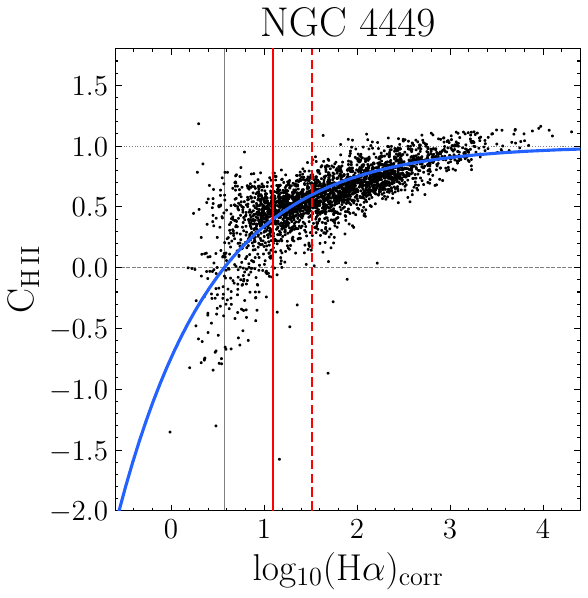}
        \hspace{0.35cm}
        \includegraphics[scale=0.54]{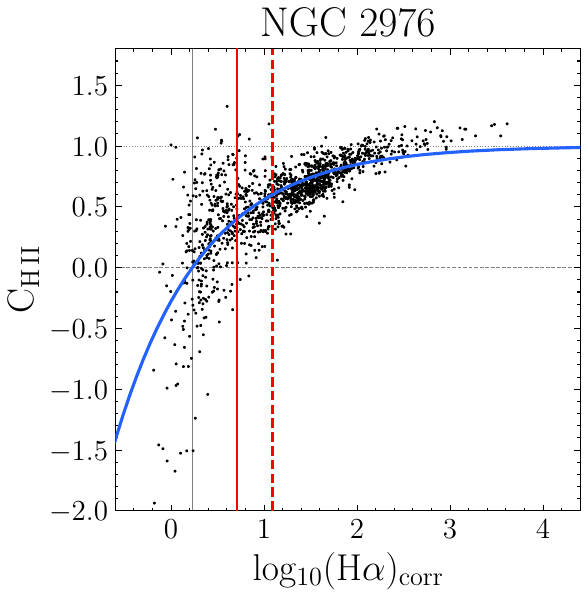}
        \hspace{0.35cm}
        \includegraphics[scale=0.54]{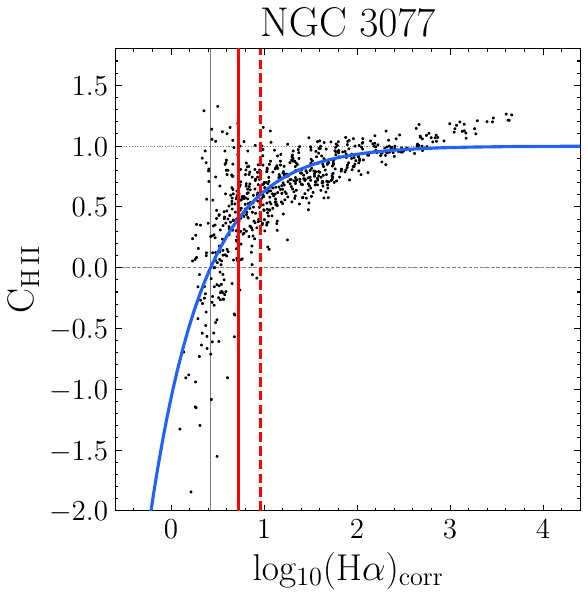}
        
         \caption{DIG models for each galaxy, where the $\rm C_{\rm H \, II}$ values and extinction corrected $\rm H\alpha$ flux are shown for each spaxel. The fit is shown with the blue line, and the red vertical line represents the $\rm H\alpha$ flux threshold for $\rm C_{\rm H \, II} < 0.4$ such that all fibres left of it are considered DIG. Solid grey vertical line represents the value of $\log_{10}(f_0)$ and the dashed grey horizontal lines the median value of $\rm C_{\rm H \, II}$ for the 100 brightest and 100 dimmest fibres. Dashed red vertical line represents the $\rm H\alpha$ flux threshold from \citet{kaplan_2016}, where the considered DIG fibres are those that satisfy $\rm C_{\rm H \, II} < 0.6$. Only fibres with a $\rm SNR > 3$ for the used lines are shown.}
         \label{fig:CHII_fit_1}
\end{figure*}

\begin{figure*}[p]
\vspace{0.5cm}
\centering
        \includegraphics[scale=0.54]{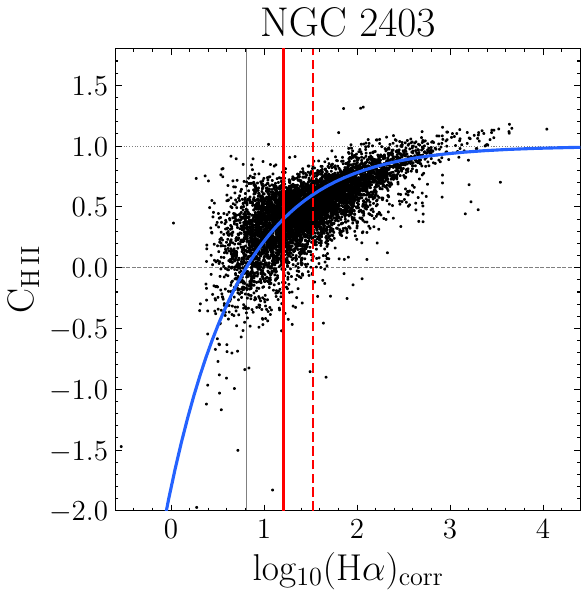}
        \hspace{0.35cm}
        \includegraphics[scale=0.54]{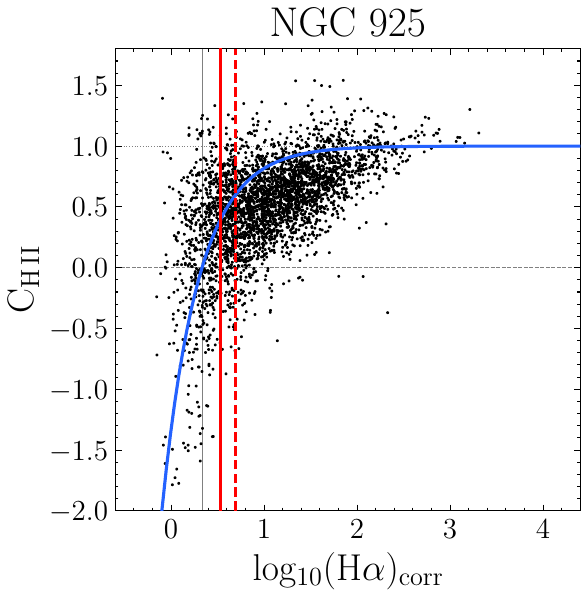}
        \hspace{0.35cm}
        \includegraphics[scale=0.54]{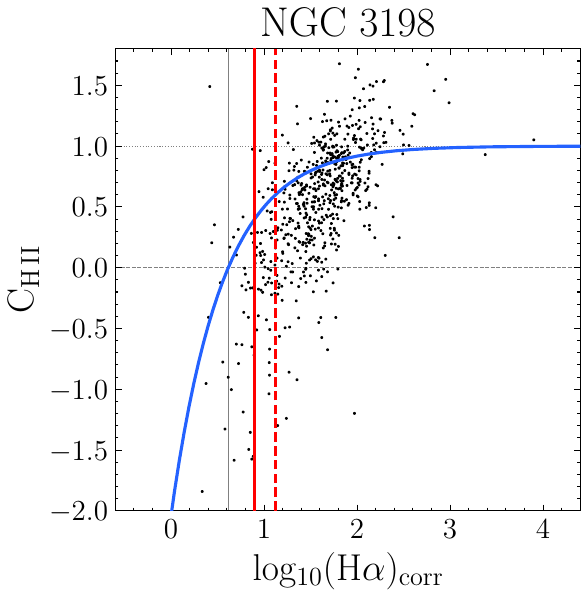}
        
        \vspace{0.3cm}

        \includegraphics[scale=0.54]{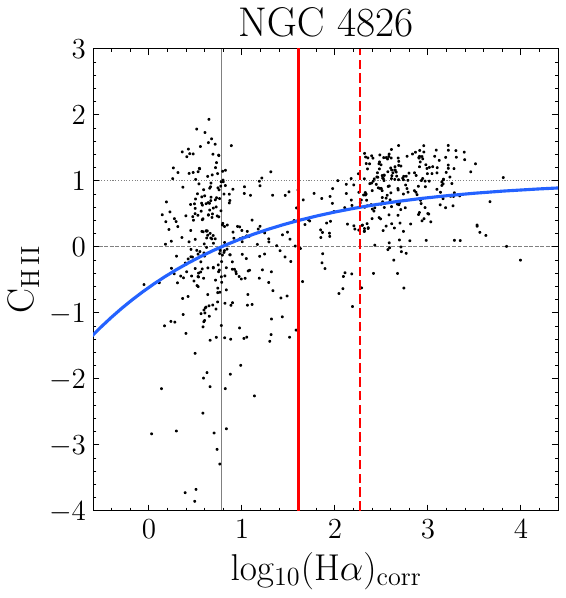}
        \hspace{0.35cm}
        \includegraphics[scale=0.54]{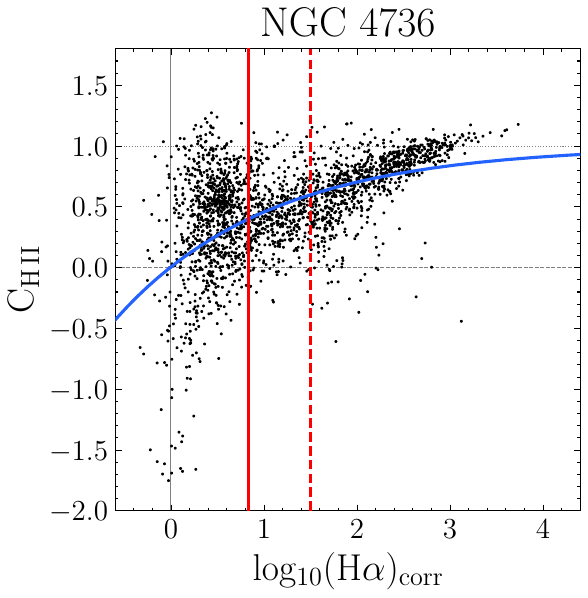}
        \hspace{0.35cm}
        \includegraphics[scale=0.54]{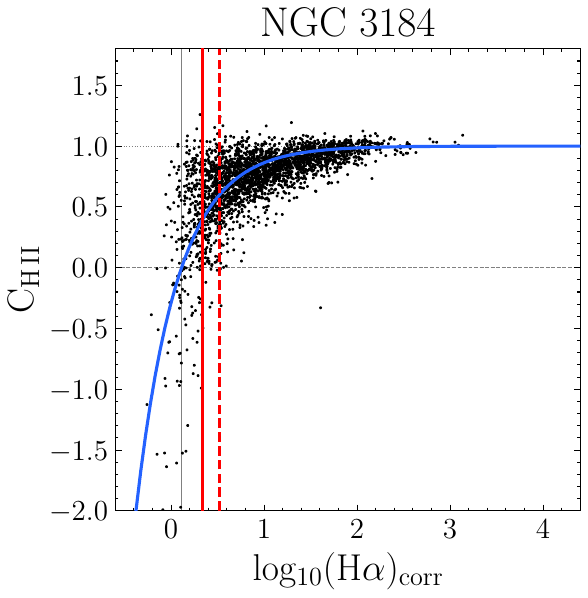}
        
        \vspace{0.3cm}
        
        \includegraphics[scale=0.54]{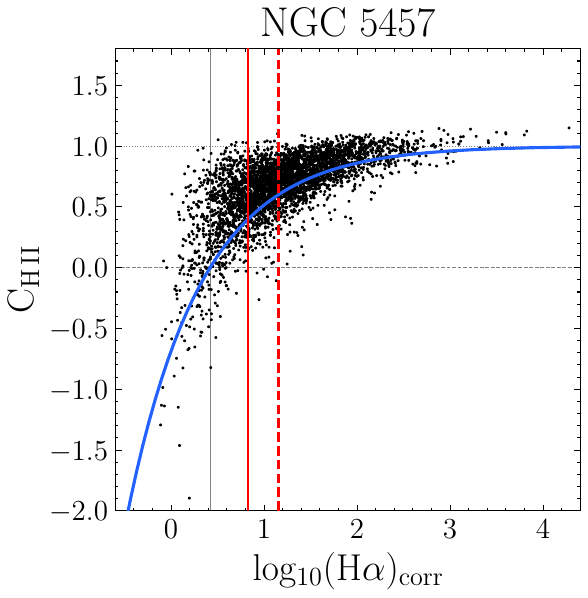}
        \hspace{0.35cm}
        \includegraphics[scale=0.54]{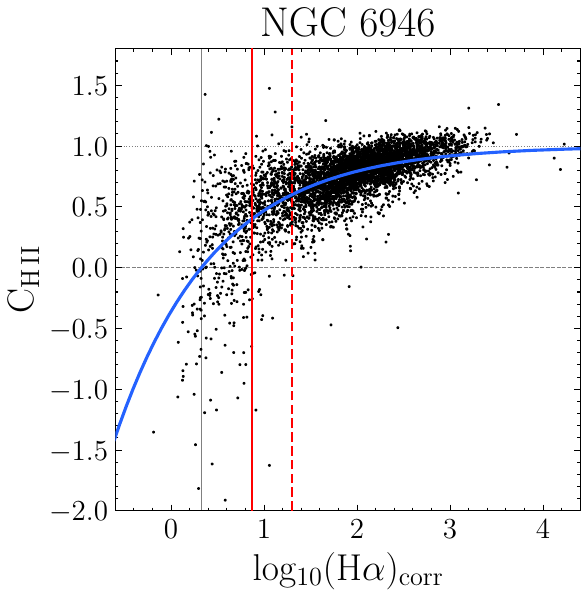}
        \hspace{0.35cm}
        \includegraphics[scale=0.54]{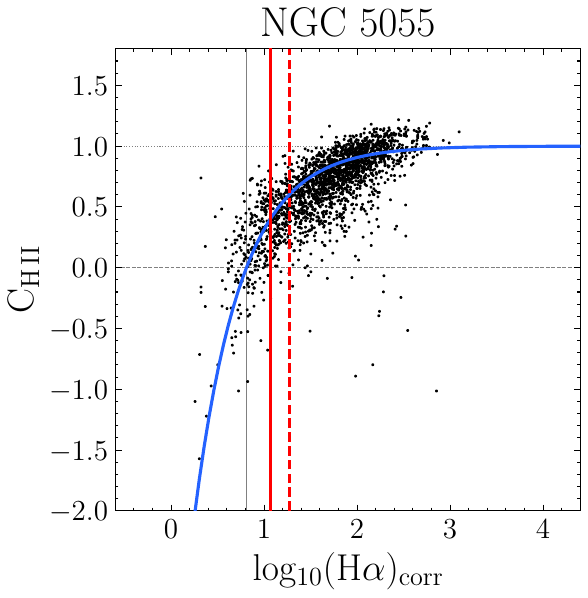}

        \vspace{0.3cm}

        \includegraphics[scale=0.54]{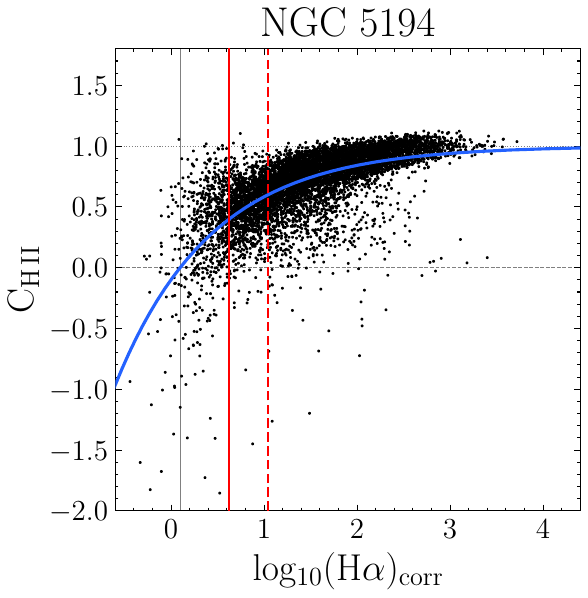}
        \hspace{0.35cm}
        \includegraphics[scale=0.54]{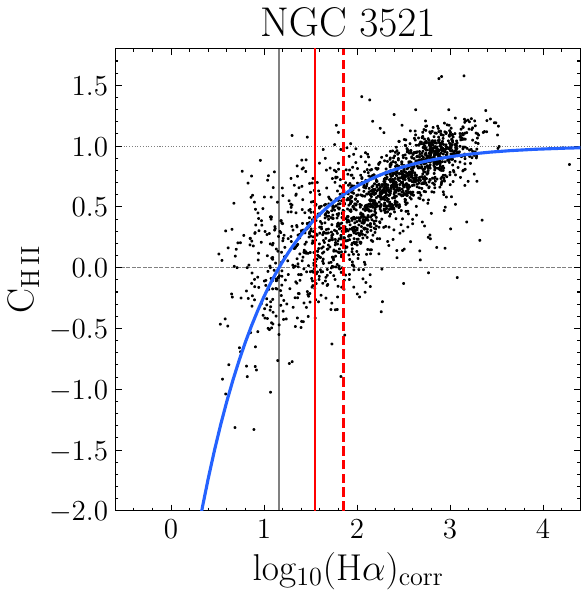}
        \hspace{0.35cm}
        \includegraphics[scale=0.54]{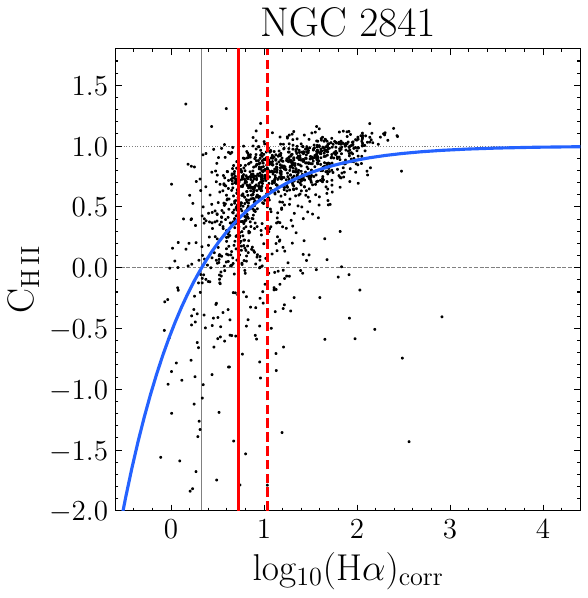}

        \caption{Same as Figure \ref{fig:CHII_fit_1}, but showing the second part of the sample}
         \label{fig:CHII_fit_2}
\end{figure*}

\begin{figure*}[t]
\centering
        \includegraphics[scale=0.54]{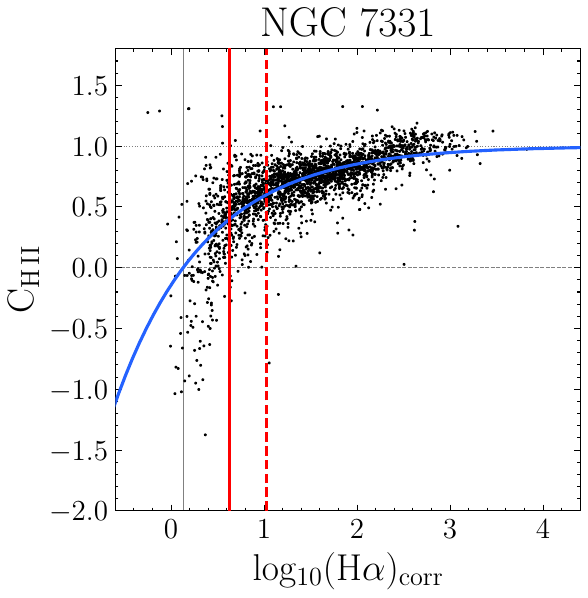}

        \caption{Same as Figures \ref{fig:CHII_fit_1} and \ref{fig:CHII_fit_2}, but showing the final part of the sample}
         \label{fig:CHII_fit_3}
\end{figure*}

All 25 galaxies in our sample were observed between the years 2017 and 2025 as described in Table \ref{tab:observation_table}, using the 2.7m Harlan Schmidt telescope at McDonald Observatory, Texas.
The Integral Field Unit (IFU) George and Cynthia Mitchell Spectrograph  \citep[GCMS,][]{Hill2008}  was used in the red setup with the low resolution grating VP1 with a resolution of 5.3 \AA. The average seeing during the observations was 1.2''. Due to the extended nature of our galaxies, we observed several partially overlapping pointings for each galaxy, as it is indicated in Table \ref{tab:observation_table}.

The GCMS is a square array of 100 $\times$ 102'', with a spatial sampling of 4.2'', and a 0.3 filling factor. The IFU consists of 246 fibres arranged in a fixed pattern. Every pointing is observed with three dither positions to ensure a 90\% surface coverage. Due to the extended nature of these galaxies, off-source sky exposures were taken for sky subtraction during the data reduction process. Every pointing was observed for 900 seconds per dither, followed by a sky exposure, and repetition of the same process until reaching a total of 45 minutes per dither. For fainter galaxies such as DDO 53 and DDO 154, we increased the exposure time up to an hour per dither to ensure a higher S/N and the detection of emission lines.

The data reduction was performed as described in  \citet{LaraLopez2021, LaraLopez2023}. The basic data reduction, bias subtraction, flat frame correction, and wavelength calibration was performed using P3D \footnote{https://p3d.sourceforge.io}. The rest of the data reduction including sky subtraction, flux calibration, combination of dithers, and mosaic generation was performed using our own routines in Python. Since we use several individual pointings for the same galaxy, astrometry is applied to each pointing. First, each pointing is converted into a collapsed data cube, then, we identify several stars and use the same star positions from the Two Micron All-Sky Survey \citep[2MASS,][]{Skrutskie2006}. Next, we apply the astrometry to each pointing using our own routines in Python, for more details see \citet{Garduno2023}. Finally, we assemble all the individual pointings and build a mosaic for our sample. 

The stellar continuum of all flux-calibrated spectra was fitted using STARLIGHT \citep{CidFernandes2005, Asari2007}, for a detailed description of this procedure see  \citet{Zinchenko2016}. In summary, to fit the continuum 45 simple stellar population (SSP) models from the evolutionary synthesis models of \citet{BruzualCharlot2003} were used, with ages from 1 Myr up to 13 Gyr and metallicities Z = 0.005, 0.02 and 0.05.  The fitted continuum was subtracted from the spectra, and the emission lines were measured using Gaussian line-profile fittings. In this paper, the individual spectra of single fibres are used, which are independent of each other. This is in contrast to other IFU data that rely on image fibres.

In Table \ref{tab:gal_data}, we gathered physical data for all galaxies in our sample. Here, for each galaxy, we report their morphology, right ascension ($\rm RA$), declination ($\rm DEC$), position angle ($\phi$), inclination ($i$) and distance ($D$) as specified in \citet{Leroy_2008}. For the stellar mass ($\log(M_{\rm star}/M_\odot)$), most values are taken from \citet{leroy_2019}, while those for Holmberg I, Holmberg II, NGC 4214, and DDO 154 are from \citet{leroy_2013};  DDO 53 from \citet{DDO53_mass}; and NGC 2366 from \citet{NGC2366_mass}. Additionally, their atomic gas mass ($\log M_{\rm H \, I}$) and molecular gas mass ($\log M_{\rm H_2}$), are mostly taken from \citet{Leroy_2008}, while NGC 4826 is from \citet{NGC4826_hi_mass}; and DDO 53, NGC 1569, NGC 2366, NGC 5457 and M81 DwB from \citet{Walter_2008_THINGS}.

As seen in Table \ref{tab:gal_data}, our sample of galaxies spans a range of stellar masses from $\sim 10^{7} M_\odot$ to $\sim 10^{11} M_\odot$, covering both low-mass and more massive galaxies, including irregular and disk galaxies. Distances to them vary from $2$ Mpc to $14.7$ Mpc, ensuring sufficient spatial resolution for the analysis of the entire sample. Additionally, our sample includes galaxies with a wide range of SFRs, spanning from 0.004 $ M_\odot \,\rm yr^{-1}$ to 6.05 $ M_\odot \, \rm yr^{-1}$ \citep{Walter_2008_THINGS}.

The Metal-THINGS survey offers a spatial resolution ranging from 40 to 300 pc, which stands out as one of its key strengths. In comparison, other major IFU surveys have lower spatial resolutions: in CALIFA the range is from 0.28 to 1.63 kpc \citep{califa_spatial_res}, in MaNGA it spans the range of 1.2 to 3.8 kpc \citep{MaNGA_spatial_res}, and SAMI from 0.21 to 4.7 kpc \citep{SAMI_spatial_res}. Only PHANGS achieves comparable resolution, ranging from 40 to 100 pc \citep{phangs_spatial_res}. This highlights the unique ability of Metal-THINGS to probe the internal structures of nearby galaxies at sub-kiloparsec scales.

\begin{figure*}[p]
\centering
        \includegraphics[scale=0.54]{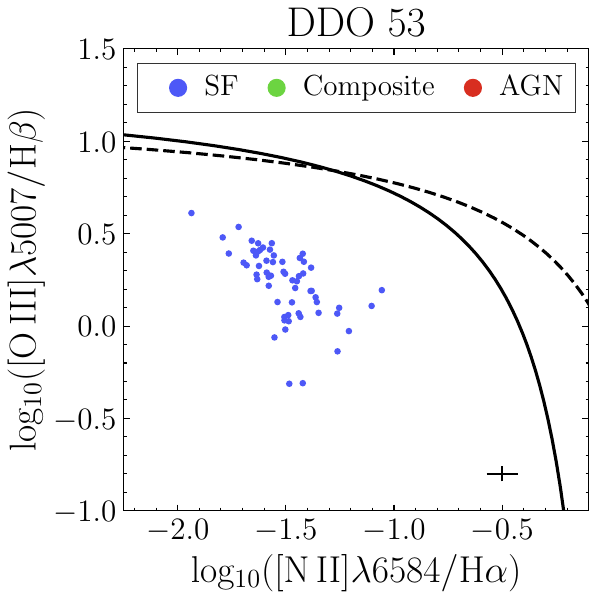}
        \hspace{0.35cm}
        \includegraphics[scale=0.54]{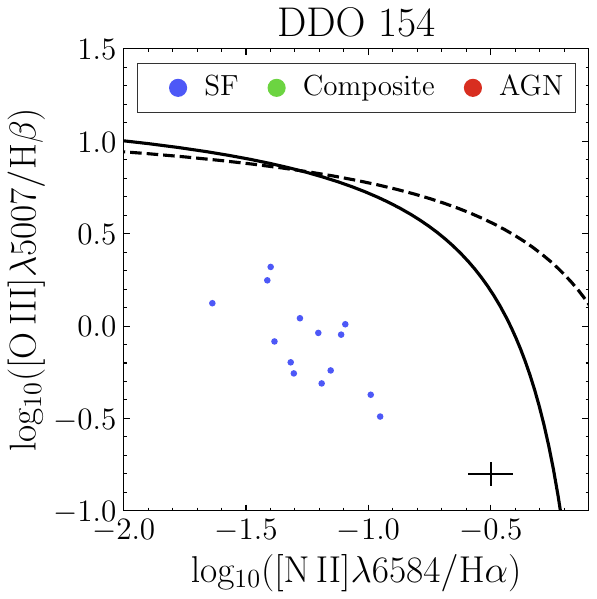}
        \hspace{0.35cm}
        \includegraphics[scale=0.54]{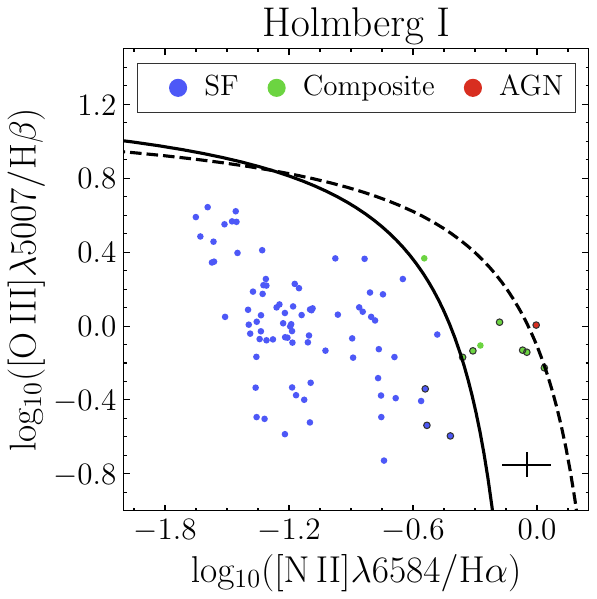}
        
        \vspace{0.25cm}

        \includegraphics[scale=0.54]{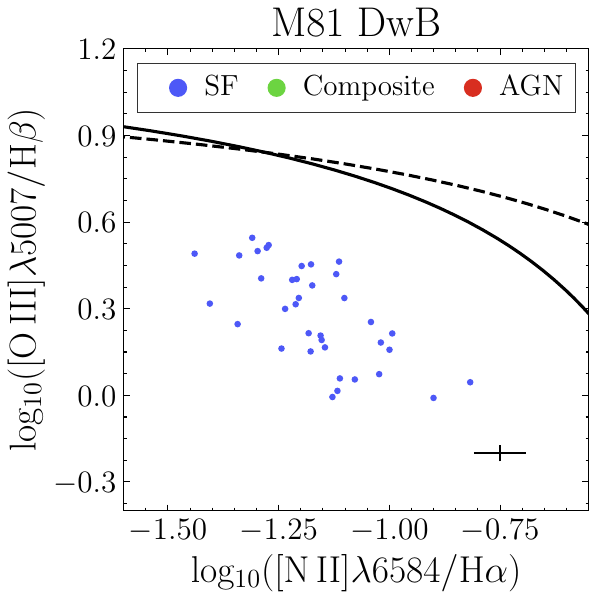}
        \hspace{0.35cm}
        \includegraphics[scale=0.54]{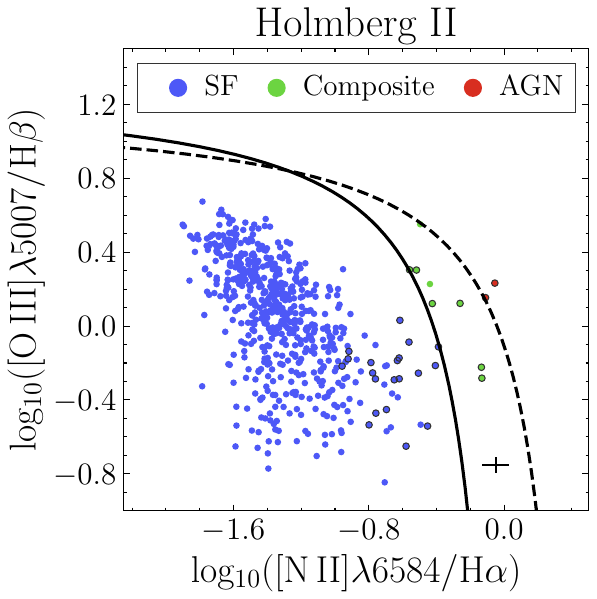}
        \hspace{0.35cm}
        \includegraphics[scale=0.54]{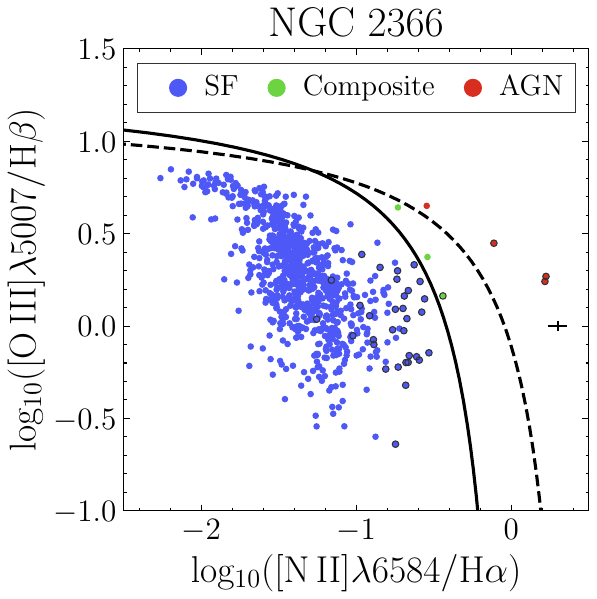}

        \vspace{0.25cm}

        \includegraphics[scale=0.54]{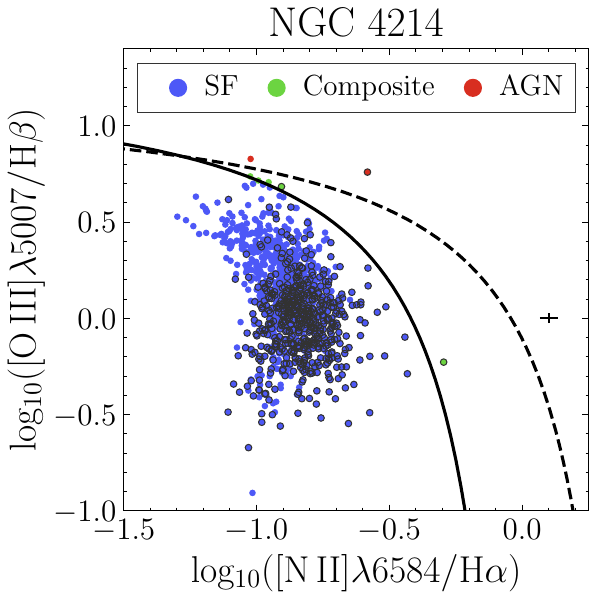}
        \hspace{0.35cm}
        \includegraphics[scale=0.54]{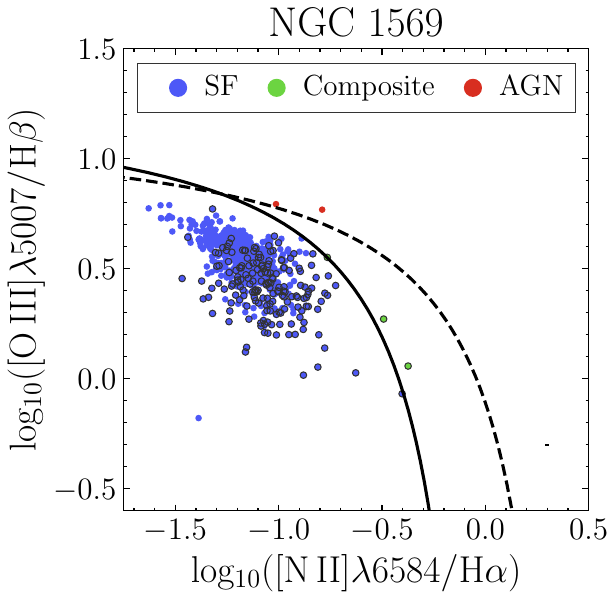}
        \hspace{0.35cm}
        \includegraphics[scale=0.54]{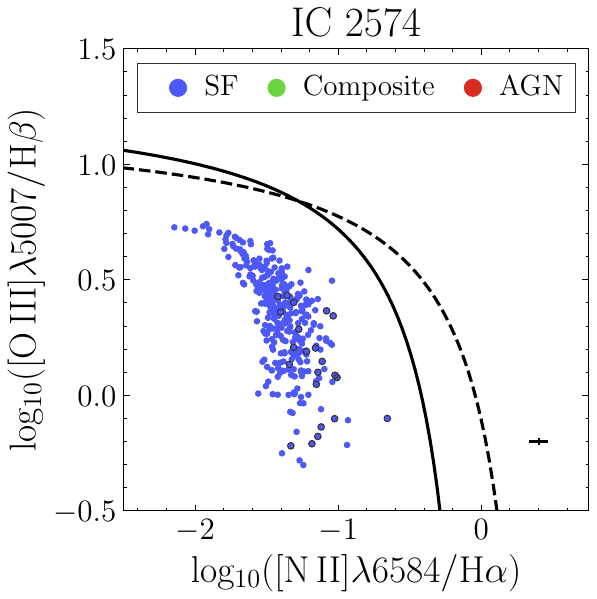}

        \vspace{0.25cm}

        \includegraphics[scale=0.54]{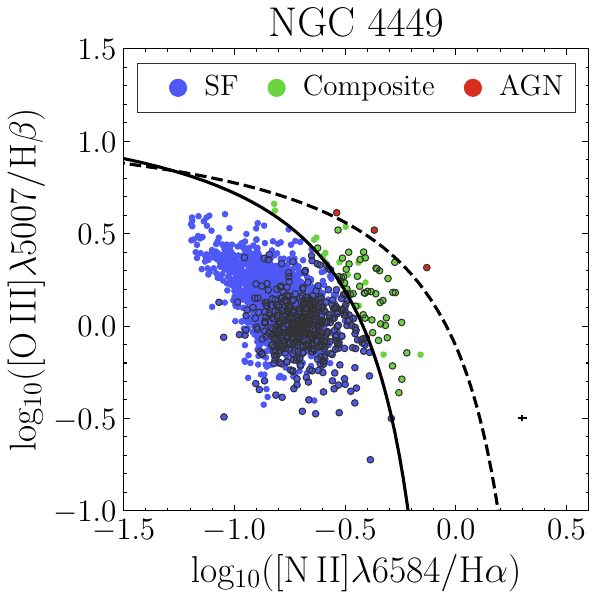}
        \hspace{0.35cm}
        \includegraphics[scale=0.54]{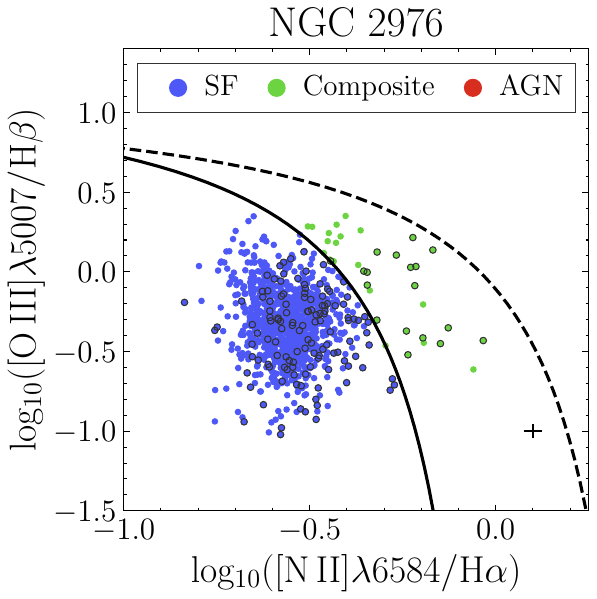}
        \hspace{0.35cm}
        \includegraphics[scale=0.54]{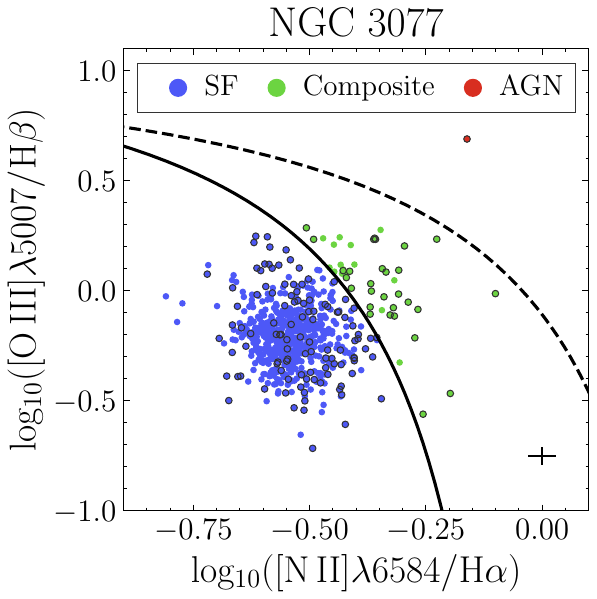}
        
         \caption{[N II] BPT diagrams for the observed galaxies, labelled accordingly, with SF-like (in blue), composite (in green) and AGN-like (in red) characteristics. Black curves, solid and dashed, represent the separation between the different regions and are defined in \citet{Kauffmann_2003} and \citet{Kewley_2001}. Additionally, the mean error in both axis is represented as a cross in the bottom right corner area of each diagram. Only fibres with a $\rm SNR > 3$ in all lines used are shown. Outlined points represent DIG fibres.}
         \label{fig:NII_BPT_1}
     \end{figure*}

\begin{figure*}[p]
\vspace{0.5cm}
\centering
        \includegraphics[scale=0.54]{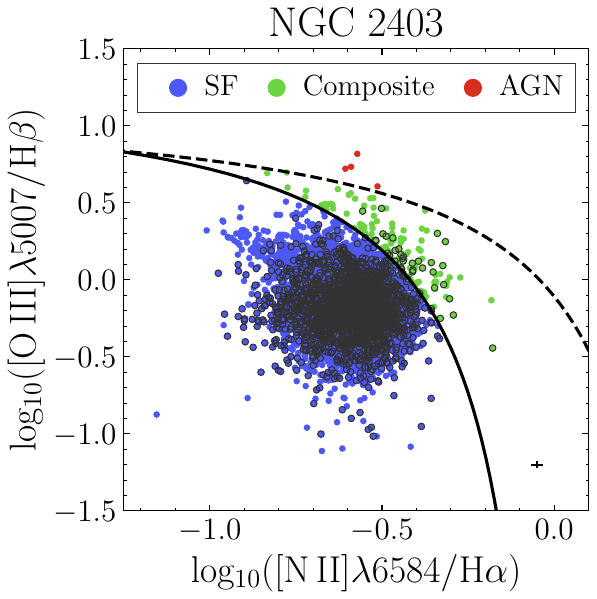}
        \hspace{0.25cm}
        \includegraphics[scale=0.54]{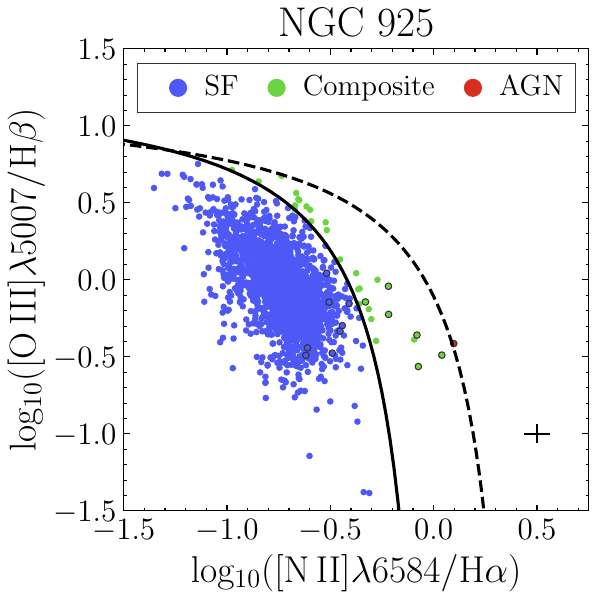}
        \hspace{0.25cm}
        \includegraphics[scale=0.54]{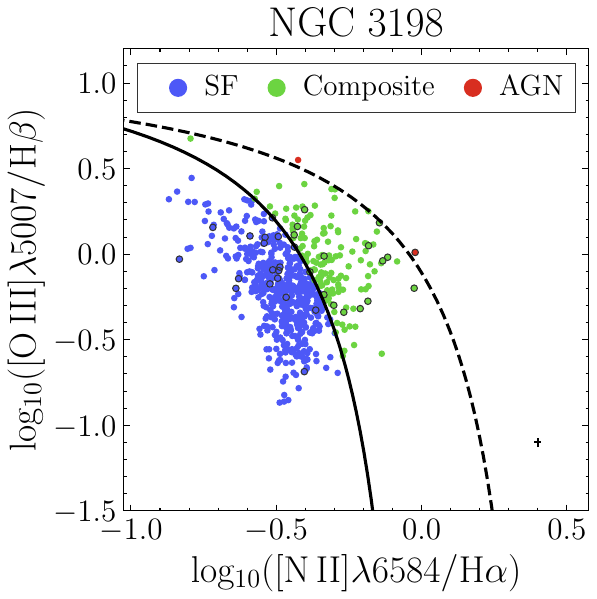}
        
        \vspace{0.3cm}

        \includegraphics[scale=0.54]{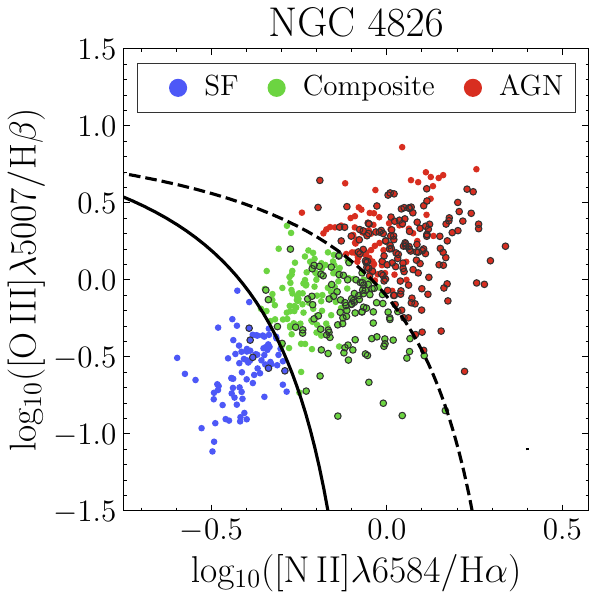}
        \hspace{0.25cm}
        \includegraphics[scale=0.54]{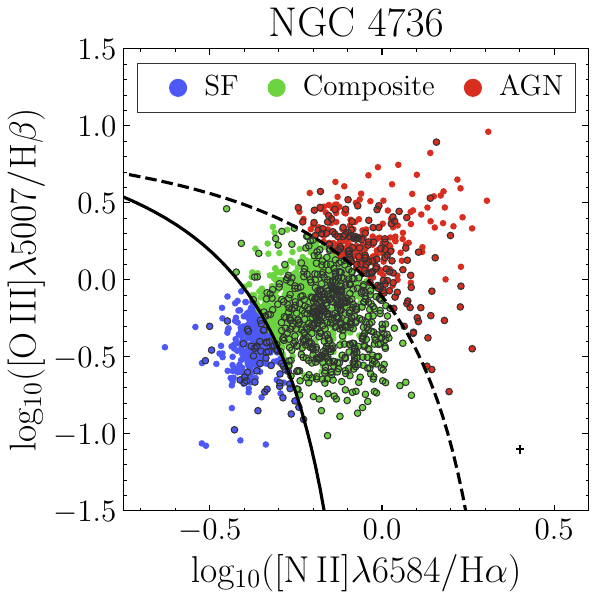}
        \hspace{0.25cm}
        \includegraphics[scale=0.54]{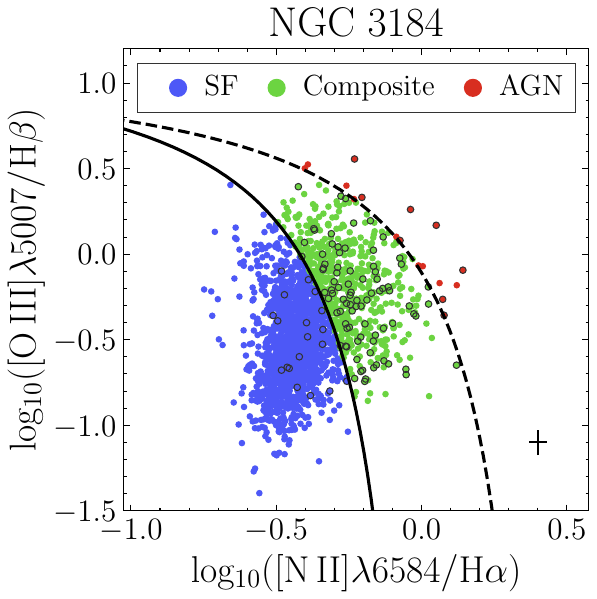}

        \vspace{0.3cm}
        
        \includegraphics[scale=0.54]{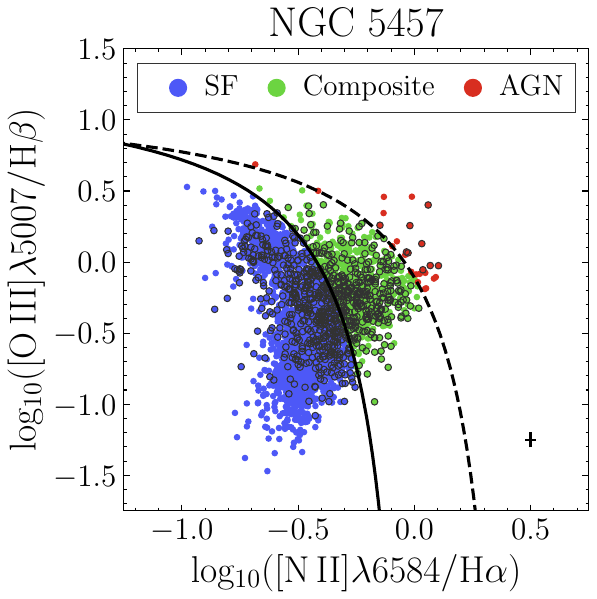}
        \hspace{0.25cm}
        \includegraphics[scale=0.54]{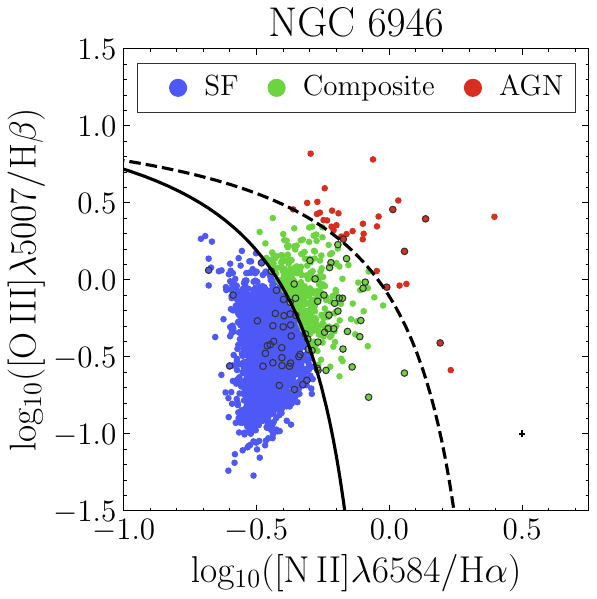}
        \hspace{0.25cm}
        \includegraphics[scale=0.54]{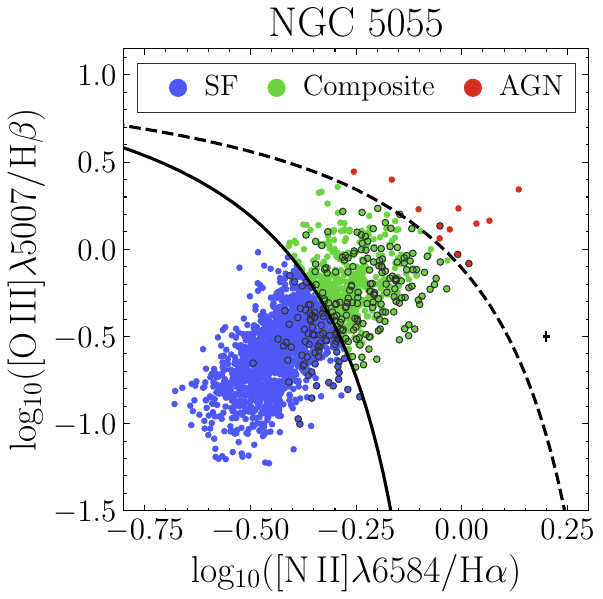}

        \vspace{0.3cm}
        
        \includegraphics[scale=0.54]{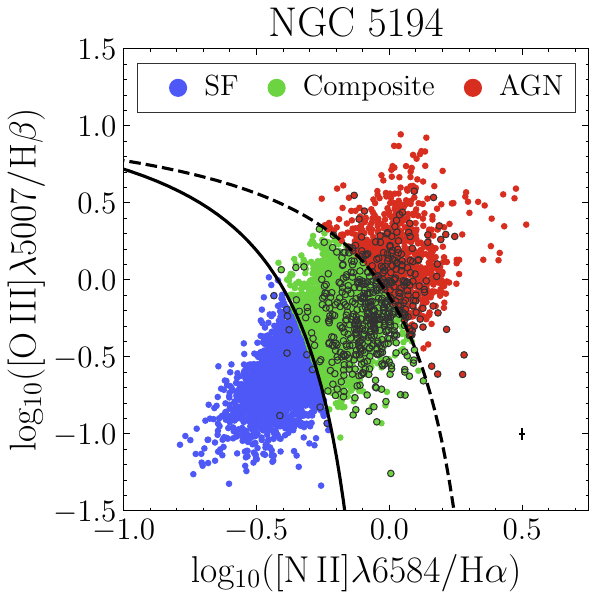}
        \hspace{0.25cm}
        \includegraphics[scale=0.54]{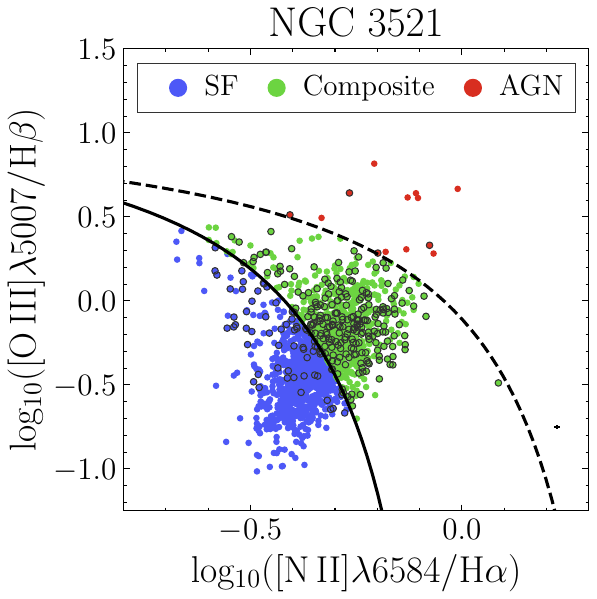}
        \hspace{0.25cm}
        \includegraphics[scale=0.54]{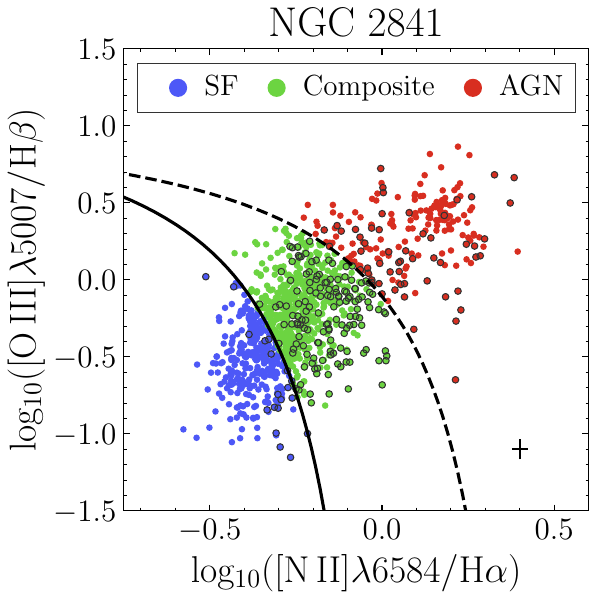}

        \caption{Same as Figure \ref{fig:NII_BPT_1}, but showing the second part of the sample}
        
         \label{fig:NII_BPT_2}
     \end{figure*}

\begin{figure*}[t]
\centering

        \includegraphics[scale=0.55]{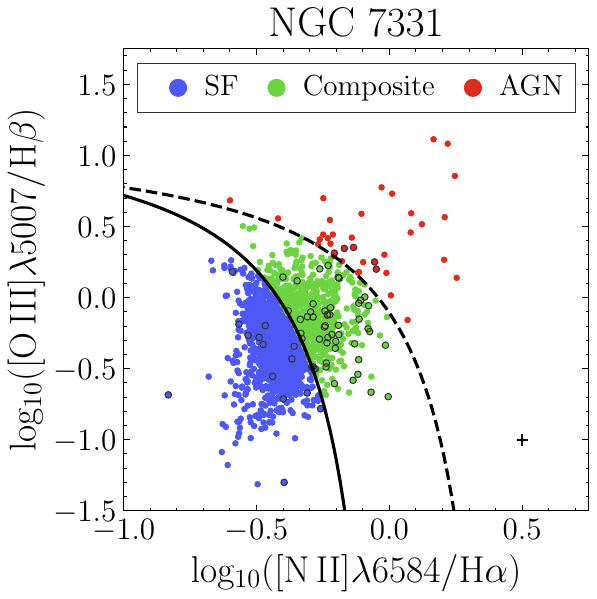}

        \caption{Same as Figures \ref{fig:NII_BPT_1} and \ref{fig:NII_BPT_2}, but showing the second part of the sample}
        
         \label{fig:NII_BPT_3}
     \end{figure*}

\section{Methodology}

\subsection{Extinction correction and DIG identification}

The extinction correction was derived using the Balmer Decrement in order to obtain the reddening coefficient C(H$\beta$). Case B recombination was assumed, with a density of 100 cm$^{-3}$ and a temperature of 10$^4$K. The predicted ratio unaffected by reddening or absorption correction of H$\alpha$/H$\beta$ is 2.86 according to \citet{Osterbrock_1989}. Hence the coefficient is given by:
\begin{align*}
    C({\rm H\beta}) &= - \frac{1}{f({\rm \lambda})} \log \left(  \frac{I(H\alpha)}{I(H\beta)} \middle/ \frac{\mathcal{F}({\rm H\alpha})}{\mathcal{F}({\rm H\beta})} \right)
\end{align*}

where $\mathcal{F}({\textrm{H}\alpha})$ and $\mathcal{F}({\textrm{H}\beta})$ are the observed fluxes, and $f({\rm \lambda})$ is the reddening curve normalised to H$\beta$ using the \citet{Cardelli_1989} extinction curve. When ${I({\rm H\alpha})}/{I({\textrm{H}\beta})}$ is lower than the theoretical value, then the adopted reddening is zero. Given that Balmer emission line fluxes were corrected for underlying stellar absorption using SED models obtained by Starlight, no further correction is applied. Extinction corrected $\rm H\alpha$ maps of the observed galaxies are shown in figures \ref{fig:Ha_maps_1} and \ref{fig:Ha_maps_2}. All fluxes were initially taken in units of $10^{-16} \; {\mathrm{erg/s/cm^2/}}$\AA.

{Diffuse Ionised Gas} (hereafter DIG), also known as the warm ionised medium (WIM), is a warm ($\sim$$10^4 \rm\, K$), low density ($\sim$$10^{-1} \rm\, cm^{-3}$) gas phase found in the interstellar medium of galaxies \citep[e.g.][]{kaplan_2016}.

The existence of this gas component provides evidence for significant ionisation of the interstellar gas outside bright, localized H \tiny{II}\normalsize{} regions \citep[][]{Reynolds_1984}. These wide spread regions are distributed throughout the galactic disc \citep{Reynolds_1983} but further research has determined that DIG extends up to two kiloparsecs above and below the plane of the spiral galaxy's disc \citep{Madsen_2006, Haffner_2009}. Its low-level emission is a possible source of contamination superimposed over emission from H \tiny{II}\normalsize{} regions.

Despite generally not being a main component of the interstellar medium, DIG is significant in star-forming galaxies, and its contribution to emission line fluxes can impact the interpretation of measured metallicities, ionisation and BPT diagrams. Because of its potential importance, the  effect of DIG on the metallicity gradients will be studied here.

To identify DIG we follow the methodology of \citet{kaplan_2016}, in which the observed ${\rm H\alpha}$ flux is described as $f({\rm H\alpha}) = C_{\rm H \, II} \, f({\rm H\alpha}) + C_{\rm DIG} \, f({\rm H\alpha})$, where $f(\rm H\alpha)$ is the $\rm H\alpha$ flux measured in a given fibre, and $C_{\rm H \rm II}$ and $C_{\rm DIG}$ are the estimated fractions of $\rm H\alpha$ emission from {H \small II\normalsize} and DIG regions, respectively, with $C_{\rm H\, II} = 1 - C_{\rm DIG}$.

First, using all fibres where the $\rm H\alpha$ has $\rm S/N > 3$, we select the 100 brightest and dimmest fibres in $\rm H\alpha$ to find the characteristic values of ([S \tiny{II}\normalsize]/$\rm H\alpha$)$_{\rm H \tiny{II} \normalsize}$ and ([S \tiny{II}\normalsize]/$\rm H\alpha$)$_{\rm DIG}$ in order to calculate an initial guess for individual $C_{\rm H\, II}$ values, as specified in \citet{kaplan_2016}. With this first estimate of $C_{\rm H\, II}$, we can solve for one value of $f_0$ and $\beta$ for each galaxy fitting the data to the following equation, weighting it with the uncertainty in the $\rm H\alpha$ line flux
\begin{equation}
    C_{\rm H \, II} = 1.0 - \left( \frac{f_0}{f({\rm H\alpha})} \right)^{\beta} \; \; \; ; \; \; \; (\text{for } f({\rm H\alpha}) > f_0)
    \label{CHII_eq}
\end{equation}
where $f_0$ is the limit value of $f(\rm H\alpha)$ such that 100\% of the flux comes from DIG and $\beta$ accounts for the variation in surface brightness of the DIG.

We now solve for a final $C_{\rm H \, II}$ with $f_0$ and $\beta$. We reuse equation \ref{CHII_eq} and each fibre's $f(\rm H\alpha)$ to calculate a final $C_{\rm H \, II}$ for each fibre where $f({\rm H\alpha}) > f_0$. In this study, we employ two different thresholds for the identification of DIG fibres, namely $C_{\rm H\, II} < 0.4$ and $C_{\rm H\, II} < 0.6$, to evaluate its contribution to metallicity gradients. The emission of fibres with $C_{\rm H\, II} < 0.4$ will be considered to be dominated by DIG, whereas a threshold of $C_{\rm H\, II} < 0.6$ is adopted from \citet{kaplan_2016} and is used to briefly compare the effect of varying the threshold on the estimation of metallicity gradients.

In the case of NGC 4736 (Fig. \ref{fig:CHII_fit_2}, second row, second column), we can see a clump of fibres in the upper left region of the $C_{\rm H \, II}$ diagram. This behaviour was further studied and we conclude that these do not belong to a particular pointing, but to outer regions of the galaxy where the concentration of DIG is greater.

\subsection{Diagnostic BPT diagrams} \label{bpt_diagnostics}

BPT diagrams are a diagnostic tool for empirically  distinguishing between active galactic nuclei (AGN) and star-forming (SF) emission using line flux ratios\break ([O III]/$\rm H\beta$, [N II]/$\rm H\alpha$, [S II]/$\rm H\alpha$, and [O I]/$\rm H\alpha$). The classification schemes used in this paper are based on the [O III]/$\rm H\beta$ versus [N II]/$\rm H\alpha$ line ratio diagrams, hereafter [N II] BPT diagram.

We selected SF-type regions following the criteria given by \citet{Kauffmann_2003} in the [N II] BPT diagram, AGN-like regions follow the criteria from \citet{Kewley_2001}, and composite regions as those in-between both demarcations.

Using the outlined classification schemes, the BPT diagrams for our galaxies are shown in Figures \ref{fig:NII_BPT_1}, \ref{fig:NII_BPT_2} and \ref{fig:NII_BPT_3}, and its respective percentage of SF, Composite and AGN is shown in Table \ref{tab:DIG_ident}. Analysing our results, we find that, as expected, low-mass galaxies are primarily composed of SF-type regions, while the percentage of AGN-like regions increases in more massive galaxies.

It is important to note that the classification of fibres as ``AGN-like'' does not necessarily indicate the presence of an AGN, as DIG, shocks or post-asymptotic giant branch (pAGB) stars can produce line ratios that mimic AGN emission when located in a BPT diagram \citep[e.g.][]{bpt_agn_paper}.

\subsection{Estimation of gas metallicities}  \label{gas_metallicity}

In this work, the gas metallicities and ionisation parameter are estimated for the extinction corrected IFS data, specifically for SF-type fibres selected from the\break [N II] BPT diagram (see Section \ref{bpt_diagnostics}), and with S/N $>$ 3 for all the required emission lines. The following notation will be used for the line ratios and excitation parameter ($P$) as defined by \citet{Pilyugin_2016}.

\begin{align}
    \label{eq:2_25}
        &S'_2 = I_{\rm [S \, II]\lambda\lambda 6717,31} / I_{\rm H\alpha} \nonumber
        &S_2 = I_{\rm [S \, II]\lambda\lambda 6717,31} / I_{\rm H\beta}  \nonumber \\
        &N_2 = I_{\rm [N \, II]\lambda\lambda 6548,84} / I_{\rm H\beta} \nonumber
        &R_2 = I_{\rm [O \, II]\lambda\lambda 3927,29} / I_{\rm H\beta} \nonumber \\
        &R_3 = I_{\rm [O \, III]\lambda\lambda 4959,5007} / I_{\rm H\beta} \nonumber
        &R_{23} = R_2 + R_3 \nonumber \\
        &P = R_{3} / R_{23} = R_3 / (R_2 + R_3)
\end{align}

We use the calibration of \citet[][hereafter, PG16]{Pilyugin_2016}, which uses the strong lines $N_2$, $S_2$ and $R_3$ as defined in Eq. (\ref{eq:2_25}). The basis of this method is that the $N_2$ ratio is an indicator of $T_{\rm e}$ in H\small II \normalsize regions, so that the $T_{\rm e}-\log (N_2)$ relation has a dependence on the excitation parameter $P$ (Eq. \ref{eq:2_25}) at high electron temperatures. The oxygen and nitrogen abundances both present clear monotonic relations with $\log(N_2)$ with both relations exhibiting a dependence on $P$ at high and low metallicities X/H (i.e. high $T_{\rm e}$) where X/H may denote the abundances $12 + \log(O/H)$ and $12 + \log(N/H)$ \citep{Pilyugin_2016}.

To account for the electron temperature $T_{\rm e}$ and the level of excitation parameter dependencies, \citet{Pilyugin_2016} divides its metallicity calibration into two branches at $\log(N_2) = -0.6$, with an upper branch $(O/H)^*_{\rm S, U}$ ($\log(N_2) \geq -0.6$) and a lower branch $(O/H)^*_{\rm S, L}$ ($\log(N_2) < -0.6$). The resulting calibrations are based on the sulphur $S_2$ ratio as noted in equation \ref{eq:2_25} (thereby the "S calibration"), and takes into account the correlation with $P$, where we denote the metallicity as ${\rm (O/H)}^* = 12 + \log({\rm O/H})$.

It should be noted that the PG16 calibration is compatible with the typical $T_{\rm e}$ metallicity scales of H \small II \normalsize regions, and in general shows a smaller scatter compared to calibrations using auroral lines such as [O III]$\lambda4363$, [N II]$\lambda5755$, and\break [S III]$\lambda6312$ \citep{Pilyugin_2016}.

The gas-phase metallicity maps estimated in this manner can be found in \ref{fig:metmaps_1} and \ref{fig:metmaps_2}. The gas metallicities of fibres identified as being dominated by DIG was estimated in the same manner, and it is flagged with green circles in the same maps.

\begin{figure*}[t]
     \centering
     \includegraphics[width=0.65\textwidth]{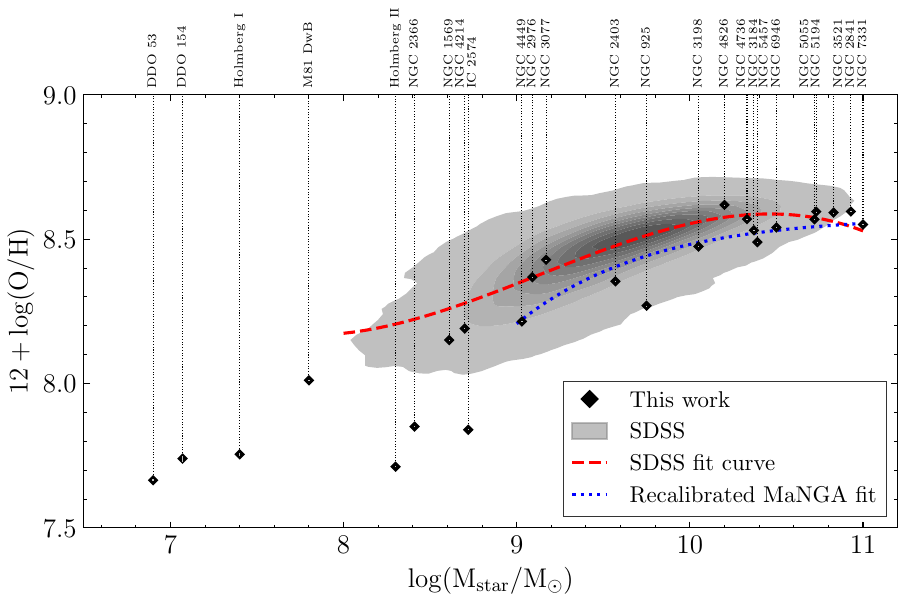}
     \caption{Mass $-$ metallicity relation of 69,830 SDSS galaxies (in the redshift range $0.04 < z < 0.2$), shown with a Kernel Density Estimate (KDE) plot as the shaded region. Markers indicate galaxies in our sample. The red dashed line represents a third degree fit to SDSS data. The dotted blue line indicates the MMR calculated for the MaNGA survey \citep{barreraballesteros_2017} using the O3N2-M13 indicator \citep{O3N2_M13_paper}, and recalibrated to the PG16 indicator (see Appendix \ref{other}).}
     \label{fig:met-stellmass}
\end{figure*}

\subsection{The Mass - Metallicity relation} \label{mass_met_rel_section}

To further characterize our sample, we located our galaxies in the so called Mass-Metallicity Relation, MMR. Figure \ref{fig:met-stellmass} illustrates this relationship using data for 69,830 galaxies from the SDSS (as the shaded region). For these galaxies, a maximum value for their redshift was set to $z = 0.2$ and their minimum to $z = 0.04$, with the intention of having a sample representative of the local universe. Emission-line fluxes of the SDSS galaxies were taken from the OSSY catalogue\footnote{\url{https://data.kasi.re.kr/vo/OSSY/index.html}} \citep{Oh_2011}.

\begin{figure*}[p]
\vspace{0.5cm}
\centering
        \includegraphics[scale=0.5]{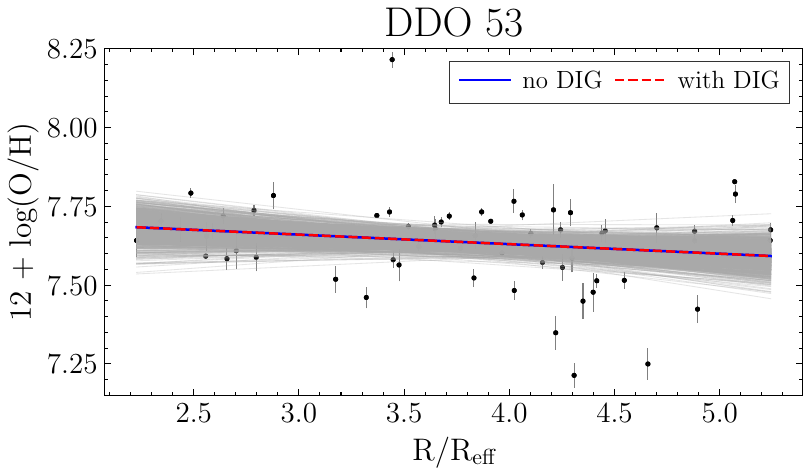}
        \hspace{2cm}
        \includegraphics[scale=0.5]{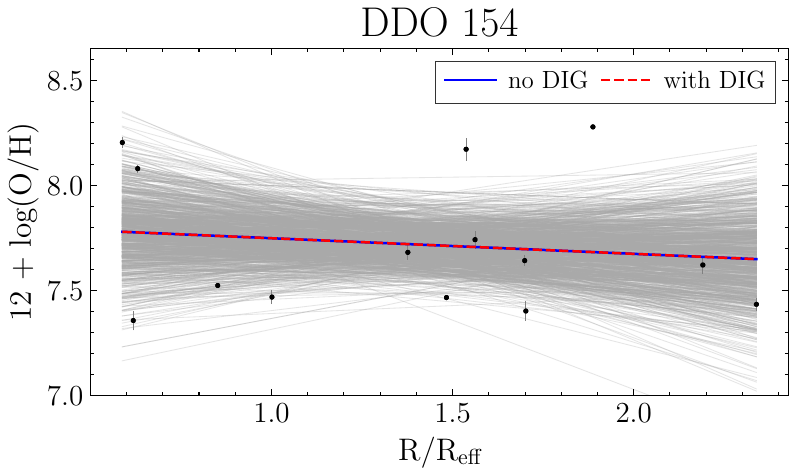}

        \vspace{0.25cm}

        \includegraphics[scale=0.5]{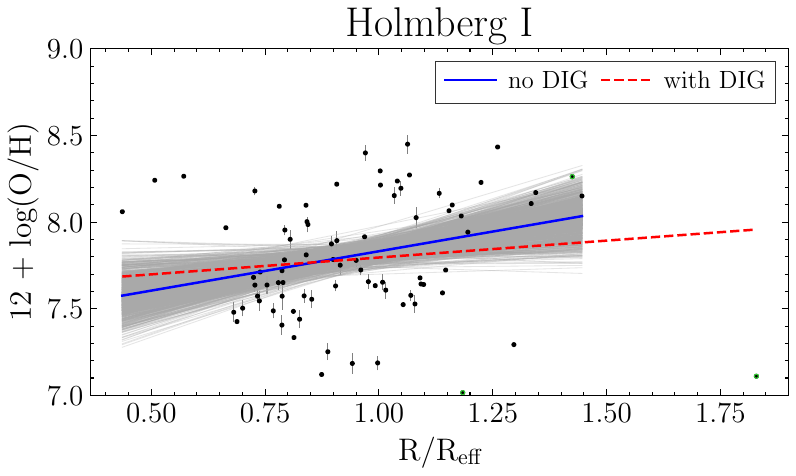}
        \hspace{2cm}
        \includegraphics[scale=0.5]{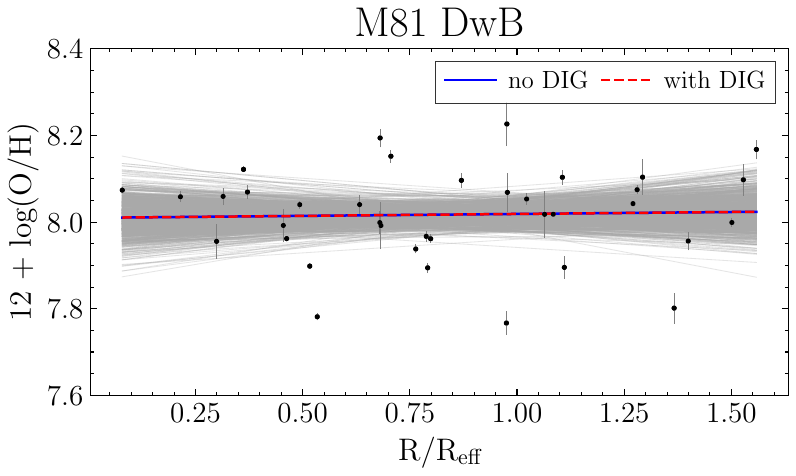}
        
        \vspace{0.25cm}

        \includegraphics[scale=0.5]{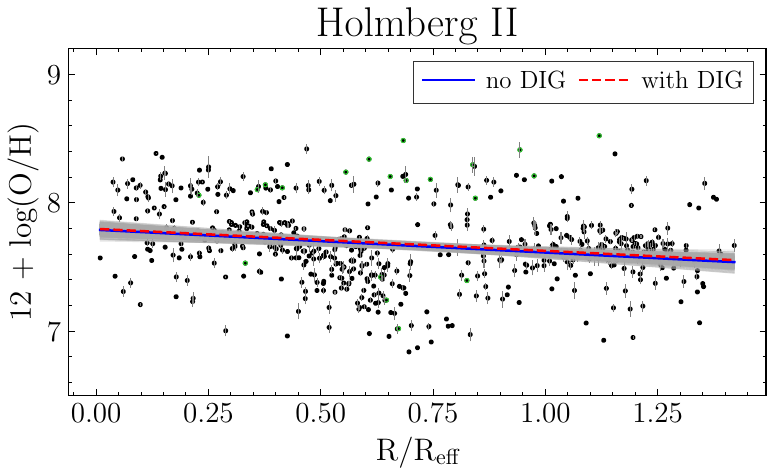}
        \hspace{2cm}
        \includegraphics[scale=0.5]{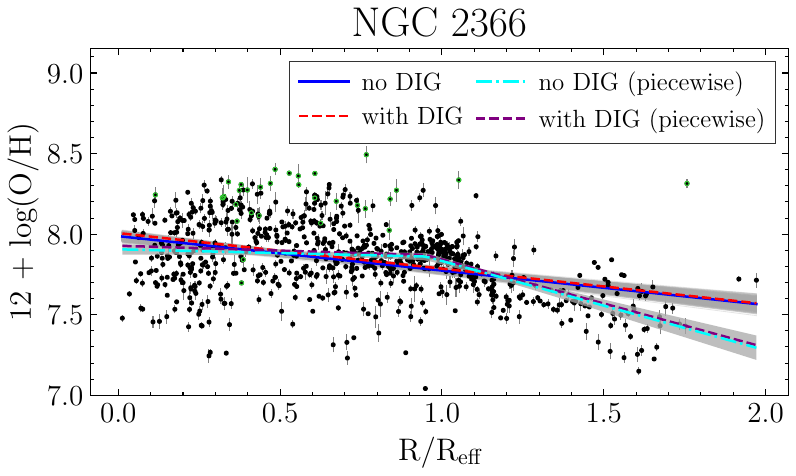}

        \vspace{0.25cm}

        \includegraphics[scale=0.5]{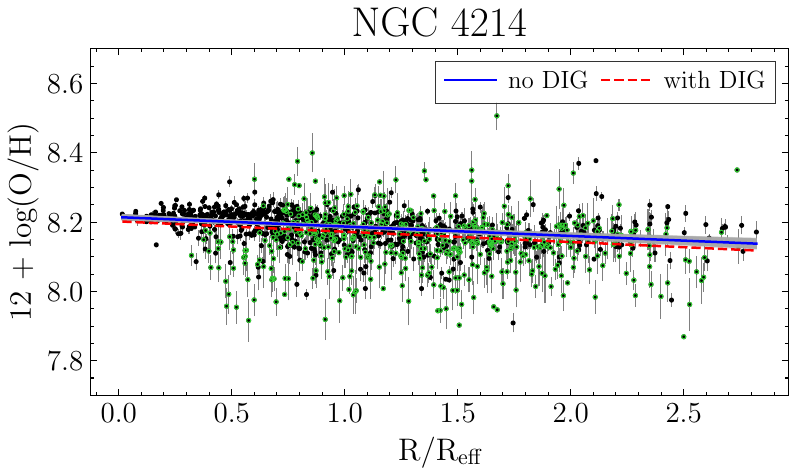}
        \hspace{2cm}
        \includegraphics[scale=0.5]{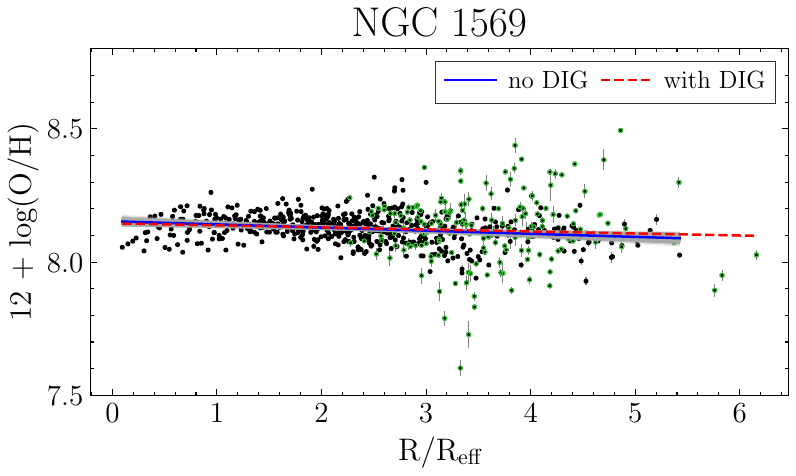}
        
        \vspace{0.25cm}

        \includegraphics[scale=0.5]{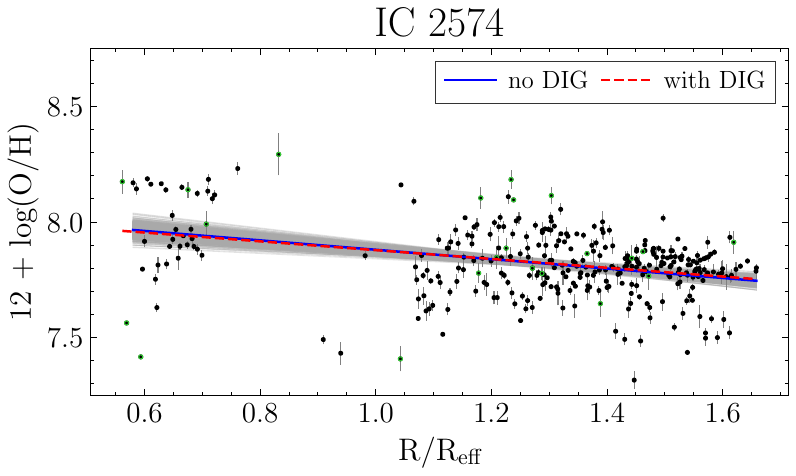}
        \hspace{2cm}
        \includegraphics[scale=0.5]{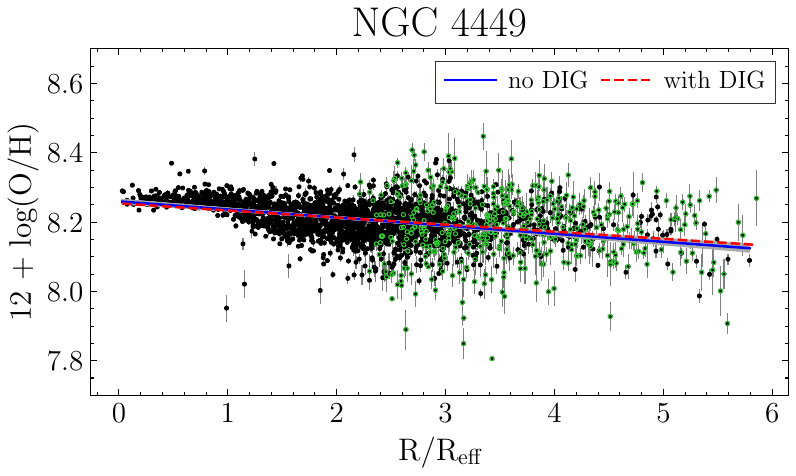}

        \caption{Metallicities and linear models for each galaxy in units of [${\rm dex}/R_{\rm eff}$], labelled accordingly. The colour code is the same for all galaxies, such that the abundances analysed with the upper branch of PG16 are represented in black and those calculated using the lower branch in blue, additionally, green outlined markers represent DIG-dominated fibres in each galaxy. The red lines are models that consider all fibres, and the blue lines represent models with no DIG fibres. Grey lines represent resulting parameters from different iterations of the MCMC fit. Effective radii in the NUV-band are used. NGC 2366 has been additionally fitted with a piecewise linear function.}
         \label{fig:met_grads_1}
\end{figure*}

\begin{figure*}[p]
\vspace{1cm}
\centering

        \includegraphics[scale=0.5]{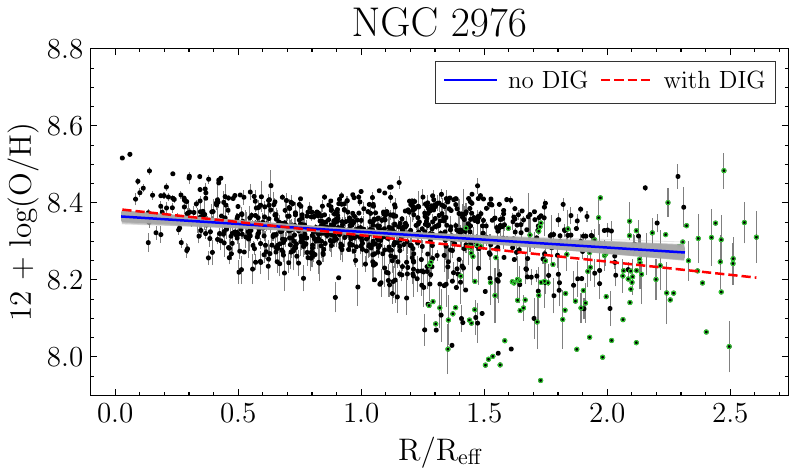} 
        \hspace{2cm}
        \includegraphics[scale=0.5]{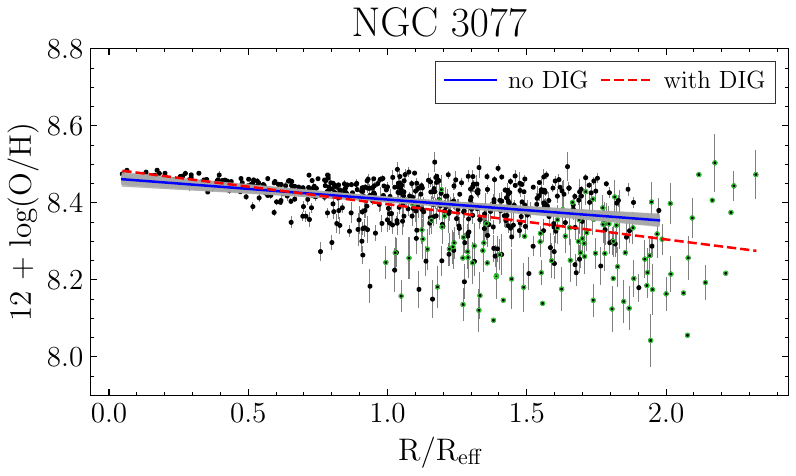}

        \vspace{0.25cm}

        \includegraphics[scale=0.5]{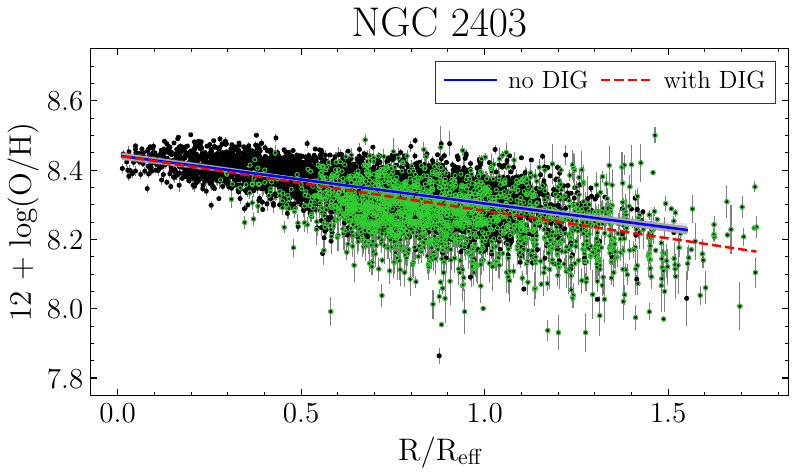}
        \hspace{2cm}
        \includegraphics[scale=0.5]{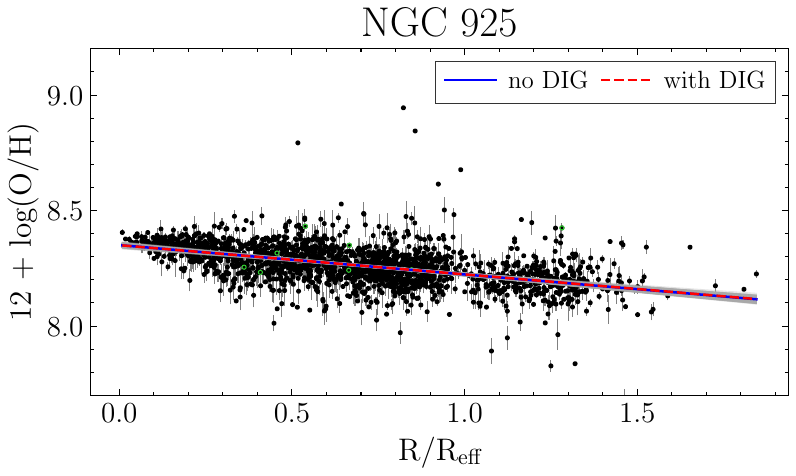}
        
        \vspace{0.25cm}
        
        \includegraphics[scale=0.5]{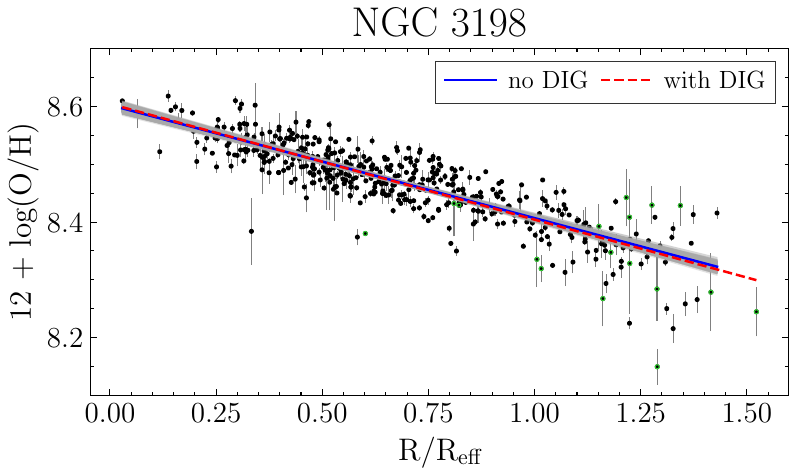}
        \hspace{2cm}
        \includegraphics[scale=0.5]{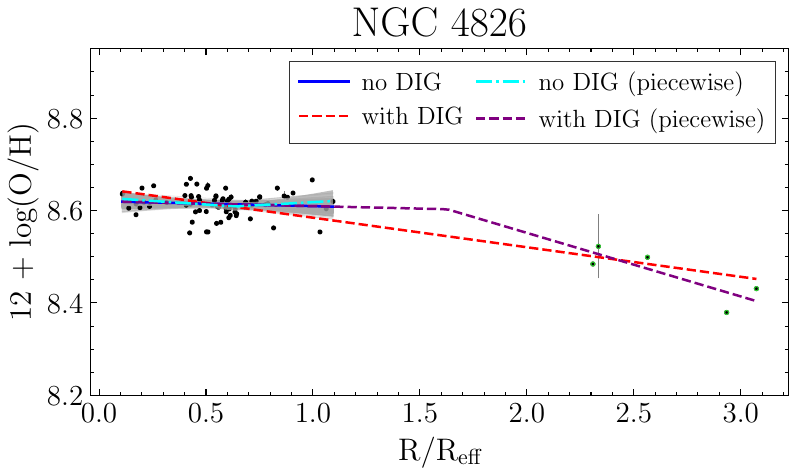}
        
        \vspace{0.25cm}

        \includegraphics[scale=0.5]{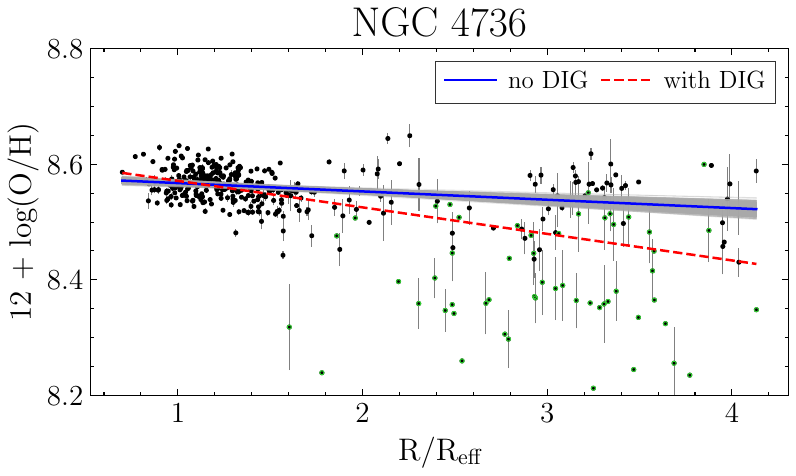}
        \hspace{2cm}
        \includegraphics[scale=0.5]{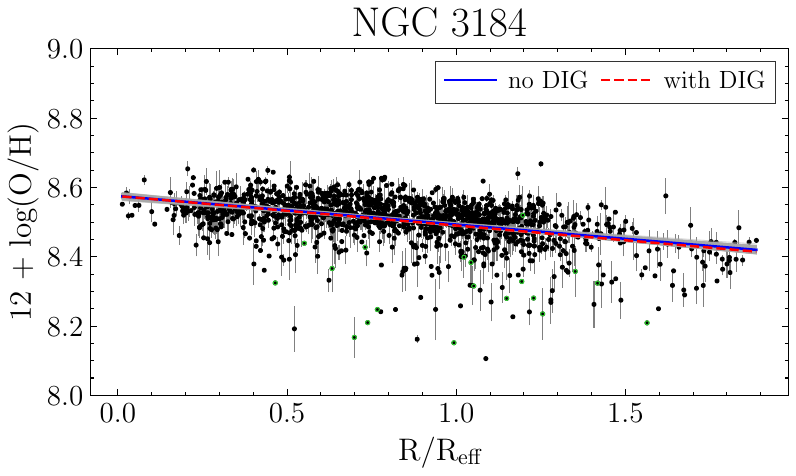}

        \vspace{0.25cm}

        \includegraphics[scale=0.5]{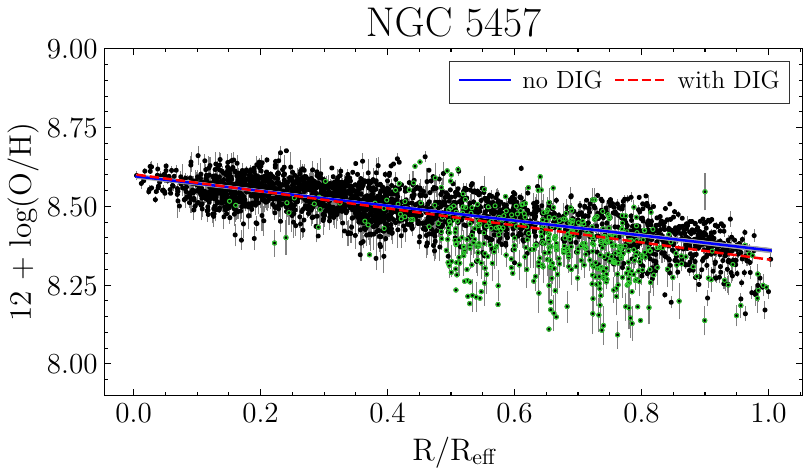}
        \hspace{2cm}
        \includegraphics[scale=0.5]{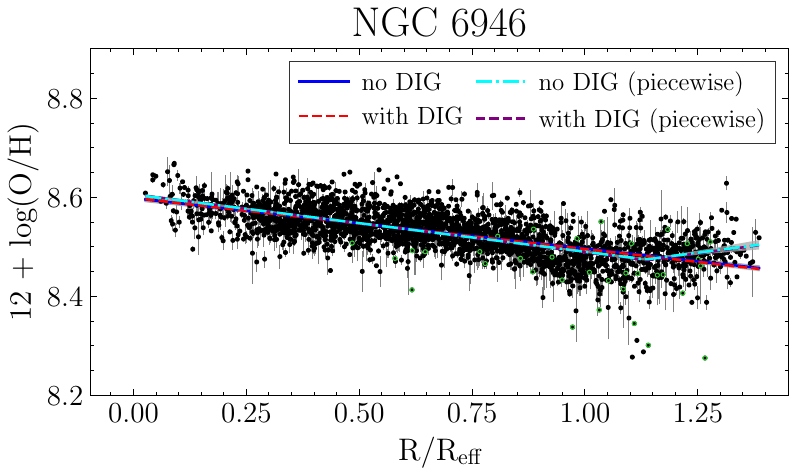}

        \caption{Same as Figure \ref{fig:met_grads_1}, but showing a second part of the sample. Effective radii in the NUV-band are used, except for NGC 3077, for which the effective radius in the r-band is used. NGC 4826 and NGC 6946 have been additionally fitted with a piecewise linear function.}
        \label{fig:met_grads_2}
        \vspace{1cm}
\end{figure*}

\begin{figure*}[!ht]
\vspace{0.2cm}
\centering

        \includegraphics[scale=0.5]{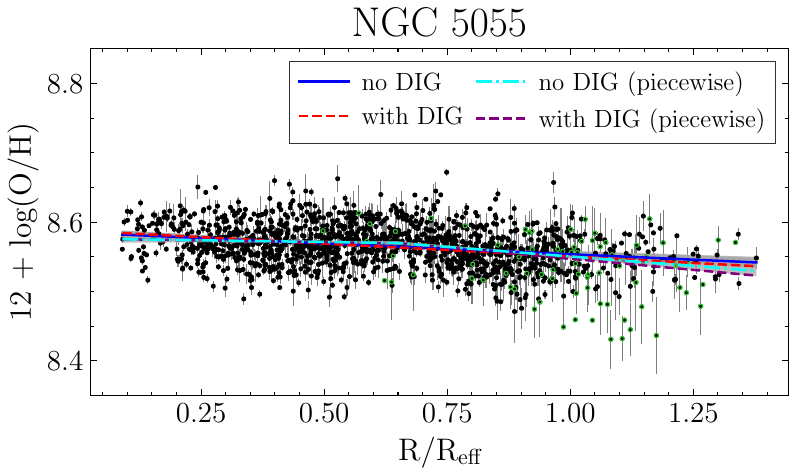}
        \hspace{2cm}
        \includegraphics[scale=0.5]{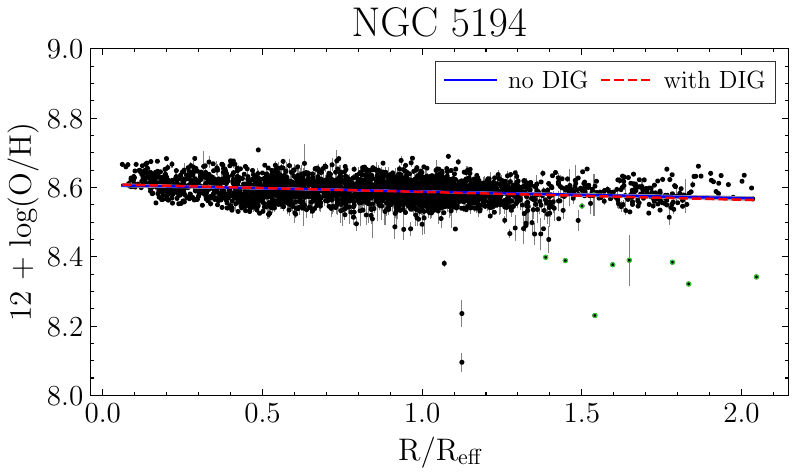}

        \vspace{0.25cm}
        
        \includegraphics[scale=0.5]{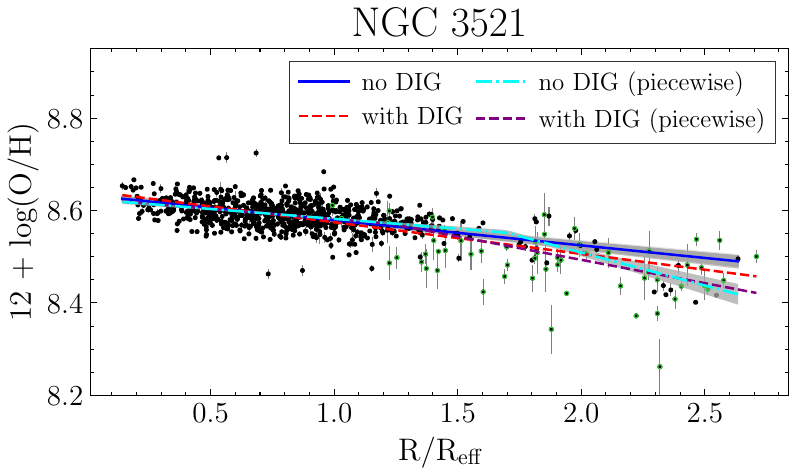}
        \hspace{2cm}
        \includegraphics[scale=0.5]{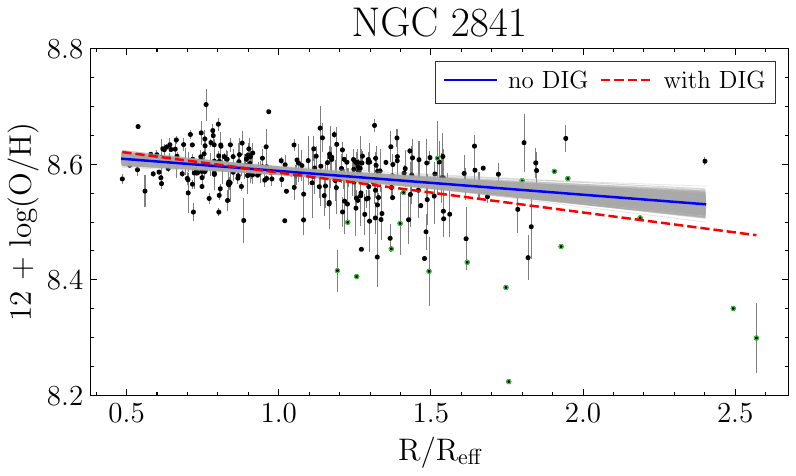}

        \vspace{0.25cm}

        \includegraphics[scale=0.5]{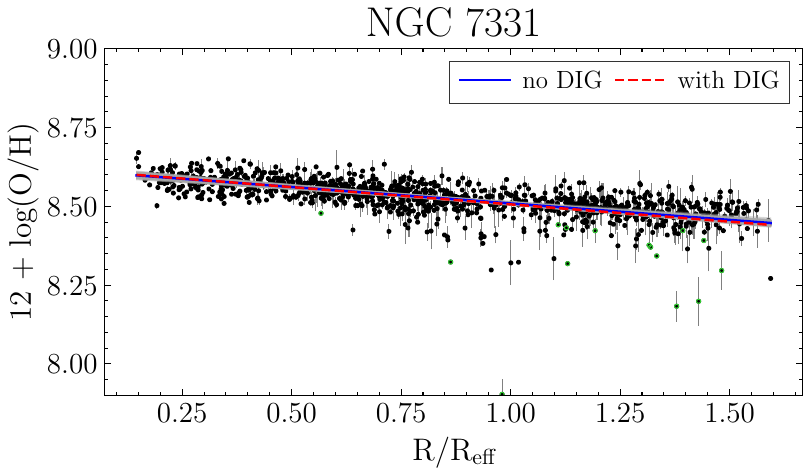}

        \caption{Same as Figures \ref{fig:met_grads_1} and \ref{fig:met_grads_2}, but showing the latter part of the sample. NGC 5055 and NGC 3521 have been additionally fitted with a piecewise linear function.}
         \label{fig:met_grads_3}
         \vspace{0.3cm}
\end{figure*}

The gas metallicities for SDSS galaxies were also estimated using the PG16 method for consistency. We used only SF-galaxies selected with the BPT diagram, and imposed a cut of a  S/N $>$3  in each one of the emission lines used. In this plot, results from our Metal-THINGS sample of galaxies are shown as black diamonds. The mean gas-phase metallicity for Metal-THINGS galaxies, as well as its uncertainty found in Table \ref{tab:all_results}, are an error weighted average of the SF fibres using the PG16 method described above.

Figure \ref{fig:met-stellmass} shows that most of our galaxies align within the contours of the SDSS population, with statistically similar metallicities. We highlight that our sample includes several galaxies with $\log(M_{\rm star}/M_\odot) < 9$, that generally have a lower metallicity than the majority of the SDSS galaxies at the same stellar masses. Not many works have explored this region of the diagram as of the time of writing this paper, some of them being very recent \cite[e.g.][]{mmr_3}. More broad IFU surveys that extend to galaxies with higher stellar masses, include MaNGA \citep[$10^9 - 10^{12} \, M_\odot$, ][]{Oyarzun_2023}, CALIFA \citep[$10^9 - 7\times10^{11} \, M_\odot$, ][]{Gonzalez_2015} or the Physics at High Angular resolution in Nearby GalaxieS (PHANGS) survey \citep[$10^{9.5} - 10^{11} \, M_\odot$, ][]{Lee_2022}. 

It is important to highlight that differences in aperture size between Metal-THINGS and SDSS can introduce systematic uncertainties in this comparison. As noted by \citet{kewley_2005}, when the SDSS optical fibre covers more than 20\% of a galaxy (which statistically corresponds to a z $\sim 0.04 $), the derived metallicity closely approximates the global value. In cases where the coverage is lower, such as nearby and/or large-angular-size galaxies, this limitation can lead to an overestimation of the global metallicity as the measurements primarily sample the more metal-rich central regions \citep[see also][]{Pilyugin24}.
In order to highlight the limited aperture effects, in Fig. \ref{fig:met-stellmass} we also show the MMR from MANGA galaxies \citep{barreraballesteros_2017}, recalibrated to the PG16 indicator for consistency (see Appendix \ref{other}). Metal-THINGS galaxies are aligned with the MANGA MMR, and within the area populated by SDSS galaxies.

The galaxies in our sample follow a trend similar to the MMR with respect to the Main Sequence (MS) of star-forming galaxies \citep[e.g.][]{mainsequence_1}, more massive galaxies tend to lie on the MS, while the less massive ones are typically found below it. This behaviour and other results will be further studied in future works using the Metal-THINGS data.

\subsection{Metallicity gradients}

In this section, we explore the gas metallicity gradients of the Metal-THINGS sample and the effect of the normalisation radius that is used. For this, only SF-type fibres were used and gas metallicities were estimated as described in Section \ref{gas_metallicity}.

Each galaxy was deprojected and the distance from each fibre to the centre in each case was estimated considering the respective position angle ($\phi$) and inclination ($i$) given in Table \ref{tab:gal_data}. To estimate metallicity gradients, a linear fit was used such that
\begin{equation}
    12 + \log({\rm O/H}) = ({\rm O/H})^* = ({\rm O/H})_0 + \nabla_{\rm O/H} \cdot r
\end{equation}
where $r = R/R_{\rm eff}$.

Historically, gas metallicity gradients have been normalised in different ways, either using $R_{25}$ \citep[e.g.][]{Magrini_2016}, where $R_{25}$ is defined as the radius from the centre of a galaxy at which the $\mu_{\rm B} = 25.0 \, mag/arcsec^2$ isophote resides, $R_{\rm eff}$ in different bands \citep[e.g.][using r-band effective radii]{MANGA_paper_mets}, or without any normalisation \citep[e.g.][using distances in kpc]{GASP_paper_radius}. To date, some works in the literature normalise metallicity gradients using effective radii across a mix of bands, even though the effect of $R_{\rm eff}$ in different colours has not been extensively studied. Here we explore this effect by using several effective radii, in the NUV-band ($R_{\rm eff, \; NUV}$), R band ($R_{\rm eff, \; r}$), $K_{\rm S}$ band ($R_{\rm eff, \; Ks}$), and $R_{25}$.

For the effective radii ($R_{\textrm{eff}}$), we use values from Valerdi et al. (in preparation), which were calculated in multiple bands using \texttt{GALFITools v0.15.2}\footnote{\url{https://github.com/canorve/GALFITools}} \citep{GALFITools_paper} by fitting one or more components to the light profile of each galaxy. This process numerically integrates all used profiles, using a more robust methodology than many previous attempts and yields smaller dispersions in the resulting radii. All photometric data used for this procedure originated from the GALEX survey for the NUV band ($1770 - 2730$ \AA) \citep{GALEX_paper}; for the r band ($6231$ \AA), the Pan-STARRS \citep{PANSTARRS_paper} and SDSS \citep{SDSS_paper} surveys; and for the Ks band ($2.159 \, \mu \textrm{m}$), the 2MASS survey \citep{2MASS_paper}.
A key point to take into consideration is that low-mass galaxies often lack emission in the redder parts of the spectrum, particularly in the near-infrared (e.g. the $\mathrm{K_s}$-band) and are barely detectable in these bands. This leads to large photometric uncertainties and, consequently, unreliable measurements of their effective radii in these wavelengths. Furthermore, the measured effective radius can vary significantly depending on the photometric band used, as there is a complex dependence of both population gradients and attenuation on galaxy morphology, stellar mass and environment \citep{wavelength_eff_rad}.

\begin{table*}[t]
\centering
\renewcommand{\arraystretch}{1.2}
\begin{tabular}{>{\rowmac}c>{\rowmac}c>{\rowmac}c|>{\rowmac}c>{\rowmac}c|>{\rowmac}c>{\rowmac}c<{\clearrow}}

\hline
\hline
1 & 2 & 3 & 4 & 5 & 6 & 7 \\

\begin{tabular}[c]{@{}c@{}c@{}} Galaxy \\ \textbf{} \\ \textbf{} \end{tabular} &
\begin{tabular}[c]{@{}c@{}c@{}} Mean \\ metallicity \\ $\rm (\overline{O/H^*})$ \end{tabular} &
\begin{tabular}[c]{@{}c@{}c@{}} $R_{\rm eff}$ \\ NUV band \\ $\rm [arcsec]$ \end{tabular} &
\begin{tabular}[c]{@{}c@{}c@{}}$ \rm \nabla(O/H) $ \\ {\small no DIG \normalsize}\\ {[}$ \rm dex/R_{eff} ${]} \end{tabular} &
\begin{tabular}[c]{@{}c@{}c@{}}$ \rm O/H_0$ \\ {\small no DIG \normalsize} \\ \textbf{} \end{tabular} &
\begin{tabular}[c]{@{}c@{}c@{}}$ \rm \nabla(O/H) $ \\ {\small with DIG \normalsize} \\ {[}$ \rm dex/R_{eff} ${]} \end{tabular} &
\begin{tabular}[c]{@{}c@{}c@{}}$ \rm O/H_0$ \\ {\small with DIG \normalsize} \\ \textbf{} \end{tabular}
\\ \hline

DDO 53      & 7.665$\pm$0.003   & 21.06   & -0.03$\pm$0.02   & 7.75$\pm$0.09   & -0.03$\pm$0.02   & 7.75$\pm$0.09 \\
DDO 154     & 7.74$\pm$0.01     & 36.75   & -0.07$\pm$0.17   & 7.82$\pm$0.26   & -0.07$\pm$0.17   & 7.82$\pm$0.26   \\
Holmberg I  & 7.755$\pm$0.006   & 68.85   &  0.45$\pm$0.19   & 7.38$\pm$0.18   &  0.20$\pm$0.17   & 7.60$\pm$0.17   \\
M81 DwB     & 8.011$\pm$0.002   & 14.79   &  0.01$\pm$0.05   & 8.01$\pm$0.04   & 0.01$\pm$0.05    & 8.01$\pm$0.04   \\
Holmberg II & 7.672$\pm$0.002   & 129.06  & -0.18$\pm$0.03   & 7.79$\pm$0.03   & -0.17$\pm$0.03   & 7.80$\pm$0.03   \\
NGC 2366    & 7.851$\pm$0.001   & 121.97  & -0.214$\pm$0.017 & 7.988$\pm$0.014 & -0.221$\pm$0.018 & 8.007$\pm$0.015 \\
NGC 4214    & 8.1907$\pm$0.0004 & 63.09   & -0.027$\pm$0.003 & 8.214$\pm$0.003 & -0.03$\pm$0.003  & 8.202$\pm$0.004 \\
NGC 1569    & 8.1281$\pm$0.0001 & 16.70   & -0.012$\pm$0.002 & 8.153$\pm$0.006 & -0.008$\pm$0.003 & 8.145$\pm$0.007 \\
IC 2574     & 7.840$\pm$0.001   & 185.88  & -0.204$\pm$0.028 & 8.084$\pm$0.037 & -0.19$\pm$0.028  & 8.069$\pm$0.038 \\
NGC 4449    & 8.2150$\pm$0.0001 & 45.00   & -0.023$\pm$0.001 & 8.26$\pm$0.003  & -0.02$\pm$0.001  & 8.254$\pm$0.003 \\
NGC 2976    & 8.3607$\pm$0.0003 & 68.46   & -0.041$\pm$0.005 & 8.366$\pm$0.006 & -0.069$\pm$0.006 & 8.384$\pm$0.007 \\
NGC 3077    & 8.4288$\pm$0.0004 & $-$     & $-$              &  $-$            &  $-$             &  $-$            \\
NGC 2403    & 8.3541$\pm$0.0001 & 220.76  & -0.138$\pm$0.004 & 8.442$\pm$0.003 & -0.159$\pm$0.004 & 8.442$\pm$0.003 \\
NGC 925     & 8.2695$\pm$0.0004 & 199.22  & -0.127$\pm$0.006 & 8.350$\pm$0.004 & -0.127$\pm$0.006 & 8.350$\pm$0.004 \\
NGC 3198    & 8.4746$\pm$0.0003 & 127.5   & -0.196$\pm$0.006 & 8.603$\pm$0.004 & -0.201$\pm$0.006 & 8.605$\pm$0.004 \\
NGC 4826    & 8.6187$\pm$0.0002 & 57.09   & -0.011$\pm$0.014 & 8.620$\pm$0.008 & -0.064$\pm$0.006 & 8.648$\pm$0.005 \\
NGC 4736    & 8.5694$\pm$0.0002 & 40.84   & -0.014$\pm$0.002 & 8.582$\pm$0.004 & -0.046$\pm$0.004 & 8.617$\pm$0.007 \\
NGC 3184    & 8.5306$\pm$0.0004 & 109.91  & -0.082$\pm$0.004 & 8.575$\pm$0.004 & -0.085$\pm$0.005 & 8.575$\pm$0.004 \\
NGC 5457    & 8.4896$\pm$0.0002 & 323.72  & -0.235$\pm$0.004 & 8.596$\pm$0.002 & -0.271$\pm$0.005 & 8.602$\pm$0.003 \\
NGC 6946    & 8.5405$\pm$0.0001 & 203.37  & -0.102$\pm$0.002 & 8.598$\pm$0.002 & -0.104$\pm$0.002 & 8.599$\pm$0.002 \\
NGC 5055    & 8.5685$\pm$0.0002 & 149.35  & -0.031$\pm$0.003 & 8.585$\pm$0.002 & -0.037$\pm$0.003 & 8.588$\pm$0.002 \\
NGC 5194    & 8.5955$\pm$0.0001 & 150.59  & -0.019$\pm$0.002 & 8.607$\pm$0.002 & -0.022$\pm$0.002 & 8.609$\pm$0.002 \\
NGC 3521    & 8.5922$\pm$0.0001 & 121.43  & -0.054$\pm$0.003 & 8.633$\pm$0.002 & -0.069$\pm$0.002 & 8.643$\pm$0.002 \\
NGC 2841    & 8.5959$\pm$0.0008 & 115.38  & -0.041$\pm$0.007 & 8.630$\pm$0.008 & -0.069$\pm$0.008 & 8.655$\pm$0.010 \\
NGC 7331    & 8.5507$\pm$0.0002 & 193.52  & -0.105$\pm$0.005 & 8.614$\pm$0.005 & -0.11$\pm$0.005  & 8.616$\pm$0.005 \\
\hline

\end{tabular}
\caption{Mean gas-phase metallicities and linear models parameters for each galaxy. Each column shows information as follows: 1) galaxy designation; 2) mean gas-phase metallicity; 3) and 6) linear model parameters and errors without DIG-dominated fibres; 7) and 8) linear model parameters and errors with all fibres. Each linear model was fitted using a Bayesian MCMC routine. For the metallicity values, the average value was calculated for each galaxy using all PG16 values for each fibre weighted by the uncertainty.}
\label{tab:all_results}
\end{table*}

We show metallicity gradients normalised using the effective radii in the NUV-band in Figures \ref{fig:met_grads_1}, \ref{fig:met_grads_2} and \ref{fig:met_grads_3} as these are the data for which Valerdi et al. (in preparation) reports more robust results, while covering a large number of galaxies in our sample. Table \ref{tab:all_results} shows the fitted coefficients, where $(\rm O/H)_0$ is the metallicity or oxygen abundance at the centre of the galaxy and $\nabla_{\rm O/H}$ is the slope of the linear relation, which is our metallicity gradient. It is also important to note that the position angle (P.A. or $\phi$) is measured counter-clockwise between the north direction in the sky and the major axis of the receding half of the galaxy. In the case of NGC 3077, the effective radius in the NUV-band could not be calculated due to bad quality data (see Valerdi et al. in preparation), and thus, the value for the r-band was used instead.

All cases were fitted with a Bayesian model using the Python packages \texttt{PyMC v5.16.2} \citep{pymc-paper} and \texttt{BAMBI v0.14.0} \citep{bambi-paper}, with 4 MCMC chains, 3000 iterations each.

Bayesian fitting methods are preferred as they incorporate prior knowledge and uncertainties directly, providing a full posterior rather than just single-point estimates, as in the least-squares method. This is especially useful when data quality varies, as is common in observations of faint galaxies. They yield more robust estimates, smaller and more realistic errors, and allow for a credible assessment of fit parameters and model reliability, making them more suited for this work.

All resulting parameters obtained for the entire galaxy sample, those with highest posterior probability, including metallicity gradients,$\nabla_{\rm O/H}$, central metallicities, $\rm (O/H)_0$, and mean gas-phase metallicity ($\rm \overline{O/H^*}$), along with their dimensions and errors, are displayed in Table \ref{tab:all_results}.

\begin{figure*}[t]
    \centering
    \includegraphics[width=0.49\textwidth]{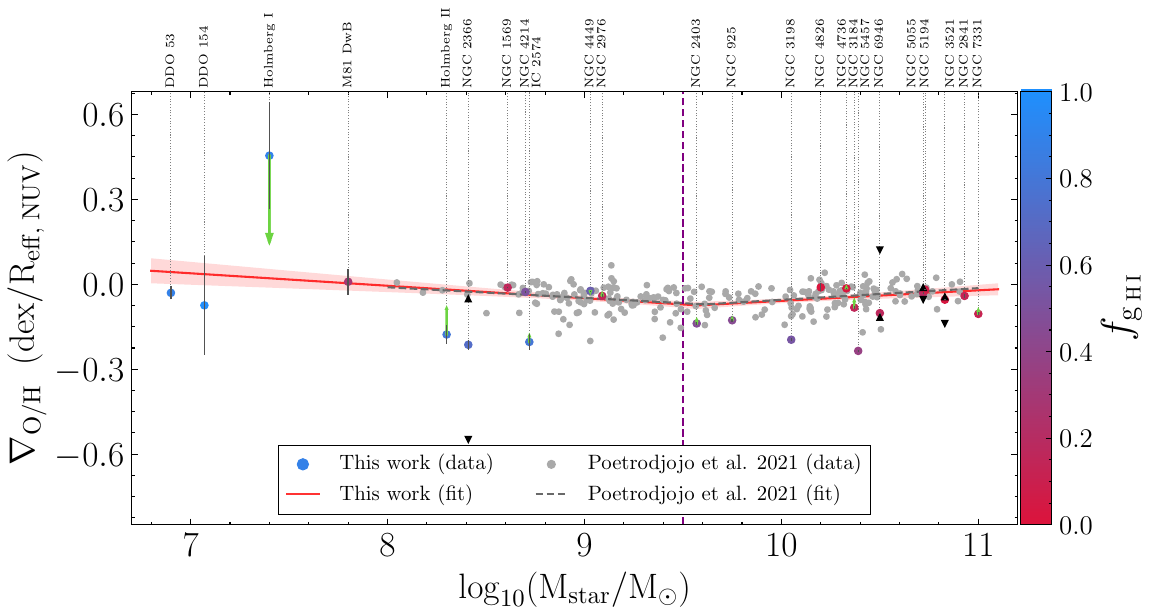}
    \includegraphics[width=0.49\textwidth]{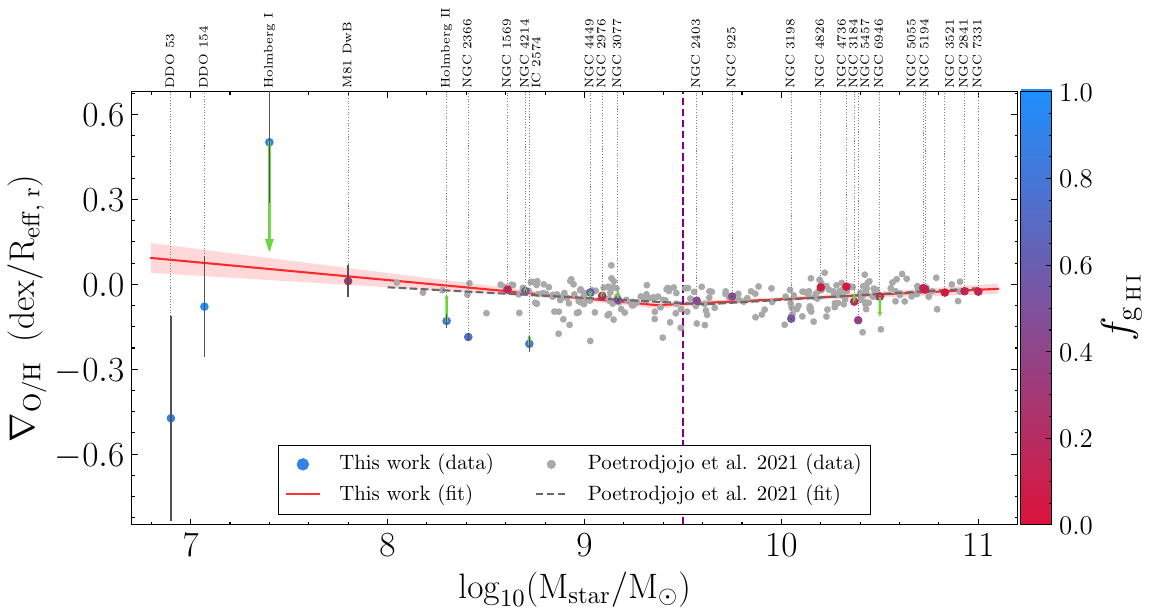}
    \includegraphics[width=0.49\textwidth]{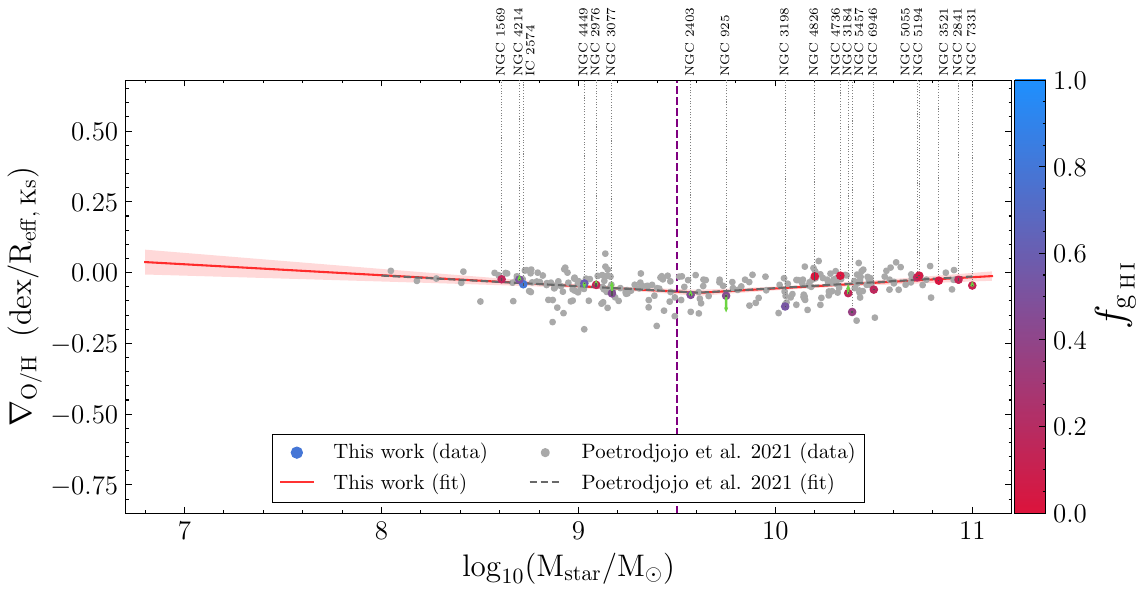}
    \includegraphics[width=0.49\textwidth]{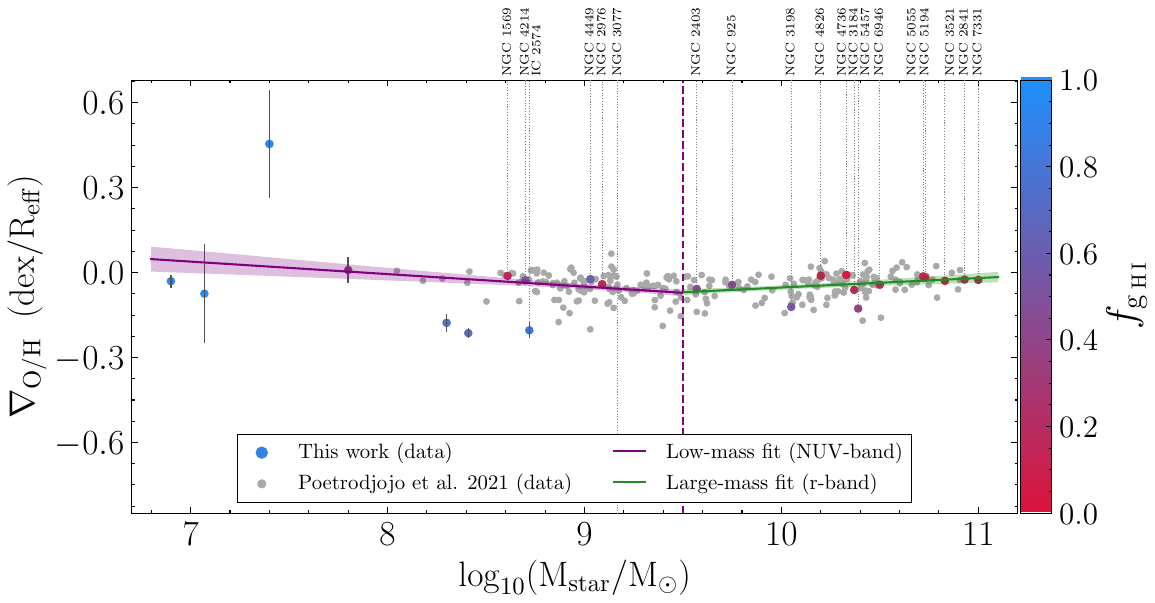}
    
    \caption{Relation between the stellar mass and the metallicity gradient using the NUV-band radius (top) and r-band radius (middle) and ${\rm K}_s$-band (bottom) for normalisation. Coloured markers correspond to the sample of galaxies used in this study, and are coloured such that it represents the fraction of atomic gas of a given galaxy against the sum of the mass in stars and atomic gas, as it is indicated in the colour bar. Grey markers correspond to the gradients of galaxies studied in \citet{Poetrodjojo_2021}, calculated using PG16's S calibration. The red line shows the piecewise linear fit to this data, which illustrates the tendencies of the gradients as mass increases. Green arrows show the deviation from values achieved using a DIG threshold of $C_{\rm H \, II} > 0.6$ as seen in \citet{kaplan_2016}, when compared to our threshold of $C_{\rm H \, II} > 0.4$. The black triangles represent the values of the gradients in the first (pointing up) and second (pointing down) intervals of the piecewise linear functions for those galaxies fitted in that manner. Galaxies with stellar masses $\log(M_{\rm star}/M_\odot) \gtrsim 9.5$ (purple dashed line) have on average disc-like morphologies \citep{Simons_2015}.}
    \label{fig:grad_mass_fit_HI_r}
\end{figure*}

\begin{table*}[t]
\centering \normalsize 
\renewcommand{\arraystretch}{1.2}
\begin{tabular}{>{\rowmac}c>{\rowmac}c>{\rowmac}c|>{\rowmac}c>{\rowmac}c>{\rowmac}c|>{\rowmac}c>{\rowmac}c>{\rowmac}c|>{\rowmac}c>{\rowmac}c>{\rowmac}c<{\clearrow}}
\hline
\hline

1 & 2 & 3 & 4 & 5 & 6 & 7 & 8 & 9 & 10 & 11 & 12  \\

Normalisation &
SAMI &
SAMI+M-T &
\multicolumn{3}{c|}{RMSE$_{\rm M_*-\nabla}$ (SAMI)} &
\multicolumn{3}{c|}{RMSE$_{\rm M_*-\nabla}$ (SAMI+M-T)} &
\multicolumn{3}{c}{RMSE$_{\rm M_*-\nabla}$ (M-T)}\\

radius&
break &
break &
&
&
&
&
&
&

\\
&
{[}$\log M_\odot$ {]} &
{[}$\log M_\odot$ {]} &
lower &
upper &
global &

lower &
upper &
global &

lower &
upper &
global
\\

\hline

$R_{\rm eff,\, NUV}$ &       & 9.6$\pm$0.2
                     &       &       &  
                     & 0.063 & 0.045 & 0.055
                     & 0.154 & 0.065 & 0.118 \\

$R_{\rm eff,\, r}$   & 9.6$\pm$0.2  & 9.4$\pm$0.2
                     & 0.044 & 0.039 & 0.042
                     & 0.067 & 0.041 & 0.053
                     & 0.196 & 0.036 & 0.149 \\

$R_{\rm eff,\, Ks}$  &       & 9.6$\pm$0.1
                     &       &       &  
                     & 0.043 & 0.039 & 0.041
                     & 0.009 & 0.038 & 0.030 \\
\hline

\end{tabular}

\caption{Break points and root-mean-square errors (RMSE) for the piecewise linear fits in the relations in Figure \ref{fig:grad_mass_fit_HI_r} separately for the SAMI and Metal-THINGS (M-T) samples along with the combination of both. Each value is calculated globally, as well as separately for data below (lower) and above (upper) each break point.}
\label{RMSE_MZR}

\end{table*}

\section{Discussion}

We find that negative metallicity gradients dominate the sample, consistent with inside-out galaxy growth. Notably, Holmberg I exhibits a positive gradient, which could be explained either by limitations in data quality or by a physical mechanism, which will be discussed in Section \ref{mass-gradient-section}. Dispersion in the data varies across the sample, with larger galaxies generally showing lower scatter along the linear fit, something that could be attributed to their more uniform star-forming regions or higher quality of data coverage.

Irregular galaxies pose a greater challenge for linear fitting due to their more complex morphology. Nevertheless, the majority of linear fits achieved relatively high confidences, as supported by the analysis of the MCMC chains. Exceptions are observed in galaxies with a poorer SNR, where confidence in gradient measurements decreases due to sparse sampling. Some galaxies show evidence of a break in their gradients, where the metallicity trend changes at a certain distance from the centre. Cases in our sample that show this behaviour are NGC 6946 and NGC 5055, where the observed metallicity gradients exhibit patterns that align well with a fit using a piecewise linear function as shown in Fig. \ref{fig:met_grads_2}.

Furthermore, the metallicity distribution in some galaxies of our sample was found to be compatible with a bimodal behaviour, and fitted with a piecewise linear model (see Fig. \ref{fig:met_grads_1} and \ref{fig:met_grads_2}). Further analysis of this phenomenon on NGC 3521 is found in \citet{leonid_2025}.

These initial results suggest no apparent correlation between a galaxy's size or morphology and its metallicity gradient, beyond the influence of data quality. However, we study the relationship of the gradient with the stellar mass and atomic gas fraction with a more robust statistical framework later in this work. Mean metallicity values align well with those reported in previous studies, including the original paper on the THINGS survey \citep{Walter_2008_THINGS}, further reinforcing the reliability of our results.

\subsection{Effect of DIG on metallicity gradients} \label{sect:effect_of_dig}

We now evaluate whether the inclusion of the DIG-dominated fibres produce important variations in the estimation of the metallicity gradients. By observing the metallicity gradients in figures \ref{fig:met_grads_1}, \ref{fig:met_grads_2} and \ref{fig:met_grads_3}, as well as the values in table \ref{tab:all_results}, we find a median difference between the value of the gradients with and without DIG fibres of $0.0055$, and $0.0057$ for the zero points. Another trend that is visible in our figures is that DIG fibres generally have  lower metallicities than fibres classified otherwise at similar distances, which is consistent with their nature. Further analysis shows that the main difference the DIG introduces is in the value of the gradients. However, it is clear that generally DIG does not seem to have a significant effect on the metallicity gradients, independently of mass or morphology of the galaxy.

However, there are two exceptions, namely Holmberg I and NGC 4736. In the case of NGC 4736, the larger difference in its gradient could be explained by the high percentage of DIG fibres, with a contribution of its irregular morphology, inner and outer rings \citep{NGC4736_ring_morph}. This galaxy also shows evidence for a low-luminosity AGN \citep{NGC4736_agn}, which could point at another possible source of this behaviour. The fact that it is still not fully mapped by the Metal-THINGS survey (see Appendix \ref{h_alpha_maps}), lays the possibility for future work exploring this particular galaxy in a deeper analysis once completely mapped. In the case of Holmberg I, the signal-to-noise ratio across the majority of the data is generally poor, resulting in a significantly reduced number of available fibres due to the SNR cut of 3. Consequently, the Bayesian linear model that was used to fit the dataset exhibits a high uncertainty. A small subset  of data points can thus introduce significant variations in the resulting fit. This is despite the fact that Holmberg I does not contain a relatively high fraction of DIG fibres.

With the exception of these two cases, we find no significant effects caused by the DIG in the gradients of the studied galaxies. In other works, such as \citet{Poetrodjojo_2019}, a more significant effect is found, which is explained by the difference in the threshold imposed to distinguish H \small II \normalsize fibres from DIG ones. In that work, all fibres with $C_{\rm H \, II} < 0.9$ were considered to originate from DIG, with a similar fraction in other similar works. In contrast to that, the condition set for this work is so that all fibres with $C_{\rm H\, II} < 0.4$ are considered to be DIG fibres.

Upon examining Figures \ref{fig:met_grads_1}, \ref{fig:met_grads_2} and \ref{fig:met_grads_3}, it is evident that fibres associated with DIG generally tend to have lower oxygen abundances compared to fibres where $\rm H\alpha$ emission is predominantly from H \small II \normalsize regions. As a result, the metallicity gradients of galaxies with a high fraction of DIG fibres tend to be marginally steeper when these data are excluded from the analysis.

\subsection{The stellar mass - metallicity gradient relation} \label{mass-gradient-section}

In this work, we successfully analysed and modelled the metallicity of a sample of galaxies with diverse values of stellar masses. As a next step we explore the relationship between a galaxy's stellar mass and its respective metallicity gradient.

Some studies indicate that metallicity gradients exhibit a break in their dependency on the mass of the galaxy. In the lower mass range, specifically at $\log(M / M_\odot) < 9.5$, the gradient becomes steeper (more negative) as the mass increases, up to a threshold of $10^{9.5} M_\odot$. At this point, there is a break in this trend, after which metallicity gradients begin to flatten, becoming shallower as the mass of a galaxy continues to increase \break \citep{Belfiore_2017, Poetrodjojo_2021}.

To determine if our results align with this trend, we compare the metallicity gradients we calculated with the data from \citet{Poetrodjojo_2021}, as they employ the same abundances calibration and incorporate a break in the mass-metallicity gradient relationship. Figures \ref{fig:grad_mass_fit_HI_r} shows the metallicity gradients obtained in this work, shown for effective radii in the NUV and r bands, respectively, against the stellar mass of each galaxy. Metallicity gradients were also calculated using the effective radii in  Ks band, although for a smaller sample (see Appendix Section \ref{other_grads_table} and Figure \ref{fig:grad_mass_fit_HI_r}), providing similar results to the r-band. Metallicity gradients using the isophotal radius $R_{25}$ were also estimated (see Appendix Section \ref{other_grads_table}), although they are not directly comparable to those of \citet{Poetrodjojo_2021}, which use $R_{\rm eff}$.

In all panels of Figure \ref{fig:grad_mass_fit_HI_r}, grey points represent data from \citet{Poetrodjojo_2021}, based on the SAMI survey, with the red line indicating the piecewise linear fit to this data along with our sample, performed with the Python package \texttt{piecewise-regression v1.5.0} \citep{piecewise_regression_paper}.

The data in these figures are colour-coded with the atomic gas fraction:
\begin{equation}
    f_{\rm g\, H\, {I}} = \frac{M_{\rm H\, {I}}}{M_{\rm H\, {I}} + M_{\rm star}}
\end{equation}
where $M_{\rm H\, {I}}$ is the atomic gas mass taken from \citet{Leroy_2008}. Since not all of our galaxies have information on their molecular gas content, a comprehensive study incorporating it could not be conducted.

As shown in both figures, the majority of galaxies analysed fall within the dispersion of the SAMI data, confirming that our results are consistent with established expectations. The exceptions to this are three dwarf galaxies DDO 53, DDO 154 and Holmberg I, which are located on the left-most side of both figures, corresponding to the lower mass range, an area where the work previously mentioned lacks data, complementing this relation. Notably, these three galaxies also exhibit large uncertainties in their metallicity gradients. For this relation, we were unable to find other works that explore this low-mass regime, including the surveys previously mentioned (SAMI, CALIFA, MaNGA, and PHANGS).

When comparing all plots, using effective radii in the NUV-band, ${\rm K}_s$-band and in the r-band, we can better understand this relationship and further contrast our results, which are, overall, consistent with previously established fits in the literature, specifically, the metallicity gradients derived from our sample closely align with values reported by \citet{Poetrodjojo_2021} who found similar trends for galaxies of comparable mass and morphology. Furthermore, this shows the robustness of our methodology and provides additional support for the general applicability of these models.

Table \ref{RMSE_MZR} shows the RMSE values for the piecewise linear fits in Figure \ref{fig:grad_mass_fit_HI_r}, calculated separately for the SAMI and Metal-THINGS samples, as well as for their combination. While the main analysis uses the combined data, computing RMSE individually reveals how Metal-THINGS data affects the SAMI fit (dashed grey line in Figure \ref{fig:grad_mass_fit_HI_r}).

\begin{figure*}[t]
\centering
    \begin{subfigure}[b]{\columnwidth}
        \includegraphics[width=\linewidth]{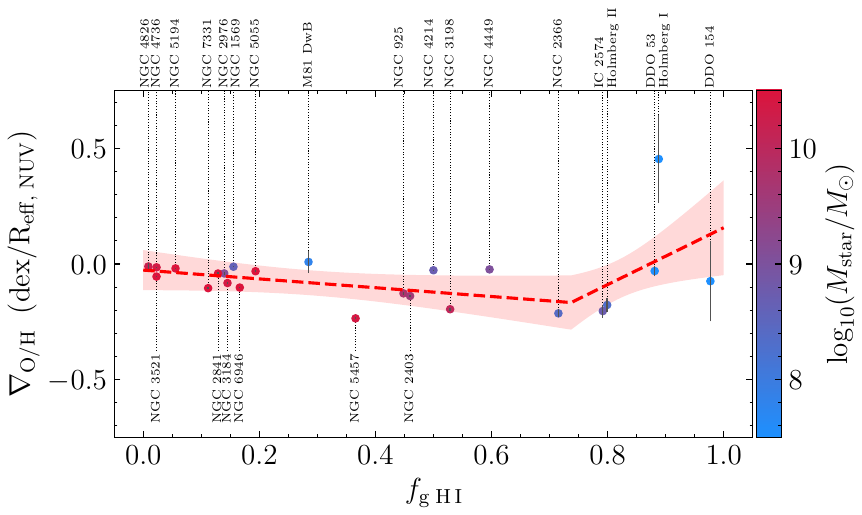}
        \caption{NUV-band effective radius}
    \end{subfigure}
    \hfill 
    \begin{subfigure}[b]{\columnwidth}
        \includegraphics[width=\linewidth]{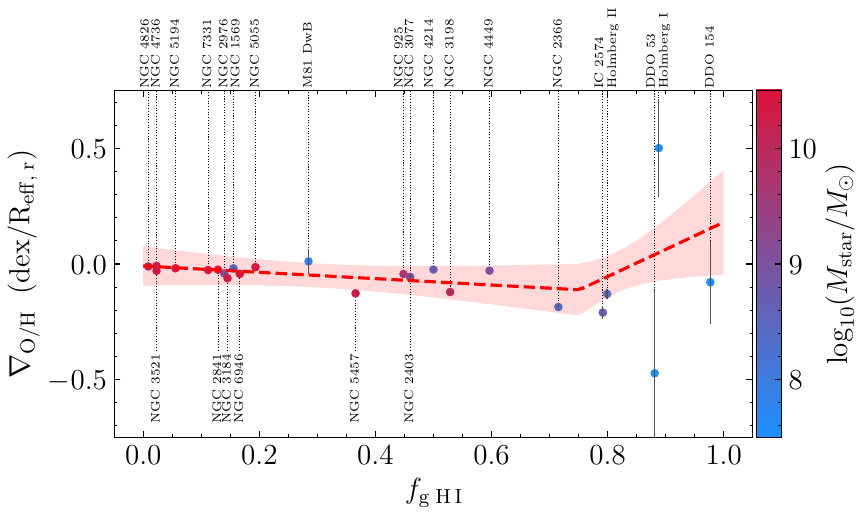}
        \caption{r-band effective radius}
    \end{subfigure}
    
    \begin{subfigure}[b]{\columnwidth}
        \includegraphics[width=\linewidth]{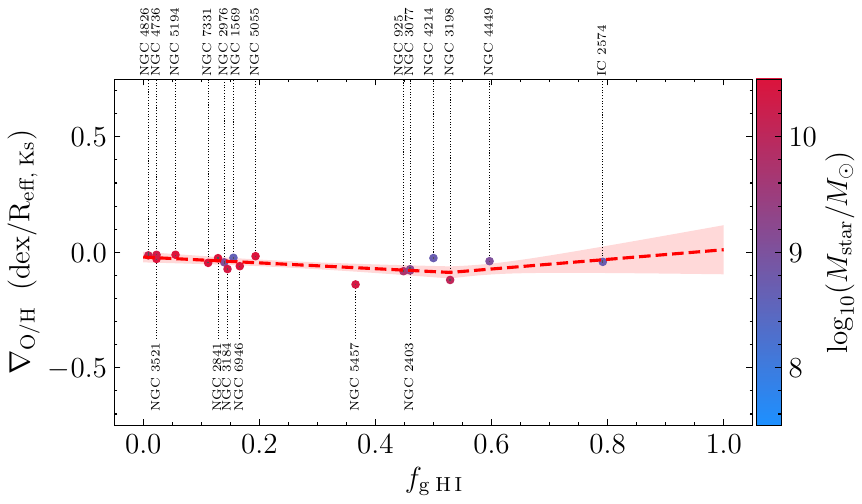}
        \caption{K$_s$-band effective radius}
    \end{subfigure}
    \hfill 
    \begin{subfigure}[b]{\columnwidth}
        \includegraphics[width=\linewidth]{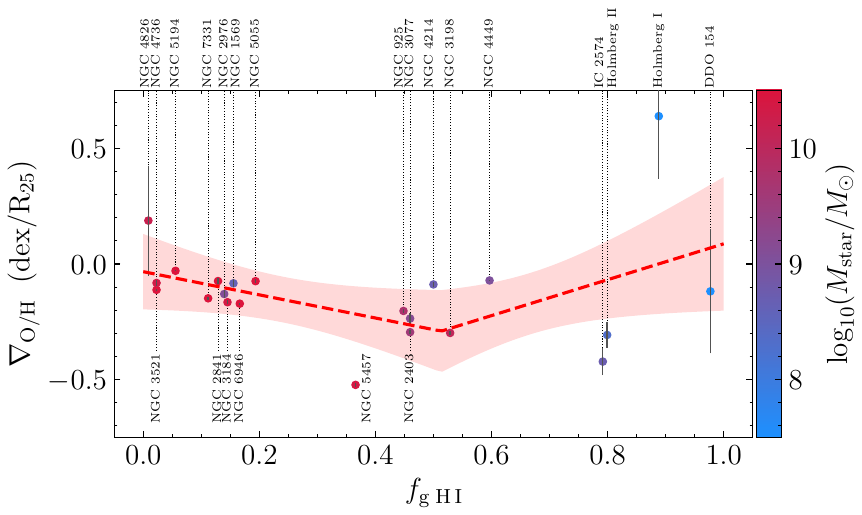}
        \caption{$R_{25}$ radius}
    \end{subfigure}

    \caption{Relation between metallicity gradients and fraction of atomic gas of a given galaxy against the sum of the mass in stars and atomic gas using all four available radii for normalisation. The colour of each marker represents stellar masses as indicated in the colour bar. The red dashed line denotes a piecewise linear fit to the data, with the fainter red region denoting its confidence interval. Each normalisation radius for the fit has been labelled accordingly.}
    \label{fig:metgrad-gasfrac}
    
\end{figure*}

\begin{table*}[t]
\centering
\renewcommand{\arraystretch}{1.2}
\begin{tabular}{>{\rowmac}c>{\rowmac}c|>{\rowmac}c>{\rowmac}c>{\rowmac}c|>{\rowmac}c<{\clearrow}}

\hline
\hline
1 & 2 & 3 & 4 & 5 & 6 \\

Normalisation &
$f_{\rm g, \, H \, I }$ & 
\multicolumn{3}{c|}{RMSE$_{\rm f_{g, HI}}$ (piecewise)} &
RMSE$_{\rm f_{g,HI}}$ (linear) \\

radius &
break &
lower &
upper &
global &
global \\

\hline
$R_{\rm eff,\, NUV}$ & 0.74$\pm$0.10 & 0.060 & 0.224 & 0.115 & 0.138 \\
$R_{\rm eff,\, r}$   & 0.75$\pm$0.10 & 0.036 & 0.331 & 0.152 & 0.148 \\
$R_{\rm eff,\, Ks}$  & 0.53$\pm$0.09 & 0.028 & 0.028 & 0.028 & 0.032 \\
$R_{25}$             & 0.51$\pm$0.13 & 0.115 & 0.328 & 0.197 & 0.237 \\
\hline

\end{tabular}
\caption{Root-mean-square error for piecewise linear fits and simple Bayesian linear fits for the relations found in Figure \ref{fig:metgrad-gasfrac}. Each value is calculated globally, as well as separately for points below (lower) and above (upper) the break point.}
\label{tab:RMSE_table}
\end{table*}

From the piecewise linear fit shown in Figure \ref{fig:grad_mass_fit_HI_r} along with the RMSE values in Table \ref{RMSE_MZR}, we can determine which normalisation band provides a better representation across different stellar mass regimes. At low stellar masses, normalising by the NUV-band effective radius is recommended, largely because dwarf galaxies can be better observed in the ultraviolet and the Reff in this wavelength is more accurate (see for instance DDO 53). Conversely, at higher stellar masses, normalisation using the r-band effective radius produces much tighter correlations with lower scatter. This result can be summarised by the following expression, where metallicity gradients were normalized using R$_{\rm eff, NUV}$ for low-mass galaxies and R$_{\rm eff, r}$ for massive galaxies (see also the bottom right panel of Fig. \ref{fig:grad_mass_fit_HI_r}).

\begin{equation}
  \nabla_{\rm O/H} =
    \begin{cases}
      & (0.35\pm0.10) - (0.04\pm0.03) \times \log(M_{\rm star}/M_\odot); \\
      & \hspace{3.5cm}\text{if } \log(M_{\rm star}/M_\odot) \lesssim 9.5 \\[0.2em]
      & (-0.39\pm0.13) + (0.03\pm0.01) \times \log(M_{\rm star}/M_\odot); \\
      & \hspace{3.5cm}\text{if } \log(M_{\rm star}/M_\odot) \gtrsim 9.5 
    \end{cases}
    \label{pw_fin}
\end{equation}
where the break indicates the stellar mass at which each normalisation is used. Here, we find that cases with $\log(M_{\rm star}/M_\odot)\sim 9.5$ show no evident signs of the preferred use of one band over the other.

In Table \ref{RMSE_MZR}, the overall RMSE measured for SAMI and the combined SAMI+Metal-THINGS sample are very similar, with a slightly higher RMSE observed in the lower mass branch for the combined sample. This increase is primarily driven by the inclusion of more low-mass galaxies in the extended sample. Such a rise in scatter is expected, as low-mass galaxies are still actively forming stars and their metallicity gradients tend to be intrinsically more variable, even showing positive gradients in some cases (e.g., Holmberg I). This trend underscores the value of the Metal-THINGS dataset in extending the stellar mass range and revealing features and behaviours not captured by SAMI alone.

We also examined how the slopes of the piecewise fits in this relation change across the three dataset combinations. Notably, the slopes exhibit small differences, mainly located in the low-mass range. When Metal-THINGS data is added to the SAMI sample, the low-mass slope steepens or remains similar depending on the band: $-0.04\pm0.01 \; \mathrm{dex}/R_{\rm eff, NUV}$ from SAMI compared to $-0.04\pm0.3 \; \mathrm{dex}/R_{\rm eff, NUV} \text{ or } -0.06\pm0.3 \; \mathrm{dex}/R_{\rm eff, NUV}$ (see Appendix \ref{fit_parameters_rel}). This highlights that the inclusion of Metal-THINGS galaxies not only broadens the stellar mass range, but may also modify the overall shape of the relation, especially where SAMI's coverage is more limited.

We can now briefly explore potential explanations for those galaxies that deviate the most from the studied trends, that is, those galaxies that fall higher or lower than previous estimates for galaxies at similar stellar mass ranges. These are the following:

(i) DDO 53, which shows a steep negative gradient with the r-band normalisation. In \citet{DDO53_exp}, they present evidence of current starburst behaviour possibly triggered by tidal disturbances by the M 81 group as a whole, or by interactions with the intergalactic medium (IGM). They also show a value for this galaxy's weighted mean metallicity compatible with that calculated in this work, which can be considered as evidence to disregard uncertainties in individual fibres as the likely reason for its particular gradient, although it cannot be excluded that this deviation could be artificial;

(ii) In DDO 154, we found evidence of an extended low surface brightness component in the outer parts, as well as no evidence of interactions with any nearby galaxy \citep{DDO154_exp1, DDO154_exp2};

(iii) Holmberg I, for which evidence of ram pressure stripping has been found within the M 81 group in addition to four expanding supershells of H \small I \normalsize and 9 new faint H \small II \normalsize regions, which are likely candidates for mechanisms that explain its positive metallicity gradient \citep{HoI_exp};

(iv) IC 2574 is a gas-rich dwarf galaxy which is currently forming stars and it does not show evidence of interactions with other galaxies, one particular finding could be that around 90\% of the galaxy's total mass is in the form of dark matter \citep{IC2574_exp1, IC2574_exp2, IC2574_exp3};

(v) NGC 2366 is described to be a green pea analogue, and shows evidence of a small satellite cloud which could have triggered episodes of star formation in the southern half of the galaxy \citep{NGC2366_exp1, NGC2366_exp2};

(vi) NGC 5457 is the dominant structure in a small group of galaxies (the M101 Group) with evidence of ongoing interactions with its lower mass companions, which are likely contributing to the growth and reshaping of its outer disk \citep{NGC5457_exp};

(vii) NGC 3198 shows no signs of interactions \citep{ngc3198_1}. This galaxy shows significant inflows of gas that fit well with the picture that its star formation is mainly fuelled by them \citep{ngc3198_2}. While this could explain the steeper gradient of this galaxy, further studies are needed to clarify this behaviour.

Another notable case is NGC 7331, which exhibits significantly different radii when measured in the r-band and NUV-band. This discrepancy arises from its prominent central bulge, which is brighter in the redder parts of the spectrum. As a result, its metallicity gradient is more sensitive to the choice of normalisation radius. This will be further discussed in Valerdi et al. (in preparation).

All previous characteristics provide plausible explanations for the various identified peculiar behaviours.

For this work, we focused on the use of a threshold for DIG identification of $C_{\rm H\, II} < 0.4$. The use of different $C_{\rm H\, II}$ thresholds, ours and $C_{\rm H\, II} < 0.6$, explores the sensitivity of metallicity gradients to the assumed contribution of the DIG. The results of this comparison can be found in Figure \ref{fig:grad_mass_fit_HI_r}, where green arrows indicate the shift in gradient values for individual galaxies between the two thresholds, such that each arrow points from the gradient calculated with DIG fibres satisfying $C_{\rm H\, II} < 0.4$ to those calculated with $C_{\rm H\, II} < 0.6$.

Overall, we find that the choice of $C_{\rm H\, II}$ threshold results in relatively small differences in the derived metallicity gradients for most galaxies in our sample, with the gradients remaining generally consistent. However, one notable outlier, Holmberg I, exhibits a significantly larger shift in its gradient, which suggests that the contribution of DIG in this galaxy may play a more prominent role in shaping its metallicity distribution.

\subsection{Metallicity gradient - atomic gas fraction relation}

Studying the relationship between the metallicity gradient and the fraction of atomic gas provides valuable information about the interplay between the gas content and the chemical evolution of a galaxy. As H \small I \normalsize gas is a key reservoir for star formation, its fraction serves as an indicator of a galaxy's evolutionary state, where galaxies with depleted atomic gas tend to be in later stages of their evolution \citep[e.g.][]{Lagos_2011}. Analysing this relation following our study of the stellar mass will allow us to examine physical mechanisms such as gas loss, accretion or feedback, that drive the evolution of these gradients across our sample of galaxies.

In Figure \ref{fig:metgrad-gasfrac} we can find the relation between the metallicity gradient and H \small I \normalsize gas fraction for our sample. We perform a piecewise linear fit to these data, where generally we first find a negative linear correlation between the value of the metallicity gradients and the fraction of atomic gas up to a fraction of $\sim0.75$ or $\sim0.5$, depending on the normalization radius, followed by a positive linear correlation at higher atomic H \small I \normalsize fractions.

At the higher end of the H \small I \normalsize fraction ($f_{\rm g \; H \, I} \geq 0.75$, when using $R_{\rm eff,NUV}$ or $R_{\rm eff,r}$), metallicity gradients show higher dispersion, with some galaxies exhibiting steeper gradients. This is consistent with these systems being gas-rich galaxies, potentially being at earlier evolutionary stages \citep[e.g.][]{Sorai_2019, Casasola_2020} or more influenced by recent inflows of metal-poor gas \citep[e.g.][]{Casasola_2004, Mancillas_2019}, which steepen these gradients. This relationship could potentially serve as an initial estimate on metallicity gradients for galaxies with estimated H \small I \normalsize gas and stellar masses. Studying cases with a higher confidence, the trend indicates that, generally, as the fraction of atomic gas of a given galaxy increases, its metallicity gradient steepens up to $f_{\rm g \; H \, I} \sim 0.75$, from which point onwards they become shallower.

In Table \ref{tab:RMSE_table}, we show the root-mean-square errors for the piecewise linear fits as well as for simple Bayesian linear fits for this relation, where we calculated this value globally and separately for points that fall below and above the break point. This helps us study the best fit values and normalisation radius for different ranges of gas content in galaxies. We generally see similar values or ones with lesser deviations for the piecewise fit, and for galaxies with a higher gas content, the gradients calculated by normalizing to the NUV effective radius exhibit values with a lower dispersion. Additionally, we find smaller uncertainties in the gradients for these galaxies in the NUV-band, with this effect being particularly evident for DDO 53. This improvement in precision highlights the potential use of this band when studying metallicity gradients in galaxies within this gas fraction regime.

In contrast, for galaxies with a lower atomic gas content, the gradients derived using the r-band tend to show a much better match in these fits, with a lower RMSE. This suggests that the r-band may provide a more accurate representation of metallicity distributions for galaxies more rich in atomic gas.

In the case of galaxies falling between these two extremes, that is, for intermediate-mass galaxies the choice of band seems to have a less significant impact on the derived metallicity gradients, as both bands yield similar results for our sample in this range.

In these cases, normalisation with $R_{25}$ shows no advantages over those performed with different effective radii. For this analysis, we exclude results for normalisation using effective radii in the K$_s$ band as our sample is too small. In our case, K$_s$-band normalisation shows a good fit for galaxies with lower gas content, while for larger fractions our sample lacks data.

Returning to Figure \ref{fig:metgrad-gasfrac}, we also find that galaxies with a lower fraction of atomic gas are generally more massive than those at higher fractions, with two more prominent exceptions in the lower end of the plot, NGC 1569 and NGC 2976. The behaviour of these galaxies could be explained as follows: (i) for NGC 1569, we found evidence of supernova explosions expelling H \small I \normalsize gas from the central regions of the galaxy and large outflows of gas extending to the north regions of the galaxy \citep{NGC1569_exp}; (ii) NGC 2976's environment contains a cloud of neutral hydrogen potentially interacting with the galaxy and shows evidence of tidal stripping \citep{NGC2976_exp}, explaining its gas deficiency. It is worth noting that NGC 1569 and NGC 2976 remain poorly studied in this regard. Further investigating this aspect would require more resolved data in optical and radio wavelengths.

\section{Summary and conclusions}

In this paper, we analyse a sample of 25 galaxies (for a total of 102 GCMS pointings) of the Metal-THINGS survey using integral field spectroscopic data. We corrected for the effects of extinction every emission line used, and identified Diffuse Ionised Gas (DIG) regions in every case. We used BPT diagnostic diagrams to classify our data as AGN, star-forming or composite type regions, and  estimated the gas metallicity for the star-forming type regions using the S calibration described in \citet{Pilyugin_2016}. Using Bayesian MCMC chains, we fit our data to linear models to acquire the metallicity gradients for the entire sample using different normalisations ($R_{\rm eff, \, r}$, $R_{\rm eff, \, NUV}$, $R_{\rm eff, \, Ks}$ and $R_{25}$).

We contrast our results with previous works and analyse each individual case of the sample. Our findings can be summarised as follows:

\begin{itemize}
    \item Negative metallicity gradients dominate our sample, which is consistent with inside-out growth in galaxies. Most results show good quality fits, but smaller irregular galaxies pose challenges to fit a model, as data quality worsens (i.e., the case of Holmberg I). Furthermore, values obtained for mean gas-phase metallicities are compatible with those found in previous works.
    \item We find no significant effects in the resulting metallicity gradients caused by the presence of emission by DIG in these galaxies, with the exception of two cases, Holmberg I and NGC 4736, which have been discussed in detail in Section \ref{sect:effect_of_dig}. Furthermore, we find that the choice of threshold for the identification of DIG fibres generally produces minimal differences in the gradients, with the exception of Holmberg I, which shows a significant shift.
    \item Our metallicity gradients are consistent with results for similar cases from previous works, more precisely that of \citet{Poetrodjojo_2021}, and present smaller dispersion. When plotted as a function of the stellar mass, we find evidence for a break in a piecewise linear fit at $\log(M_{\rm star}/M_\odot) \simeq 9.5$, such that galaxies at lower masses have steeper gradients with increasing mass, and galaxies at higher masses show increasingly shallower gradients.
    \item We find a bimodal correlation on the value of the metallicity gradient of a galaxy and its H \small I \normalsize gas, such that, as the fraction of H \small I \normalsize gas in a galaxy increases, its gradient becomes  steeper in a linear manner up to a gas fraction of $\sim$ $0.75$. From atomic gas fractions of $0.75$ onwards, the dispersion of this relation increases, as galaxies with higher quantities of H \small I \normalsize tend to be dwarfs/irregular in early stages of their evolution or affected by different processes in their environment. This relationship could serve as a tool for a first estimate on the metallicity gradient of a galaxy by studying its H \small I \normalsize content.
    \item Analysing our results for metallicity gradients using effective radii in both the r-band and NUV-band, we can safely confirm that in the case of galaxies with higher atomic gas content ($f_{\rm g, H\, I} \gtrsim 0.75$), the use of the NUV-band should be preferred, while in galaxies more deficient in gas ($f_{\rm g, H\, I} \lesssim 0.75$), the use of the r-band (or K$_s$-band) is suggested. Cases falling close to this value show no evident signs of the preferred use of one band over the other.
    \item Using the stellar mass for the same task, we find that galaxies with $\log(M_{\rm star}/M_\odot) \gtrsim  9.5$ exhibit smaller dispersion when normalised using r-band effective radii. This is likely because the brightness of these galaxies is more intense in this band and thus better characterised. Conversely, galaxies with $\log(M_{\rm star}/M_\odot) \lesssim  9.5$ show smaller dispersion when using the effective radii in the NUV. 

    \item We found that the galaxies NGC 1569 and NGC 2976 exhibit lower atomic gas mass fractions when compared to other galaxies of similar stellar masses, which is potentially due to environmental effects that expel gas from these systems \citep{NGC1569_exp, NGC2976_exp}.
\end{itemize}

The Metal-THINGS survey, with its high-resolution IFU data, proves especially well-suited for studying the interplay between gas content and the metallicity structure. Future work expanding this dataset, or combining it with other IFU surveys, have the potential to enable more comprehensive studies of how internal processes and environmental factors shape chemical enrichment across different galaxy types and evolutionary stages.

\begin{acknowledgements}
We thank the referee for his/her thoughtful comments that improved the manuscript, and our Editor for her guidance.
\\

MALL acknowledges support from the Ramón y Cajal program RYC2020-029354-I funded by MICIU/AEI/10.13039/ 501100011033 by ESF+, and the  Spanish grant PID2021-123417OB-I00, funded by MCIN/AEI/10.13039/501100011033/FEDER, EU.
\\

SPO acknowledges support from the Comunidad de Madrid Atracción de Talento program via grant 2022-T1/TIC-23797, and grant PID2023-146372OB-I00 funded by MICIU/AEI/10.13039/501100011033 and by ERDF, EU.
\\

LSP acknowledges support from the Research Council of Lithuania (LMTLT, No. P-LU-PAR-25-8).
\\

JCN ackowledges support by the Evolution of Galaxies project, of reference PID2021-122544NB-C41, within the Programa estatal de fomento de la investigación científica y técnica de excelencia del Plan Estatal de Investigación Científica y Técnica y de Innovación of the Spanish Ministry of Science and Innovation/State Agency of Research MCIN/AEI.
\\

VC acknowledges funding from the INAF Mini Grant 2022 program ``Face-to-Face with the Local Universe: ISM’s Empowerment (LOCAL)''.
\\

MEDR acknowledges support from {\it Agencia Nacional de Promoci\'on de la Investigaci\'on, el Desarrollo Tecnol\'ogico y la Innovaci\'on} (Agencia I+D+i, PICT-2021-GRF-TI-00290, Argentina).
\\

JF acknowledges financial support from the UNAM-DGAPA-PAPIIT IN111620 grant, Mexico
\end{acknowledgements}

\bibliographystyle{aa.bst} 
\bibliography{bibliography} 

\clearpage

\appendix

\section{H$\alpha$ maps} \label{h_alpha_maps}

\begin{figure}[H]
\onecolumn
\centering
        \includegraphics[scale=0.395]{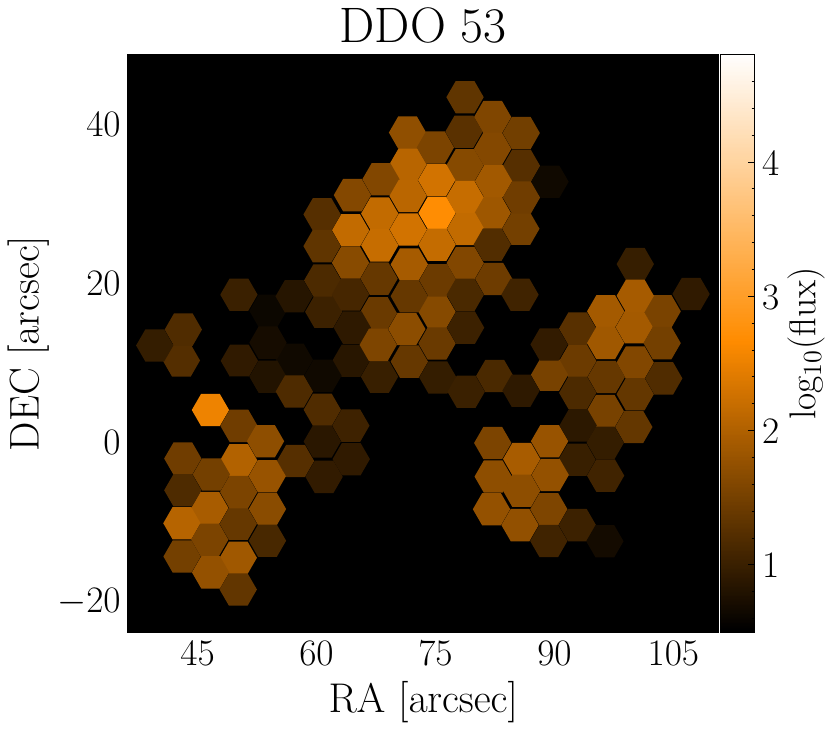}
        \hspace{0.1cm}
        \includegraphics[scale=0.395]{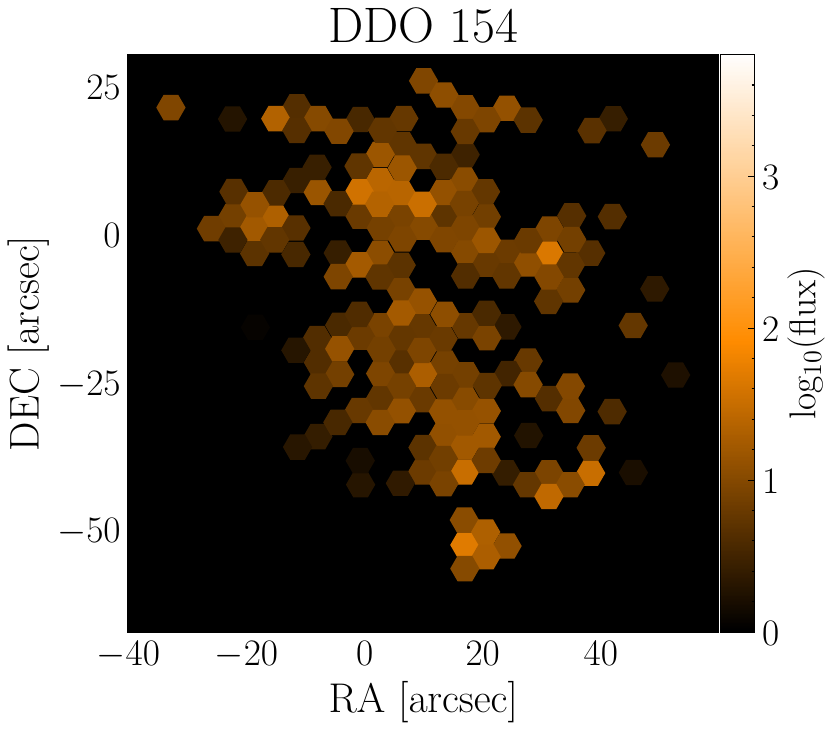}
        \hspace{0.1cm}
        \includegraphics[scale=0.395]{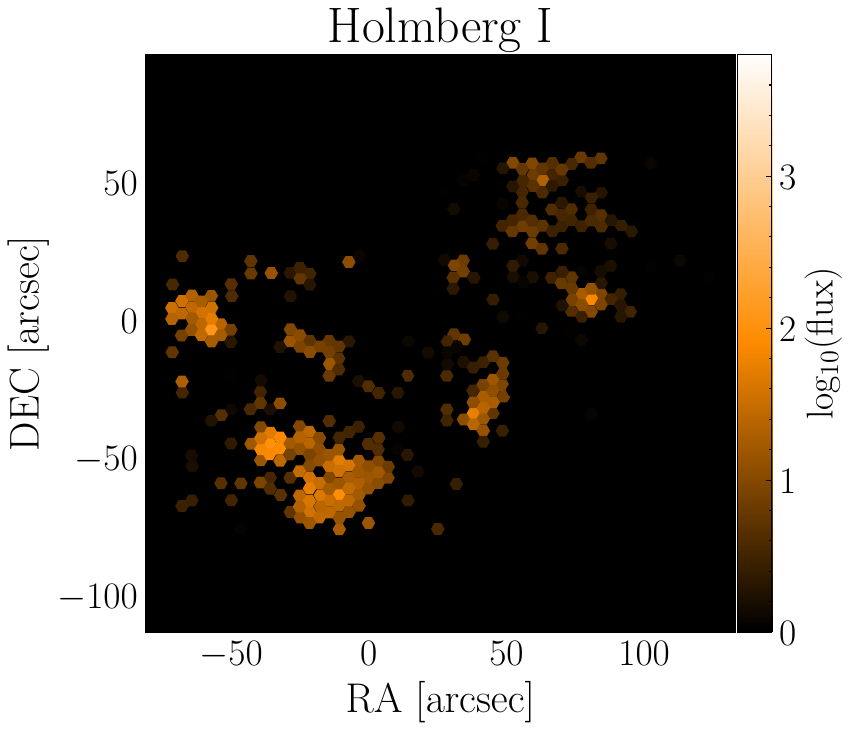}

        \vspace{0.75cm}
        
        \includegraphics[scale=0.395]{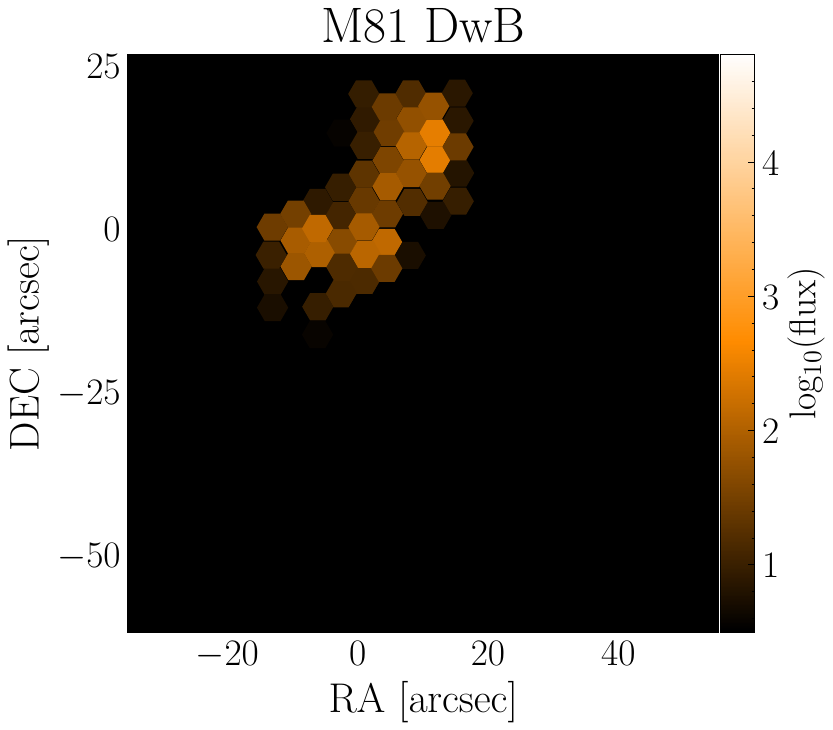}
        \hspace{0.1cm}
        \includegraphics[scale=0.395]{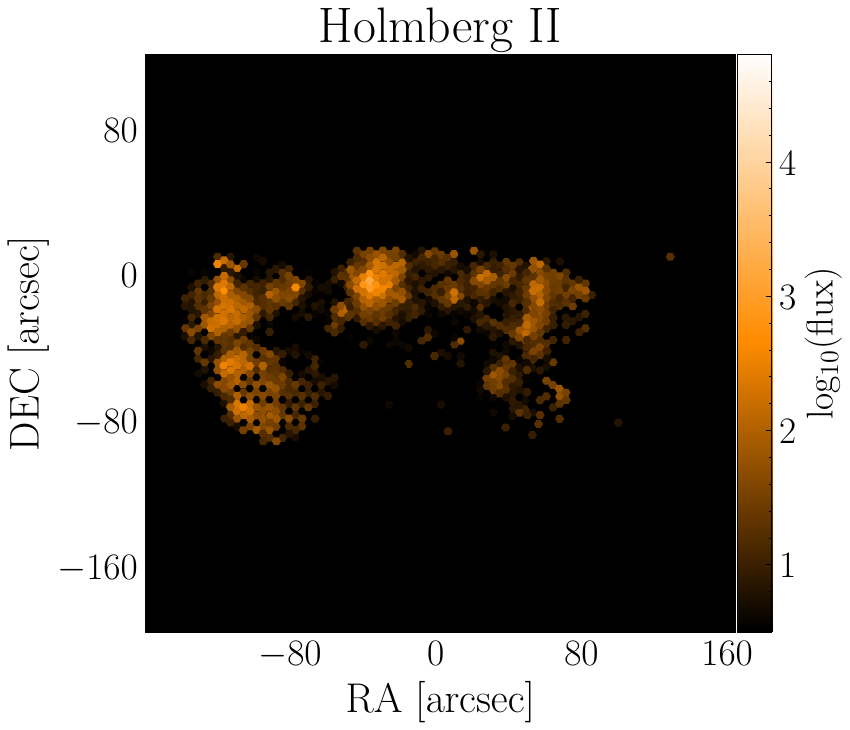}
        \hspace{0.1cm}
        \includegraphics[scale=0.395]{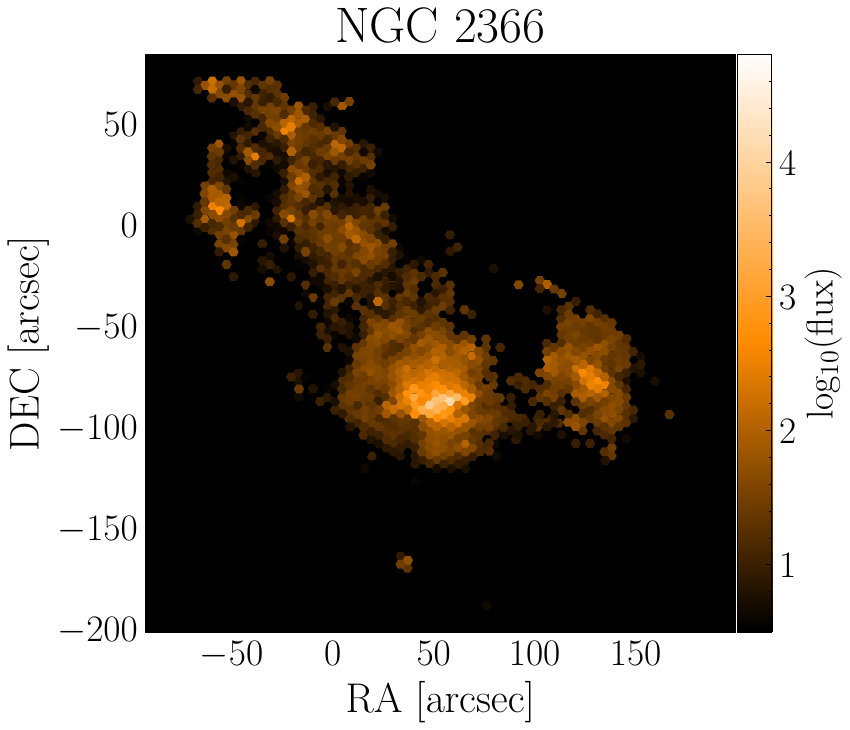}
        \vspace{0.75cm}

        \includegraphics[scale=0.395]{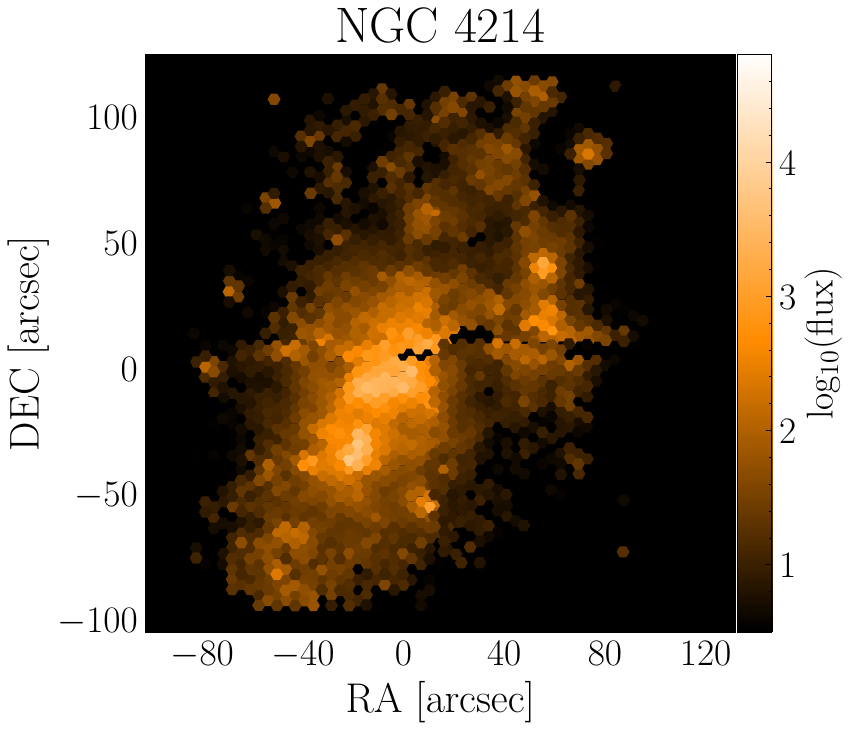}
        \hspace{0.1cm}
        \includegraphics[scale=0.395]{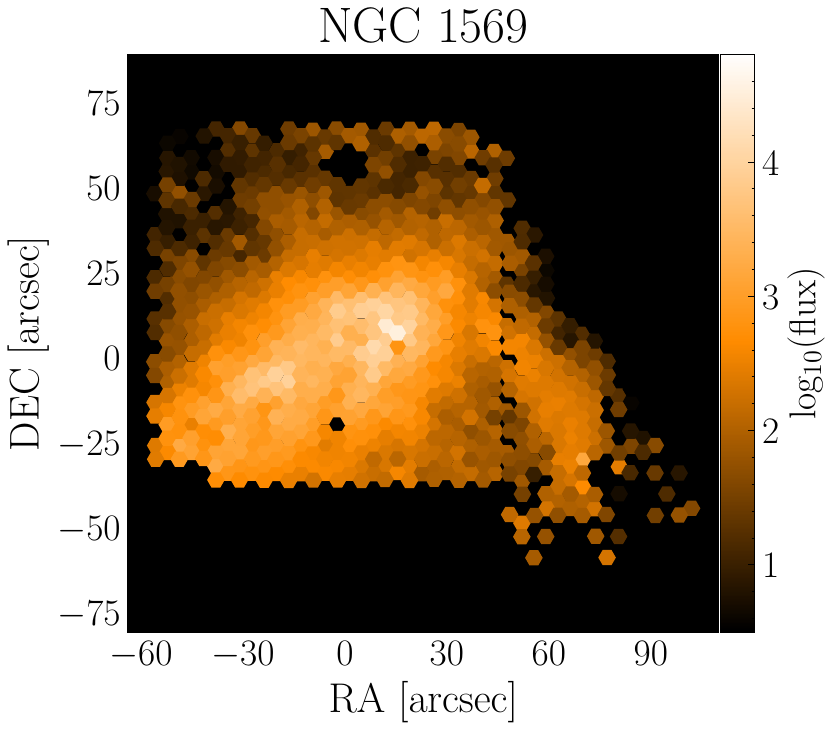}
        \hspace{0.1cm}
        \includegraphics[scale=0.395]{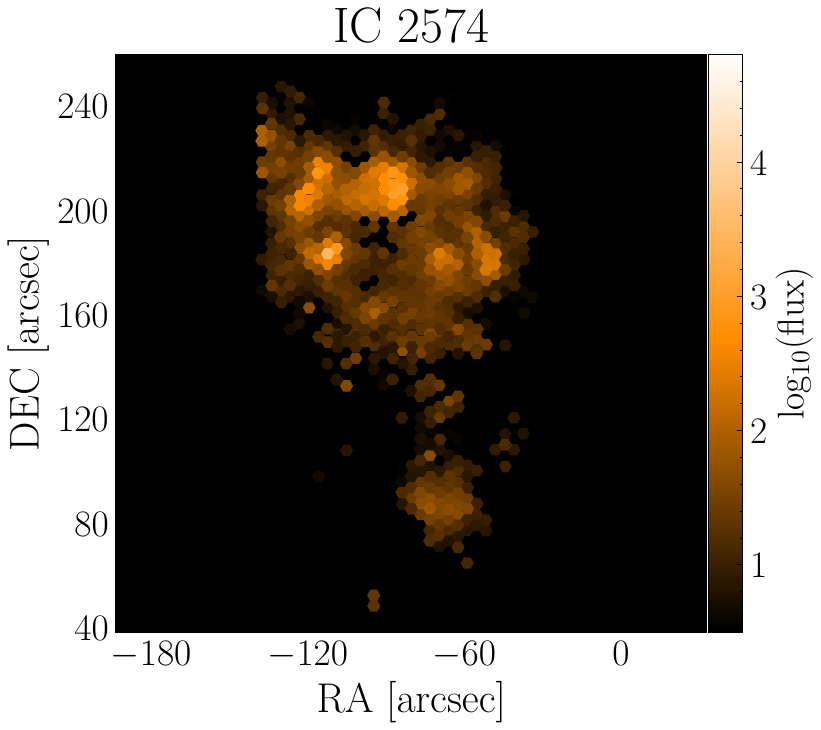}

        \vspace{0.75cm}

        \includegraphics[scale=0.395]{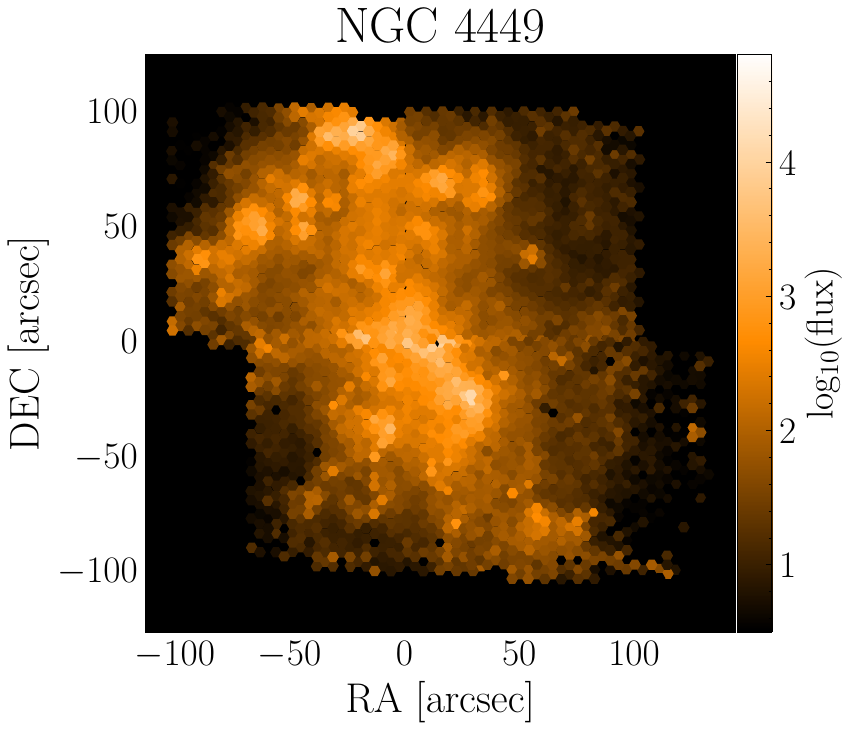}
        \hspace{0.1cm}
        \includegraphics[scale=0.395]{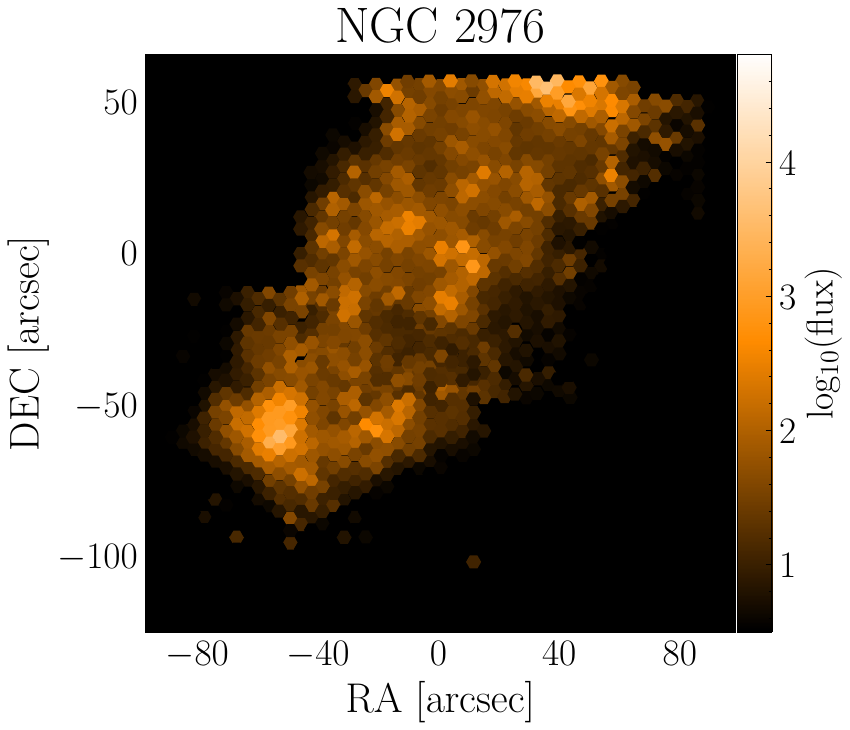}
        \hspace{0.1cm}
        \includegraphics[scale=0.395, align = center]{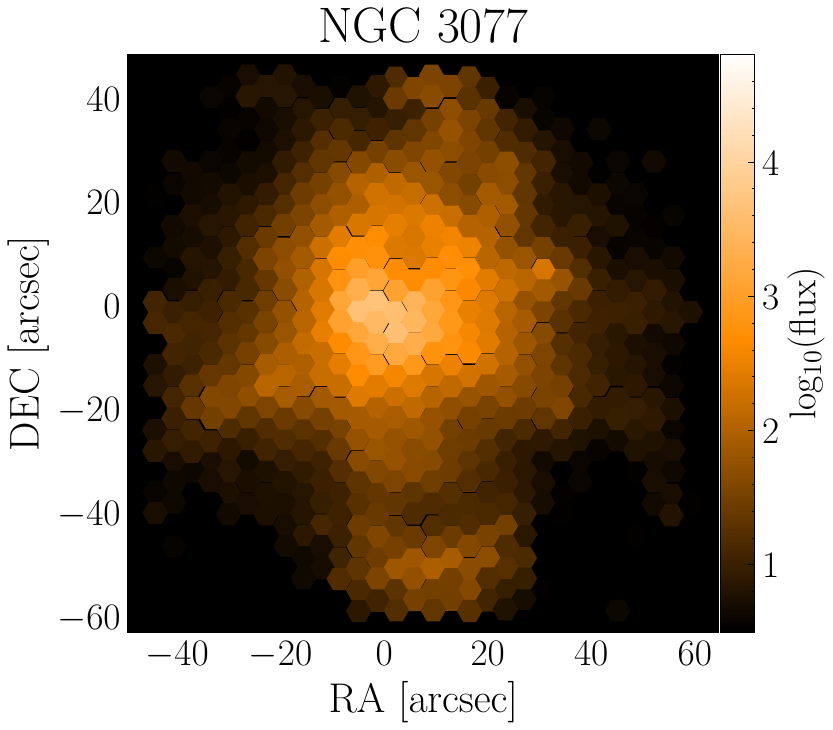}

        \caption{Extinction corrected H$\alpha$ maps for the observed galaxies, labelled accordingly. Only fibres with a $\rm SNR > 3$ for the $\rm H\alpha$ and $\rm H\beta$ lines are shown. DDO 154 and Holmberg I have different flux scales to account for their lower surface brightness. All fluxes are given in units of $10^{-16} \; {\rm erg/s/cm^2/\AA}$.}
        \label{fig:Ha_maps_1}
        \twocolumn
\end{figure}

\begin{figure*}[h]
\vspace{0.5cm}
\centering

        \includegraphics[scale=0.395]{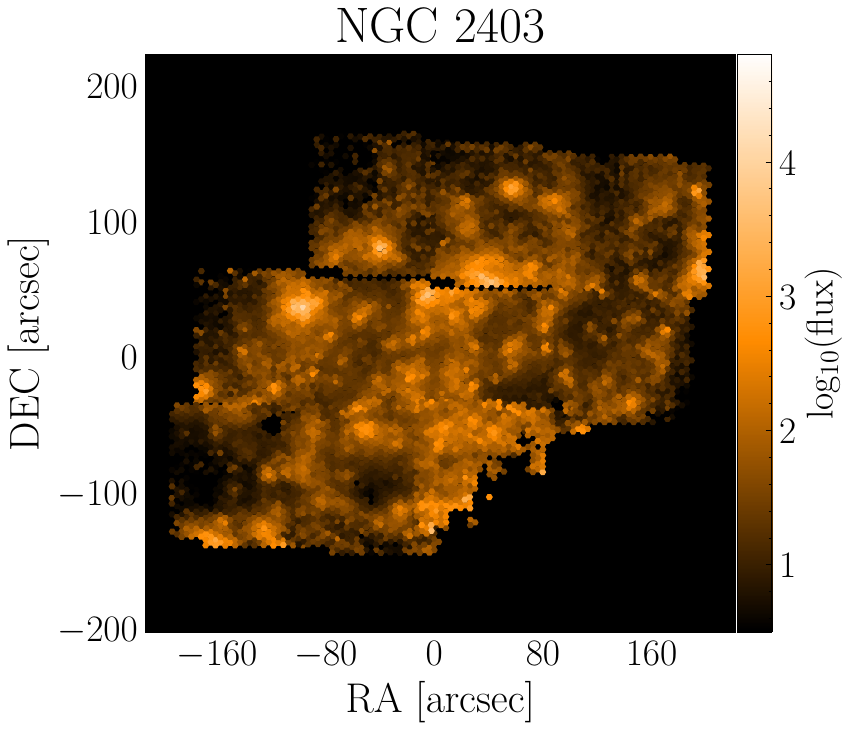}
        \hspace{0.1cm}
        \includegraphics[scale=0.395]{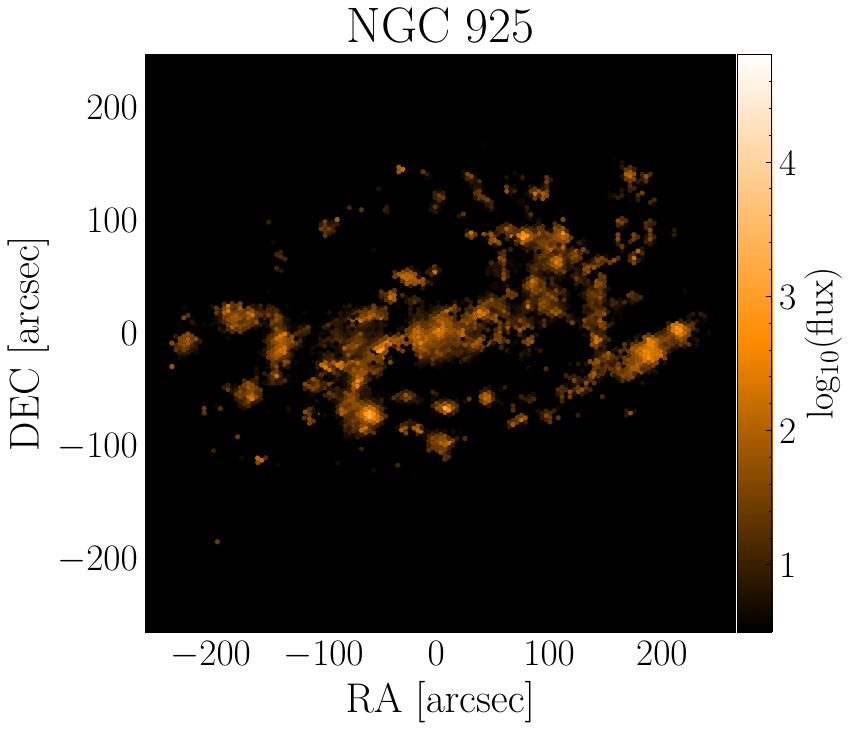}
        \hspace{0.1cm}
        \includegraphics[scale=0.395]{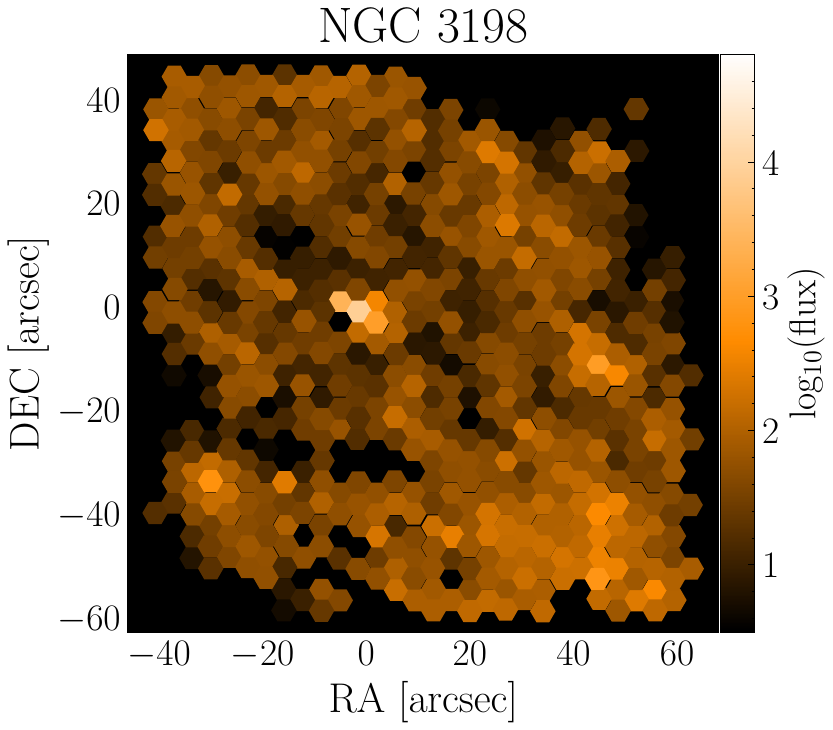}
        
        \vspace{0.25cm}

        \includegraphics[scale=0.395]{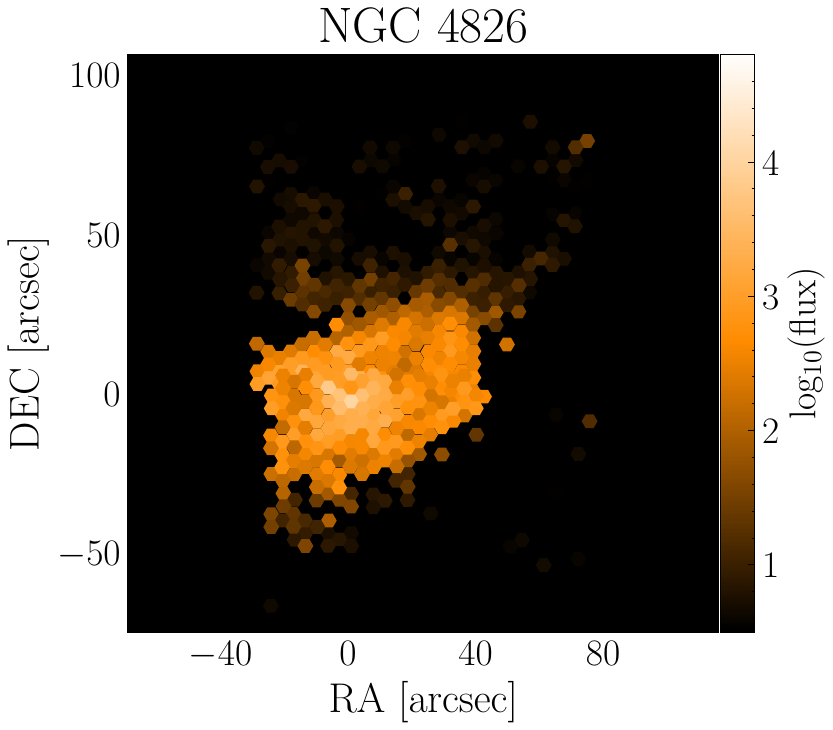}
        \hspace{0.1cm}
        {\includegraphics[scale=0.395]{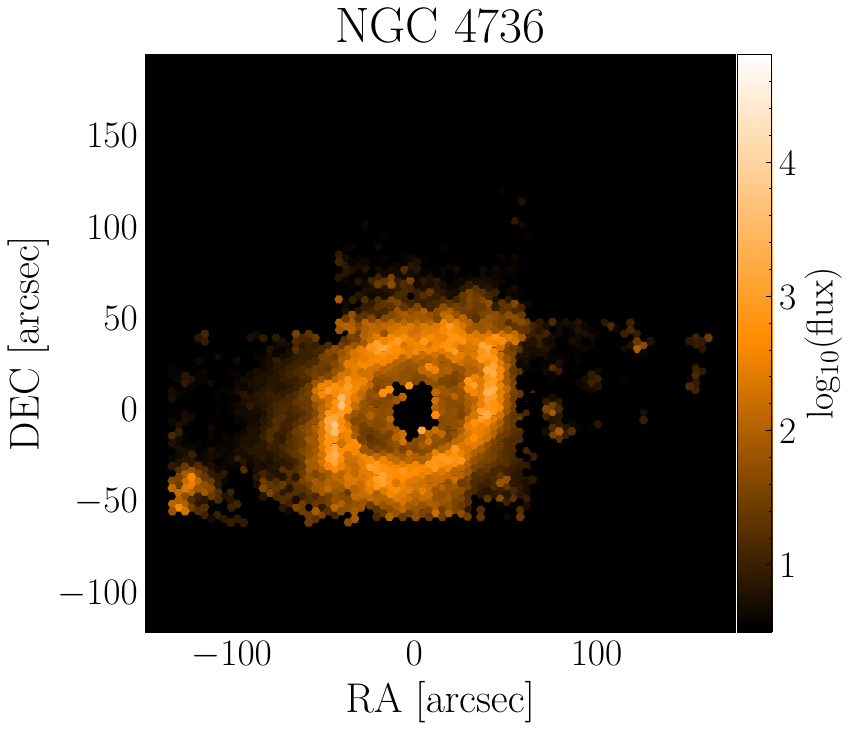}}
        \hspace{0.1cm}
        \raisebox{0.5cm}{\includegraphics[scale=0.395]{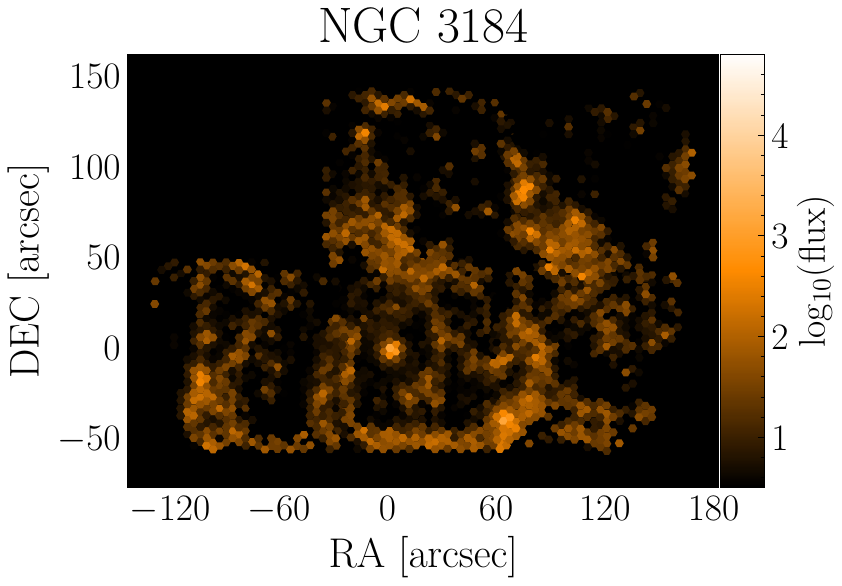}}

        \vspace{0.25cm}

        \raisebox{1cm}{\includegraphics[scale=0.395]{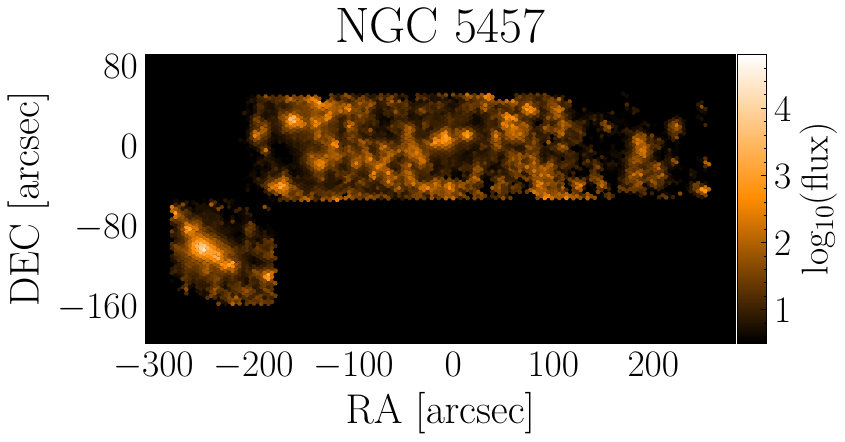}}
        \hspace{0.1cm}
        \includegraphics[scale=0.395]{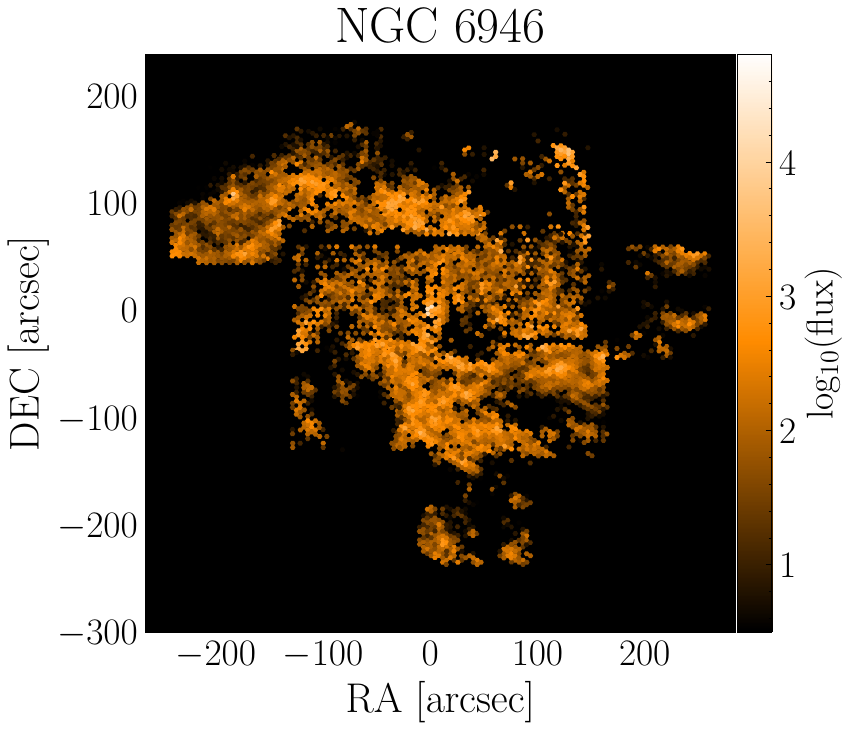}
        \hspace{0.1cm}
        \raisebox{1cm}{\includegraphics[scale=0.395]{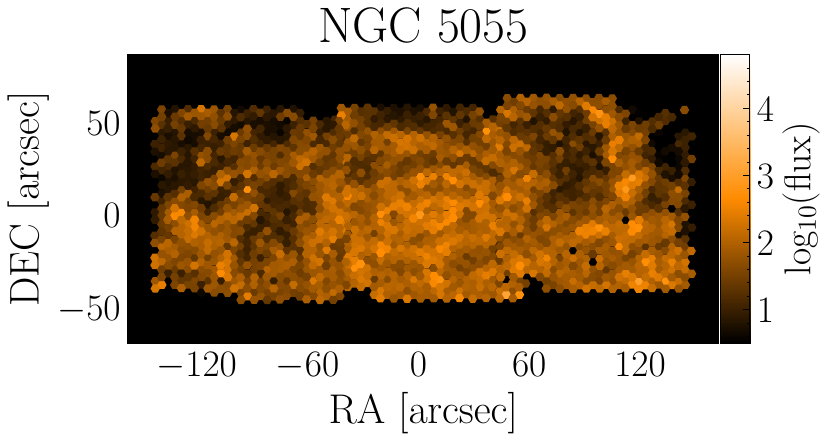}}

        \vspace{0.25cm}

        \includegraphics[scale=0.395]{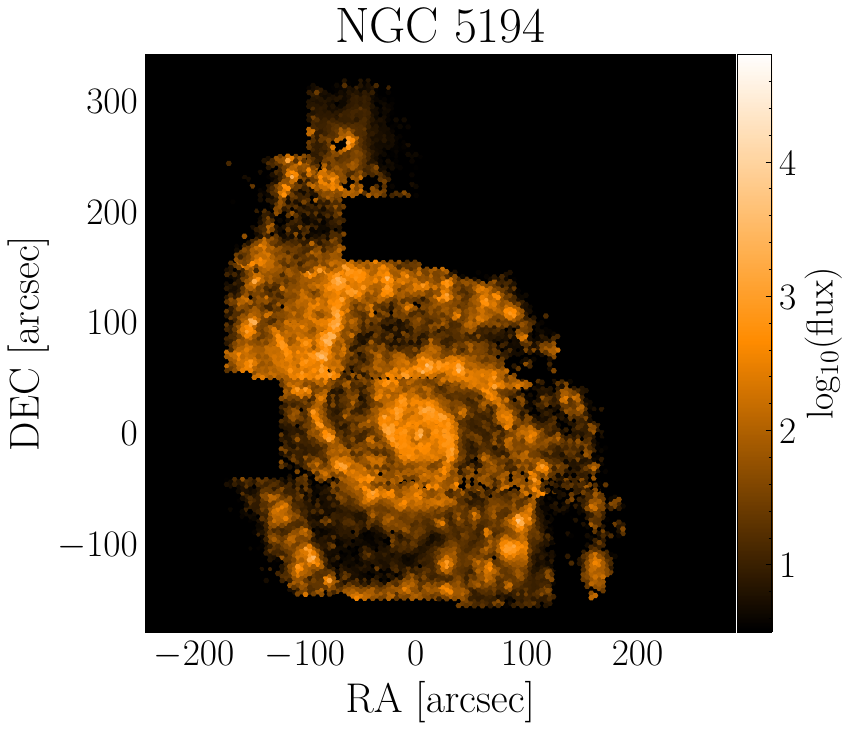}
        \hspace{1cm}
        \includegraphics[scale=0.395]{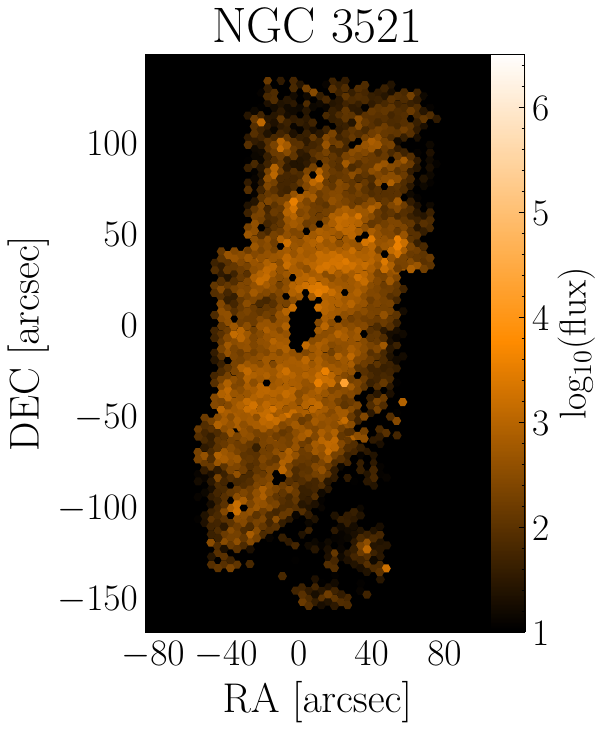}
        \hspace{1cm}
        \includegraphics[scale=0.395]{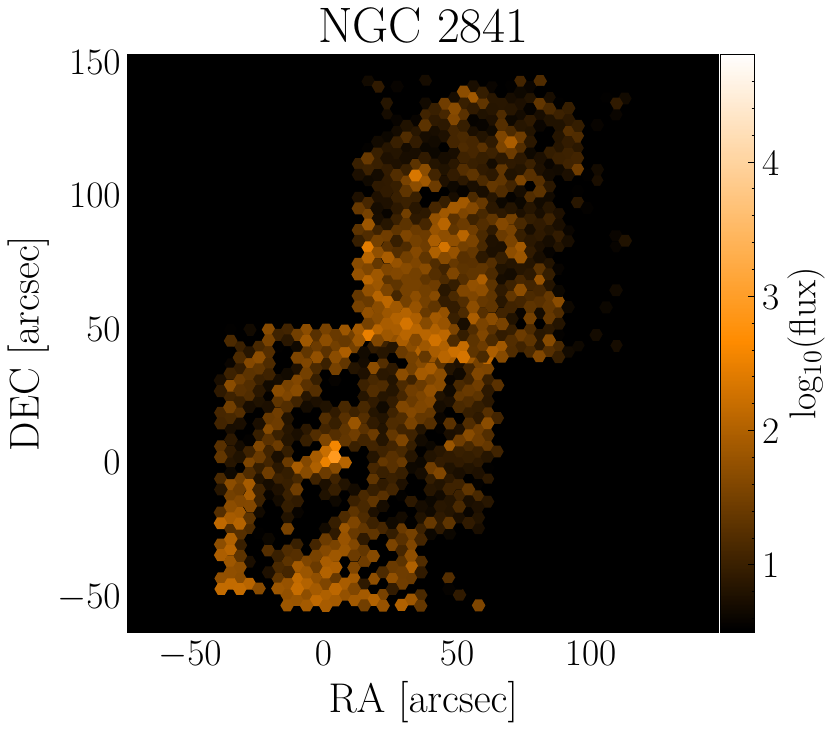}

        \caption{Same as Figure \ref{fig:Ha_maps_1}, but showing the second part of the sample}
        \label{fig:Ha_maps_2}
\end{figure*}

\clearpage
\onecolumn

\begin{figure}[h]

\vspace{0.5cm}
\centering

    \includegraphics[scale=0.395]{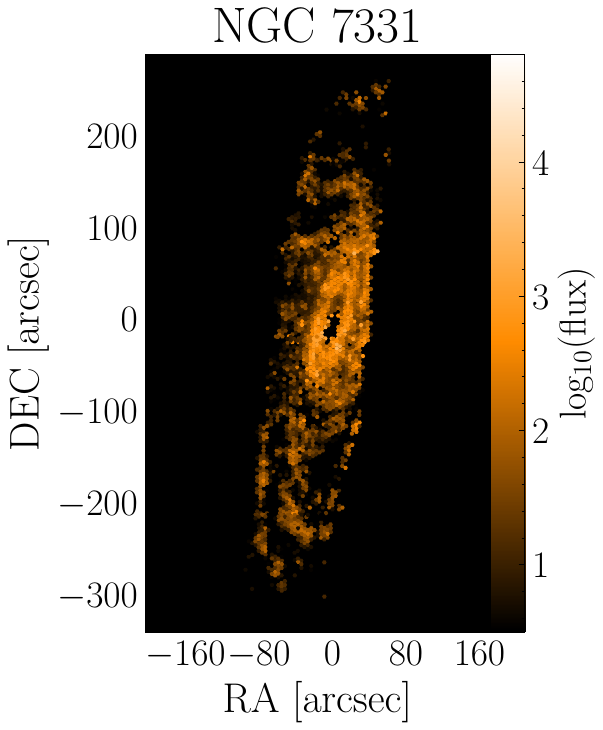}

    \caption{Same as Figures \ref{fig:Ha_maps_1} and \ref{fig:Ha_maps_2}, but showing the final part of the sample}
    \label{fig:Ha_maps_3}
        
\end{figure}

\vspace{-0.4cm}
\section{[N II] BPT maps}
\vspace{-0.4cm}

\begin{figure}[H]

\centering
        \includegraphics[scale=0.39]{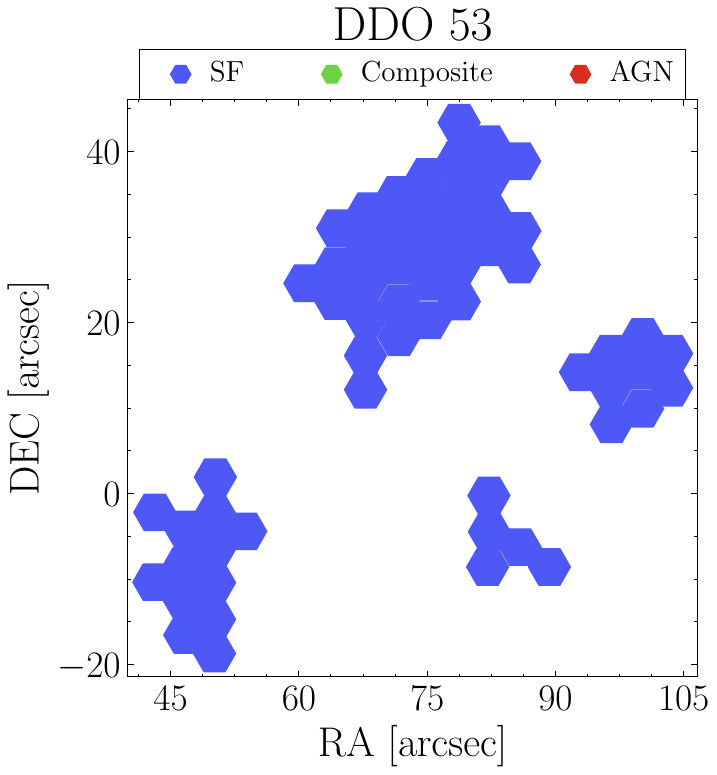}
        \hspace{0.75cm}
        \includegraphics[scale=0.39]{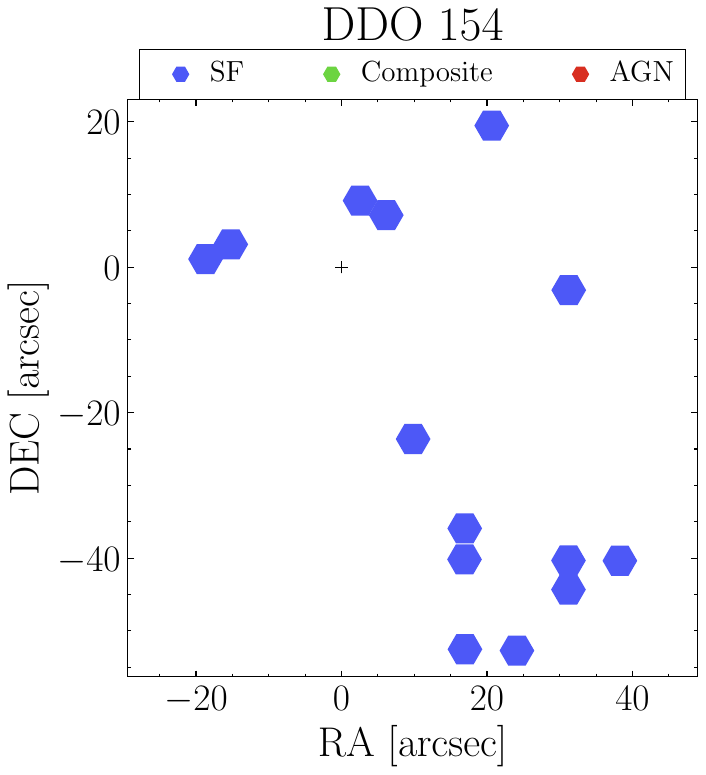}
        \hspace{0.75cm}
        \includegraphics[scale=0.39]{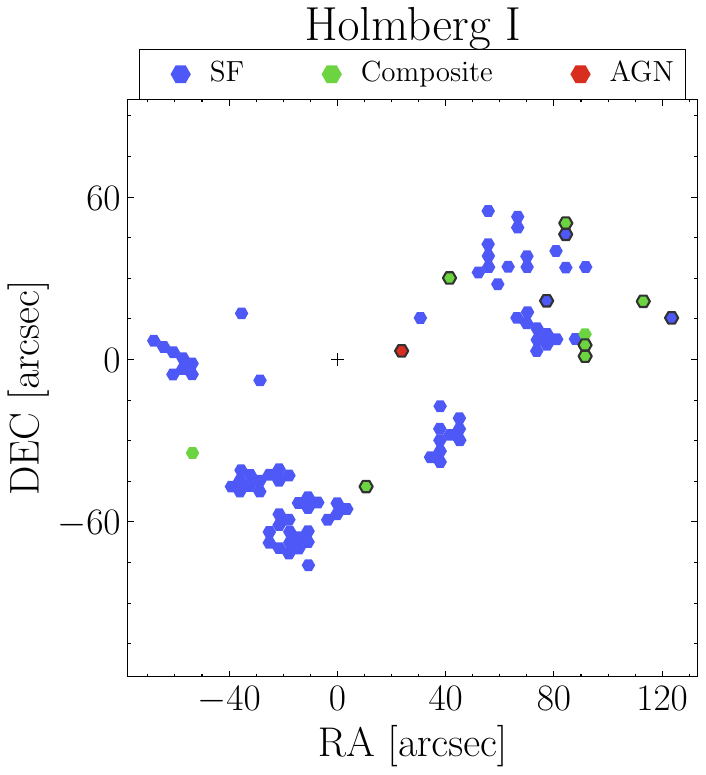}

        \vspace{0.25cm}

        \includegraphics[scale=0.39]{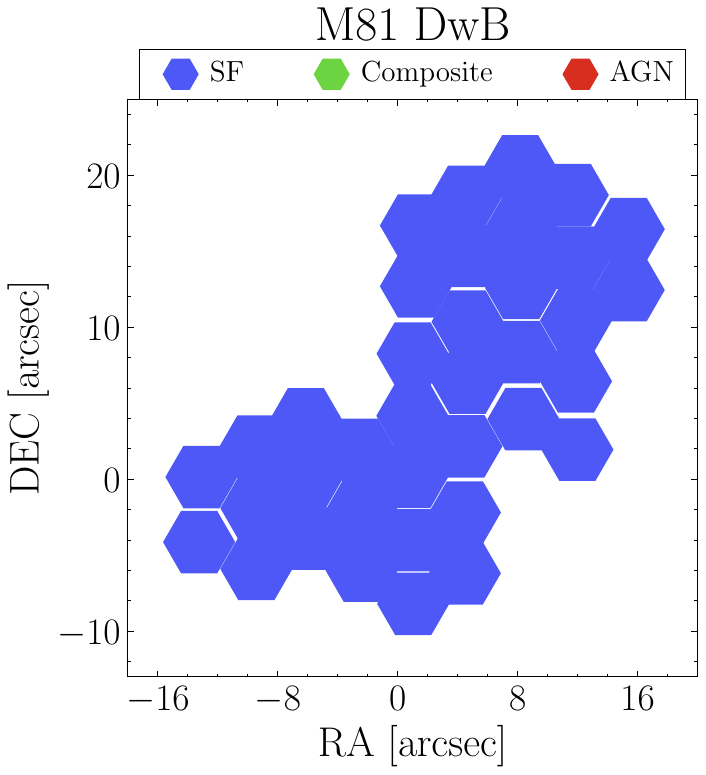}
        \hspace{0.75cm}
        \includegraphics[scale=0.39]{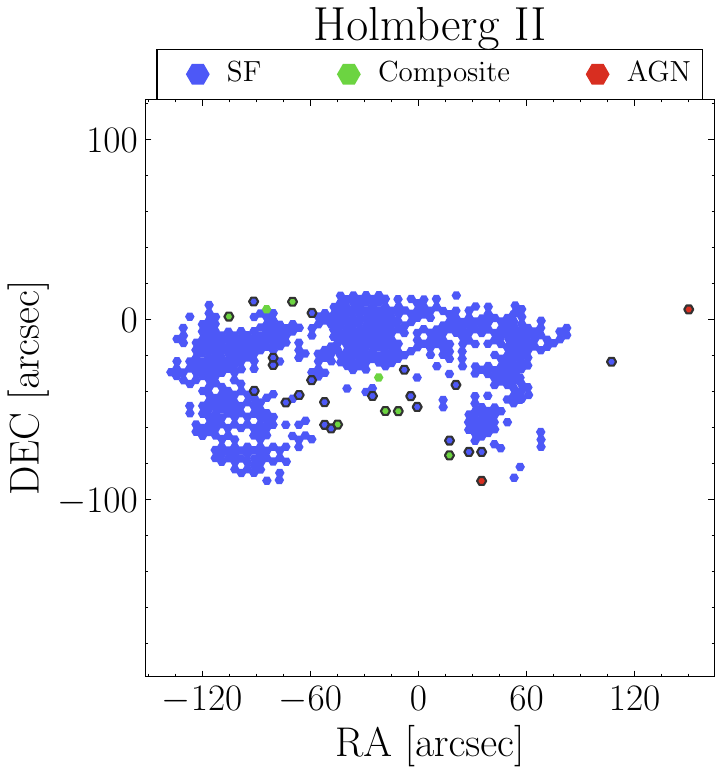}
        \hspace{0.75cm}
        \includegraphics[scale=0.39]{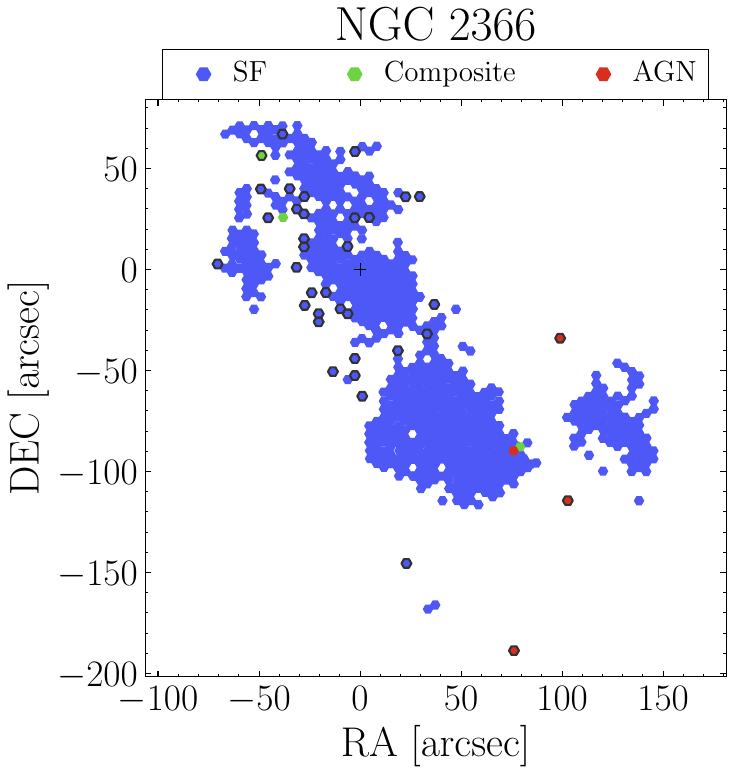}

        \vspace{0.25cm}

        \includegraphics[scale=0.395]{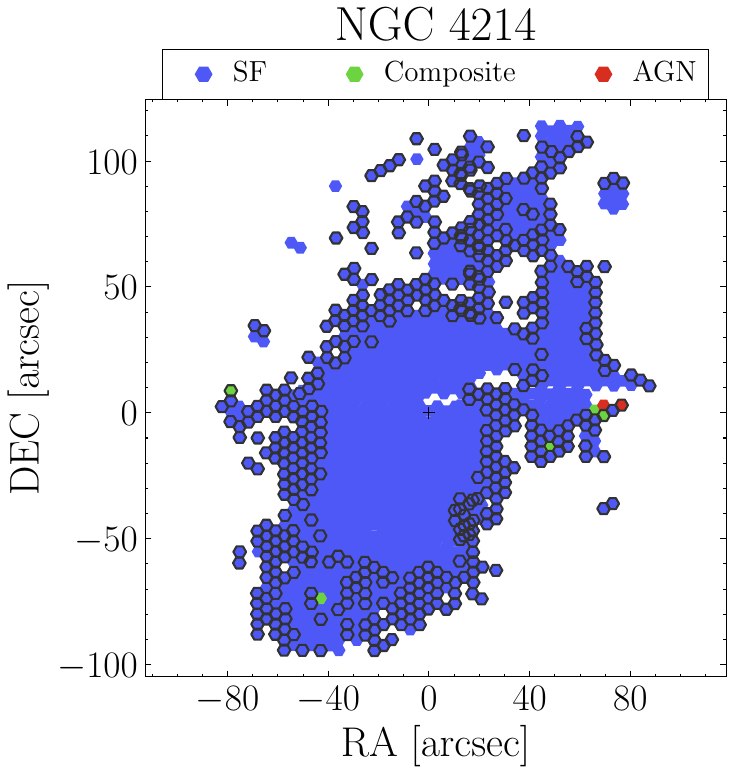}
        \hspace{0.75cm}
        \includegraphics[scale=0.39]{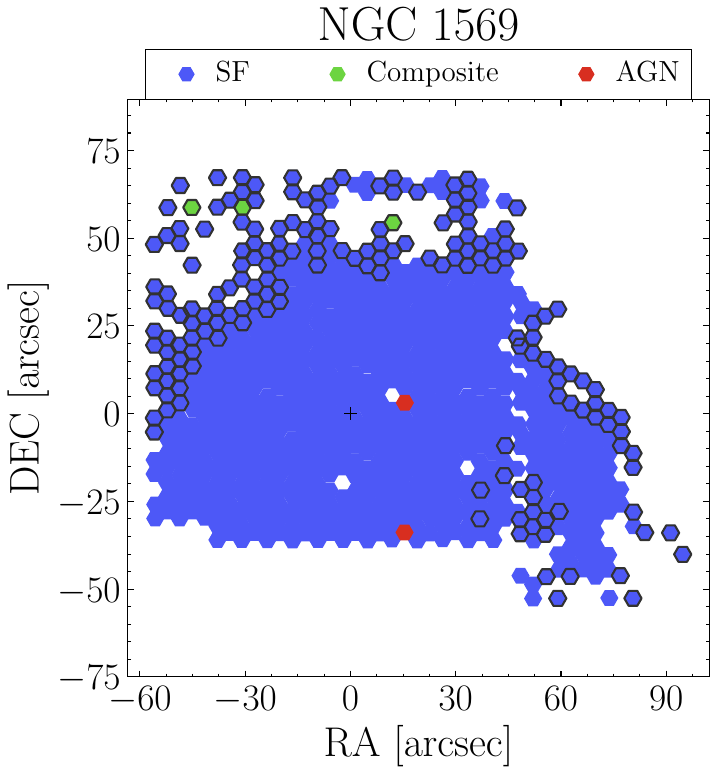}
        \hspace{0.75cm}
        \includegraphics[scale=0.39]{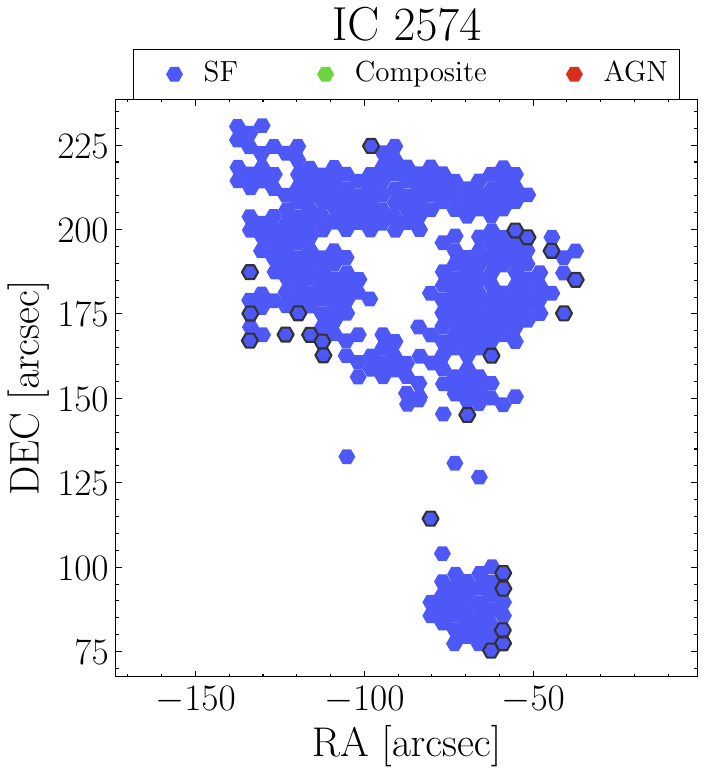}

        \caption{[N II] BPT diagnostic maps for the observed galaxies, labelled accordingly, with SF-like (in blue), composite (in green) and AGN-like (in red) characteristics. Only fibres with a $\rm SNR > 3$ in all lines used are shown. Fibres identified as DIG have been flagged in dark grey.}
        \label{fig:NII_BPT_1_map}
        
\end{figure}

\twocolumn

\clearpage

\begin{figure*}[h]
\vspace{0.5cm}
\centering

        \includegraphics[scale=0.395]{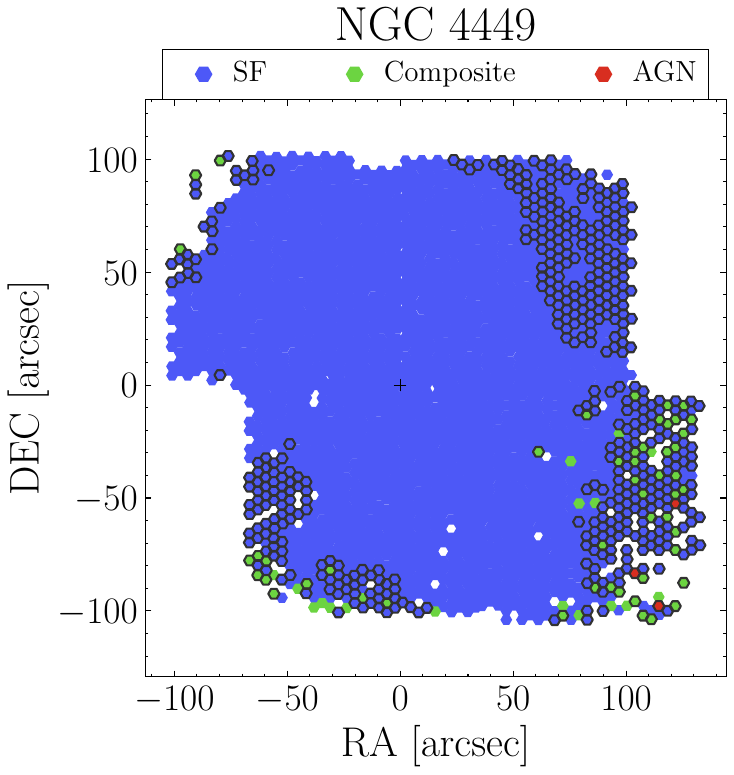}        
        \hspace{0.75cm}
        \includegraphics[scale=0.39]{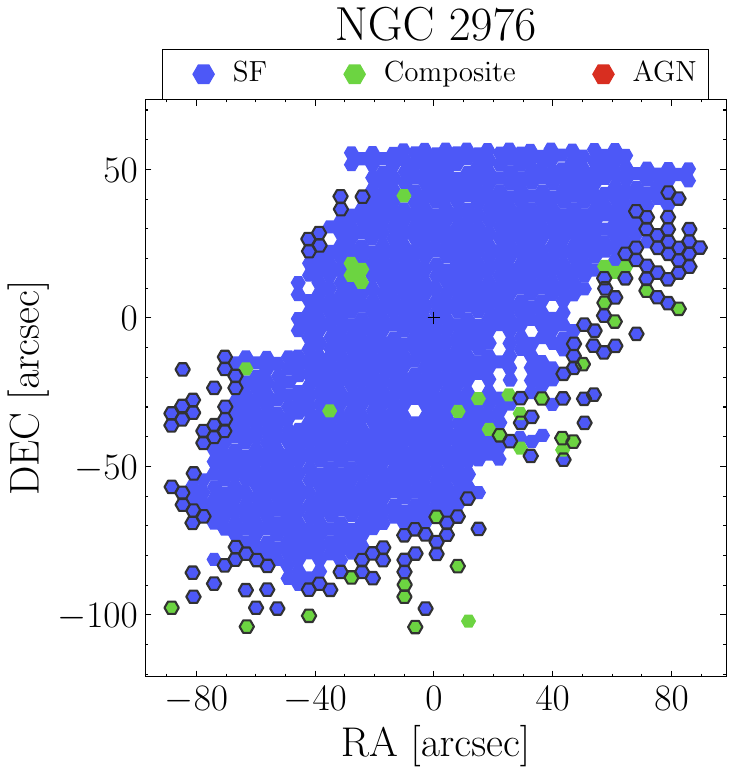}
        \hspace{0.75cm}
        \includegraphics[scale=0.39]{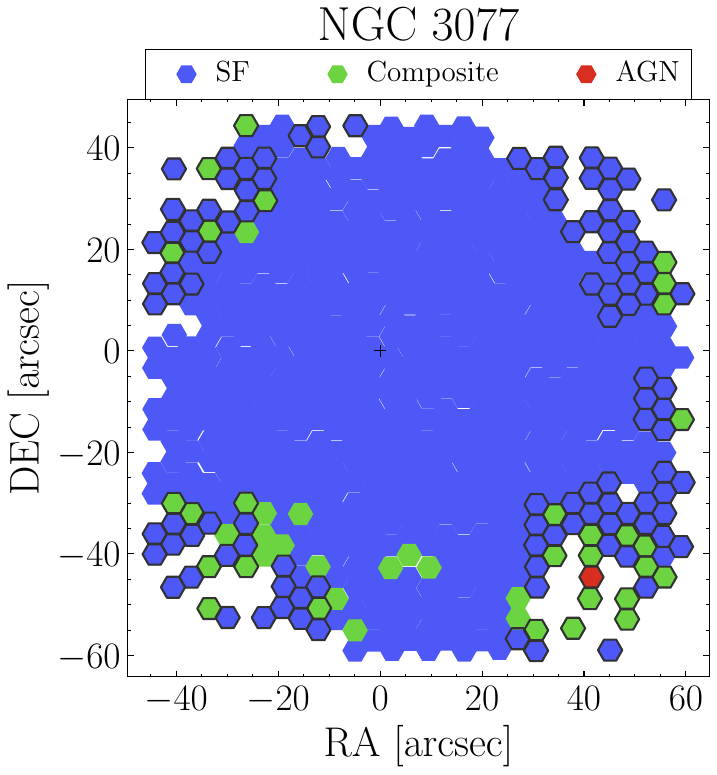}

        \vspace{0.25cm}
        
        \includegraphics[scale=0.39]{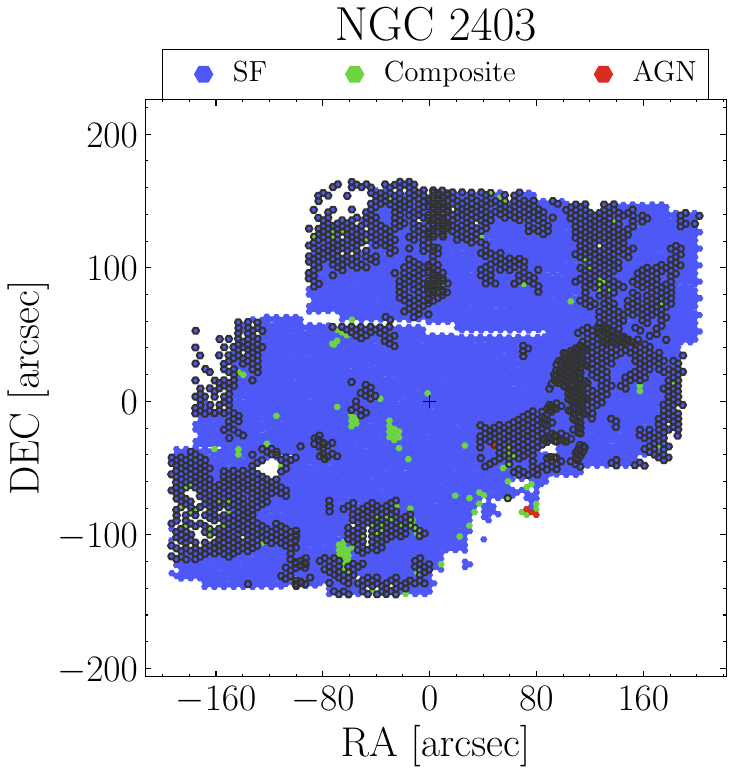}
        \hspace{0.75cm}
        \includegraphics[scale=0.39]{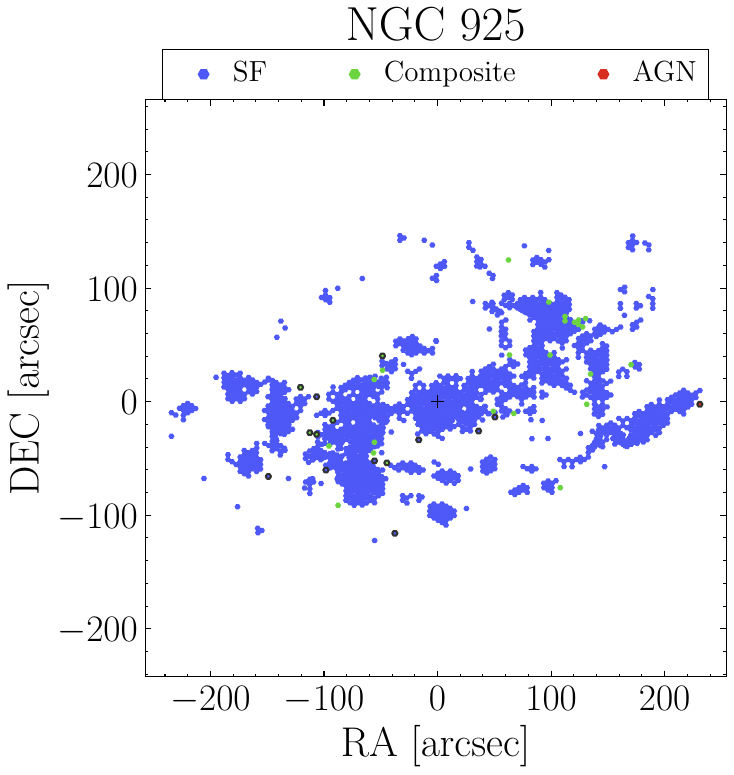}
        \hspace{0.75cm}
        \includegraphics[scale=0.395]{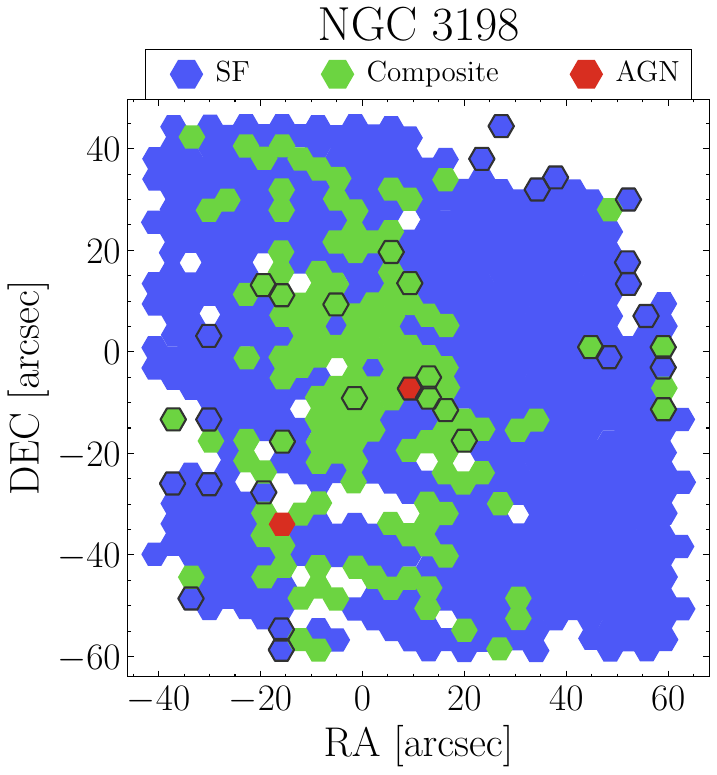}

        \vspace{0.25cm}

        \includegraphics[scale=0.395]{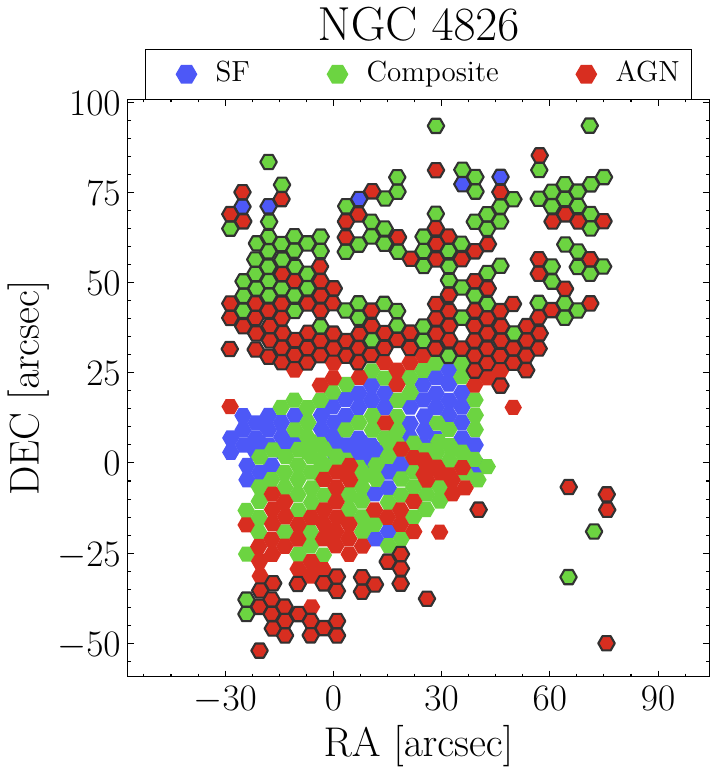}
        \hspace{0.75cm}
        \includegraphics[scale=0.395]{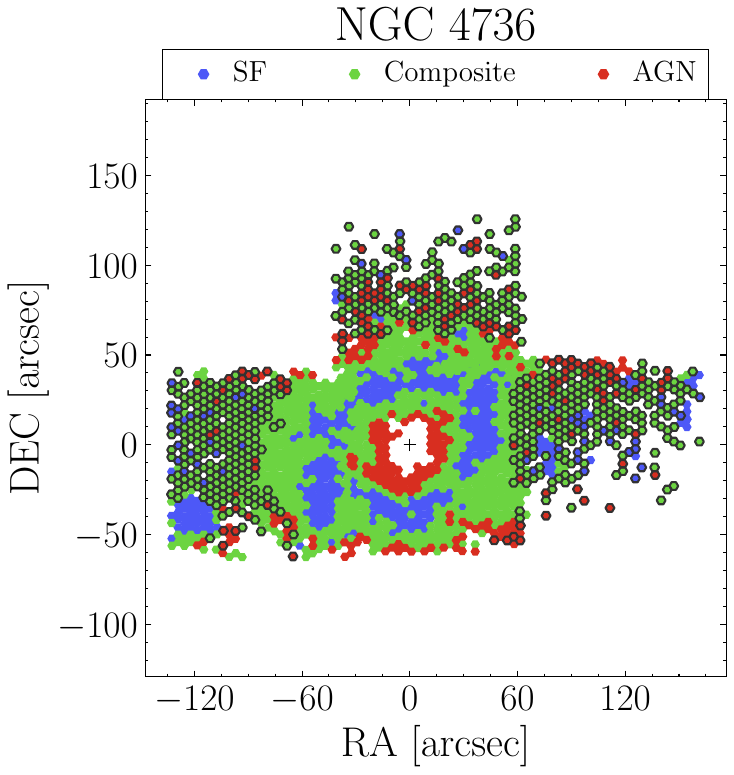}
        \hspace{0.75cm}
        \raisebox{0.5cm}{\includegraphics[scale=0.39]{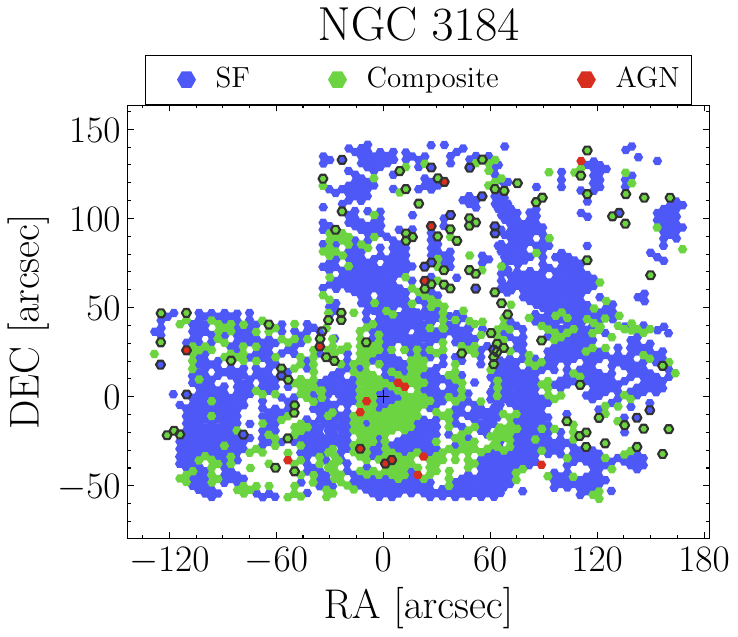}}

        \vspace{0.25cm}

        \raisebox{1cm}{\includegraphics[scale=0.395]{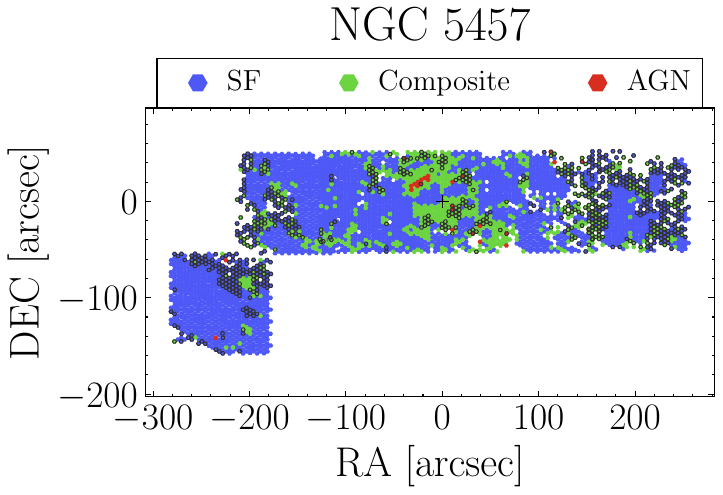}}
        \hspace{0.75cm}
        \includegraphics[scale=0.395]{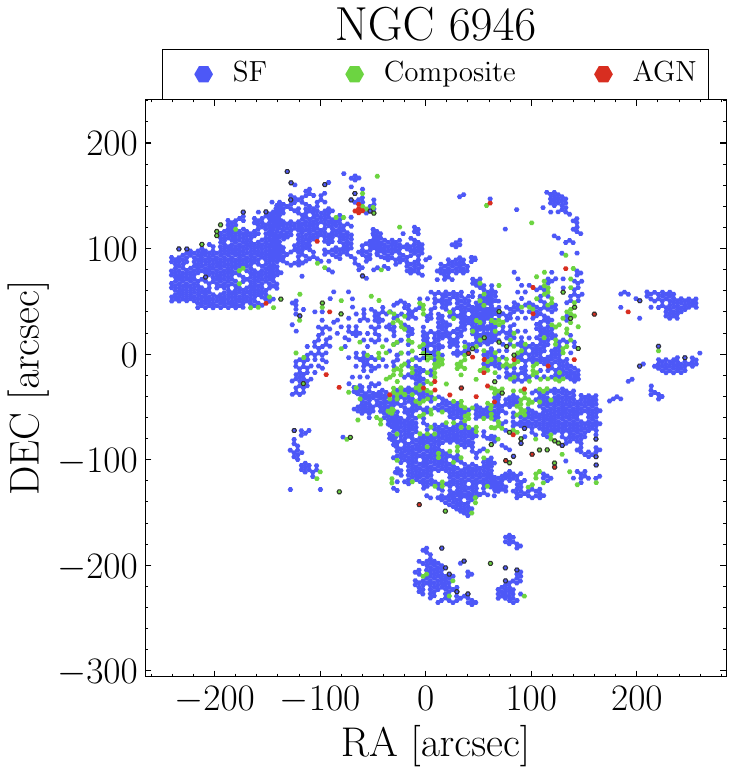}
        \hspace{0.75cm}
        \raisebox{1cm}{\includegraphics[scale=0.395]{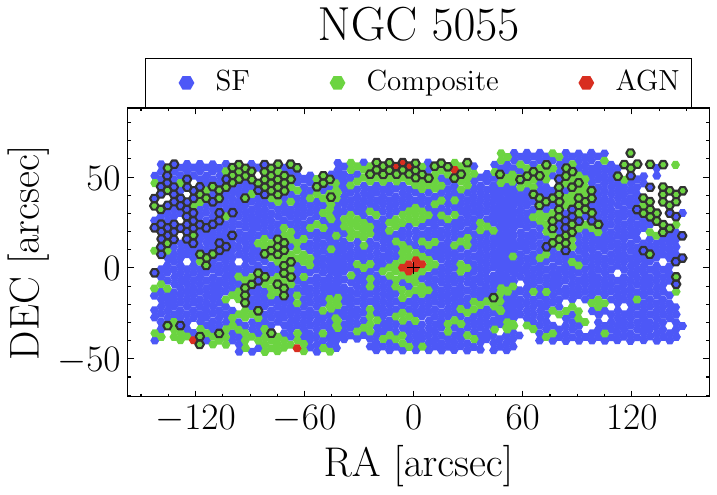}}

        \caption{Same as Figure \ref{fig:NII_BPT_1_map}, but showing the second part of the sample}
        \label{fig:NII_BPT_2_map}
\end{figure*}

\clearpage
\onecolumn

\begin{figure}[h]
\centering

        \includegraphics[scale=0.395]{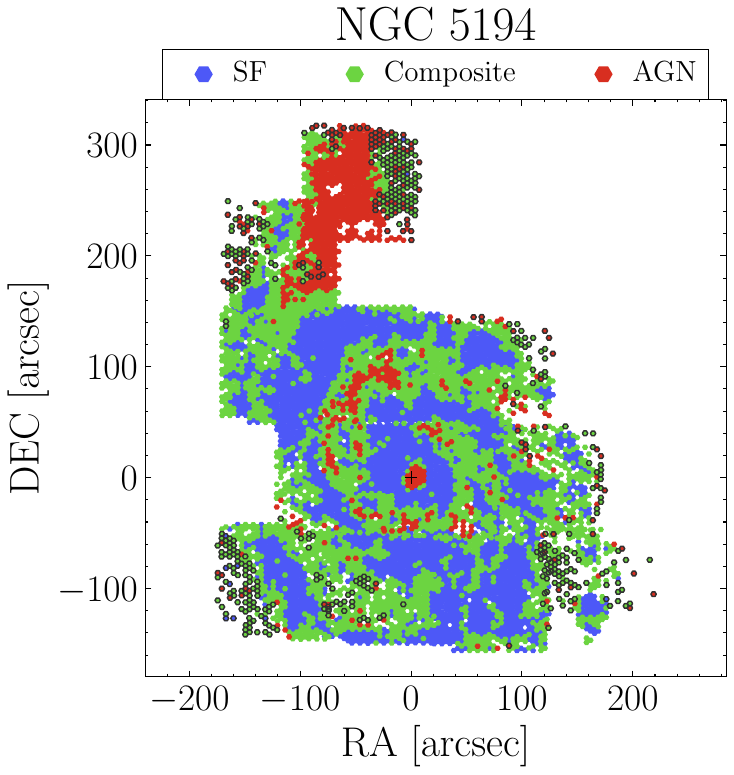}
        \hspace{1cm}
        \includegraphics[scale=0.395]{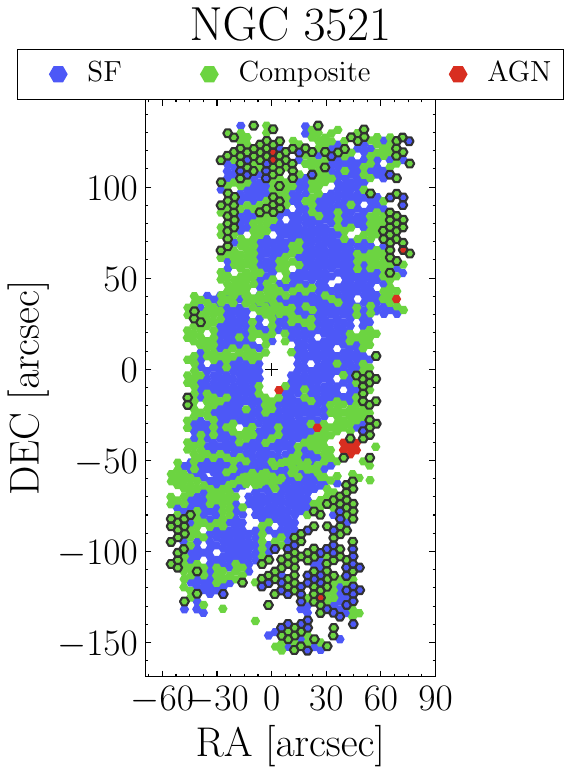}
        \hspace{1cm}
        \includegraphics[scale=0.395]{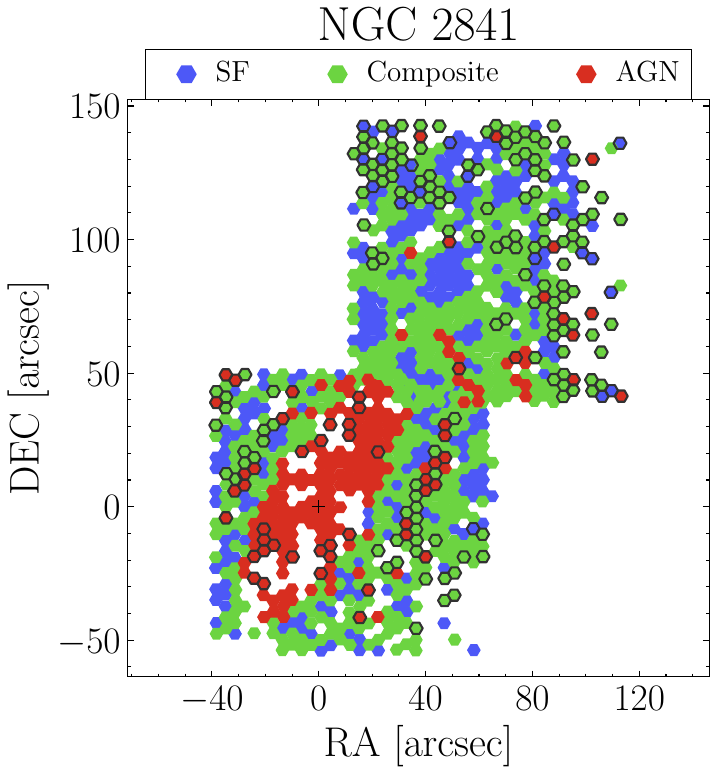}

        \vspace{0.25cm}
        
        \includegraphics[scale=0.395]{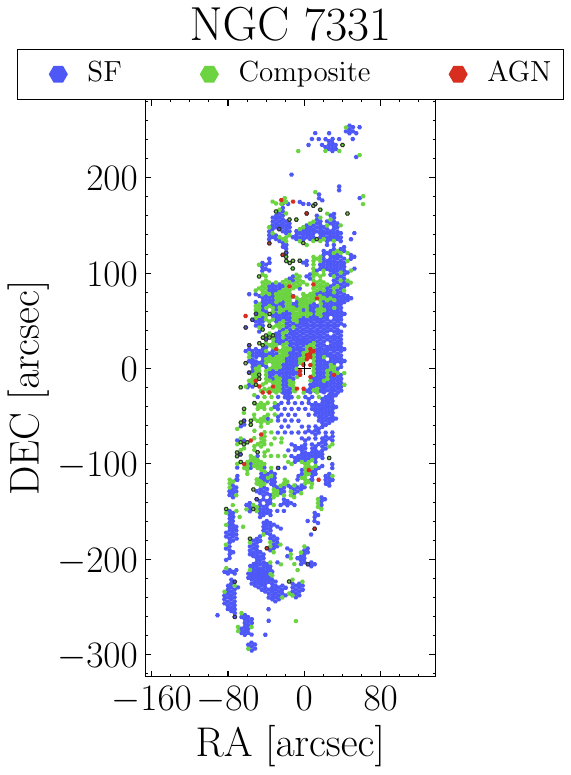}
        
        \caption{Same as Figures \ref{fig:NII_BPT_1_map} and \ref{fig:NII_BPT_2_map}, but showing the final part of the sample}
        \label{fig:NII_BPT_3_map}
\end{figure}

\vspace{-0.7cm}
\section{Metallicity maps}
\vspace{-0.4cm}

\begin{figure}[H]
\centering
        \includegraphics[scale=0.39]{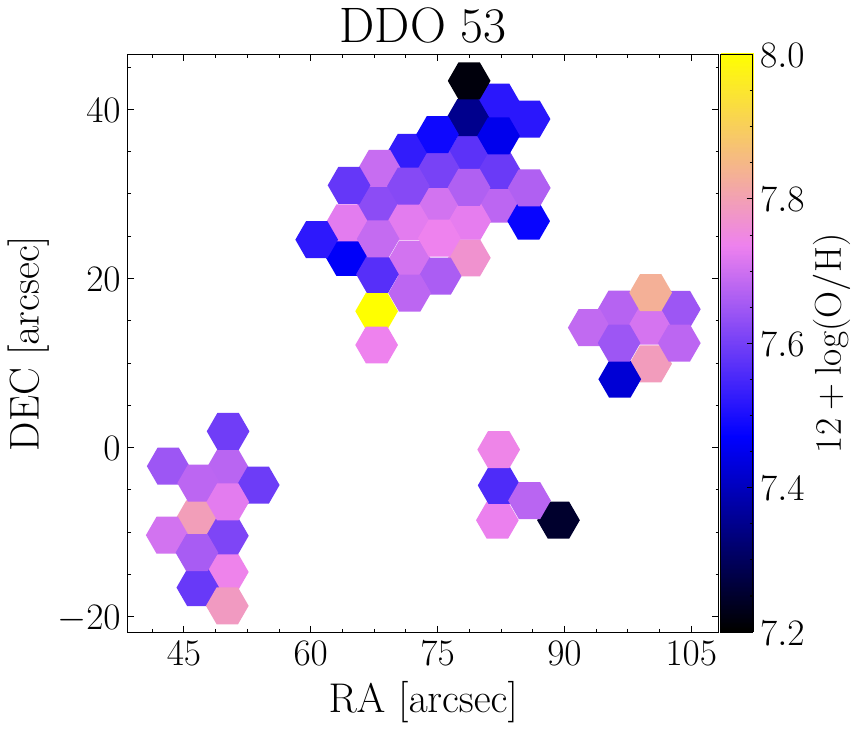}
        \hspace{0.1cm}
        \includegraphics[scale=0.39]{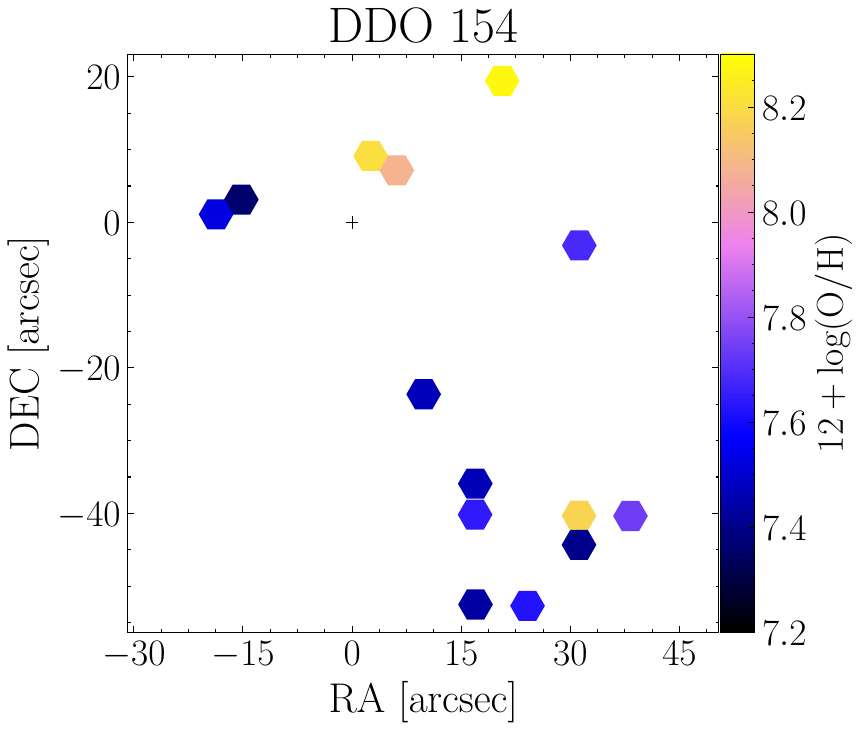}
        \hspace{0.1cm}
        \includegraphics[scale=0.39]{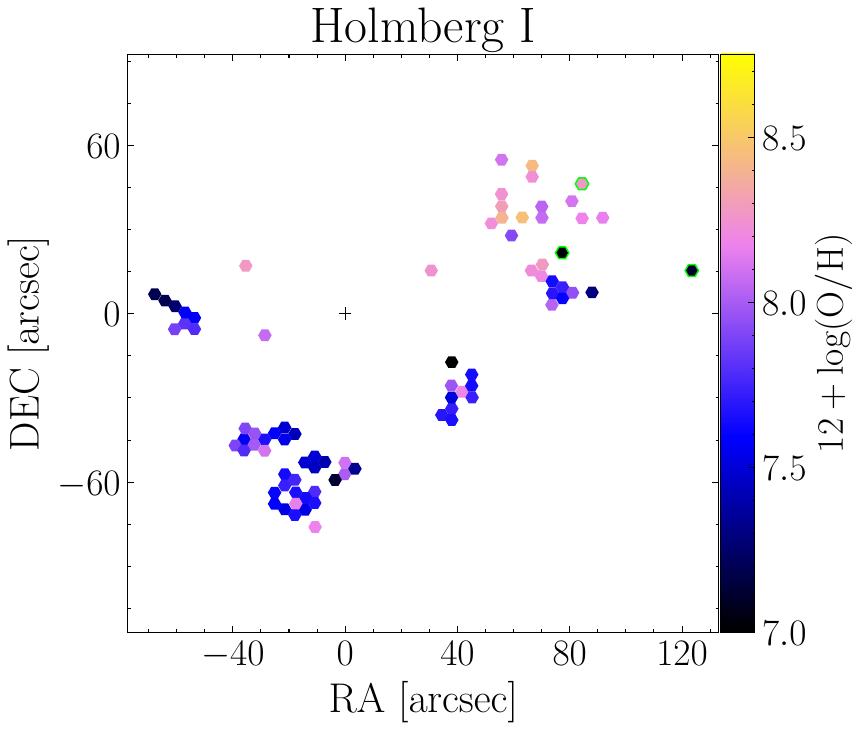}

        \vspace{0.75cm}

        \includegraphics[scale=0.39]{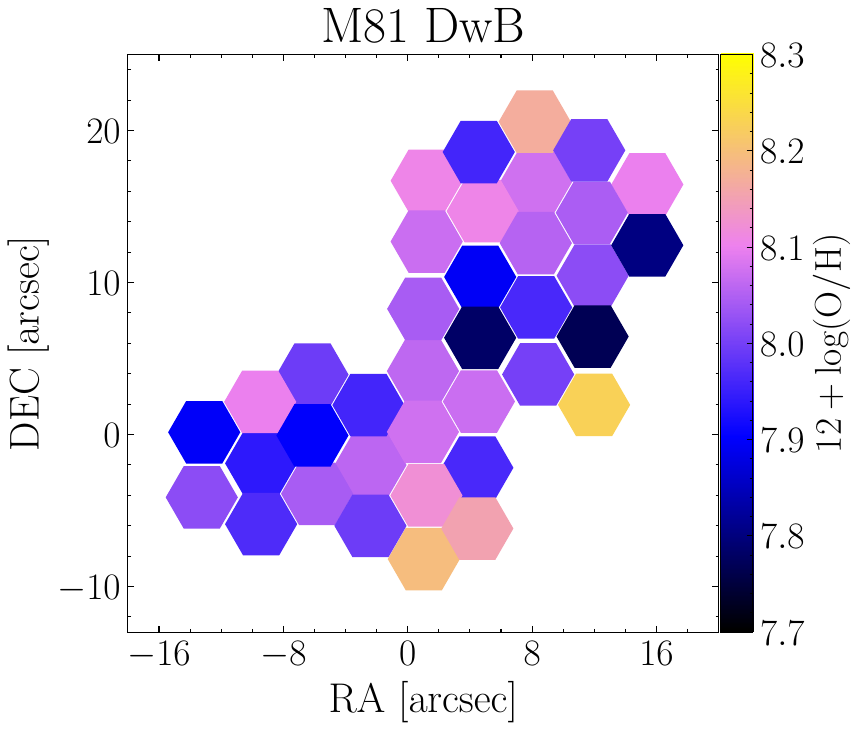}
        \hspace{0.1cm}
        \includegraphics[scale=0.39]{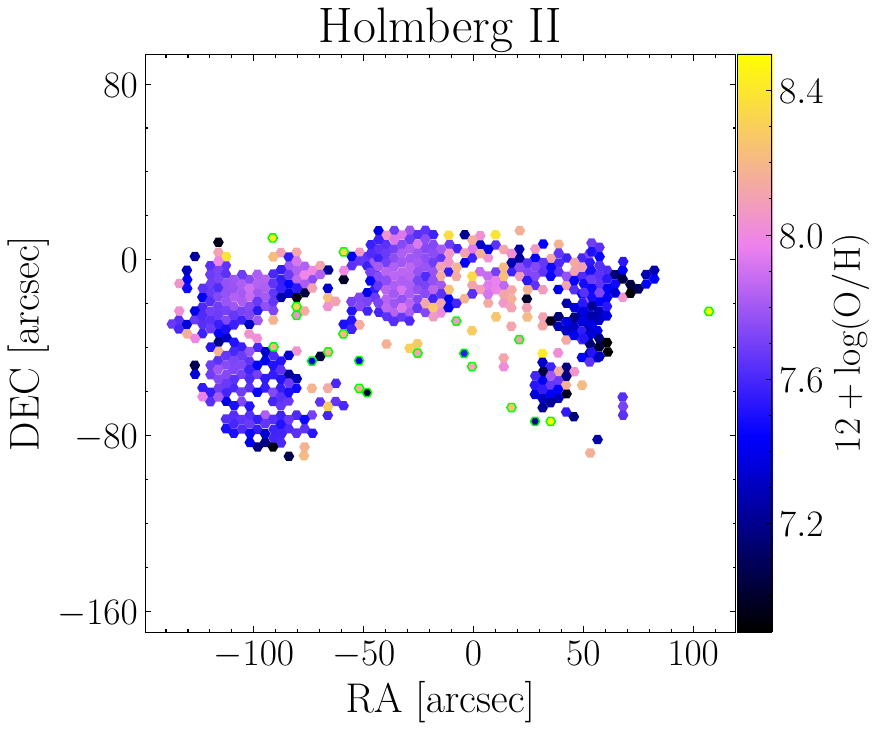}
        \hspace{0.1cm}
        \includegraphics[scale=0.39]{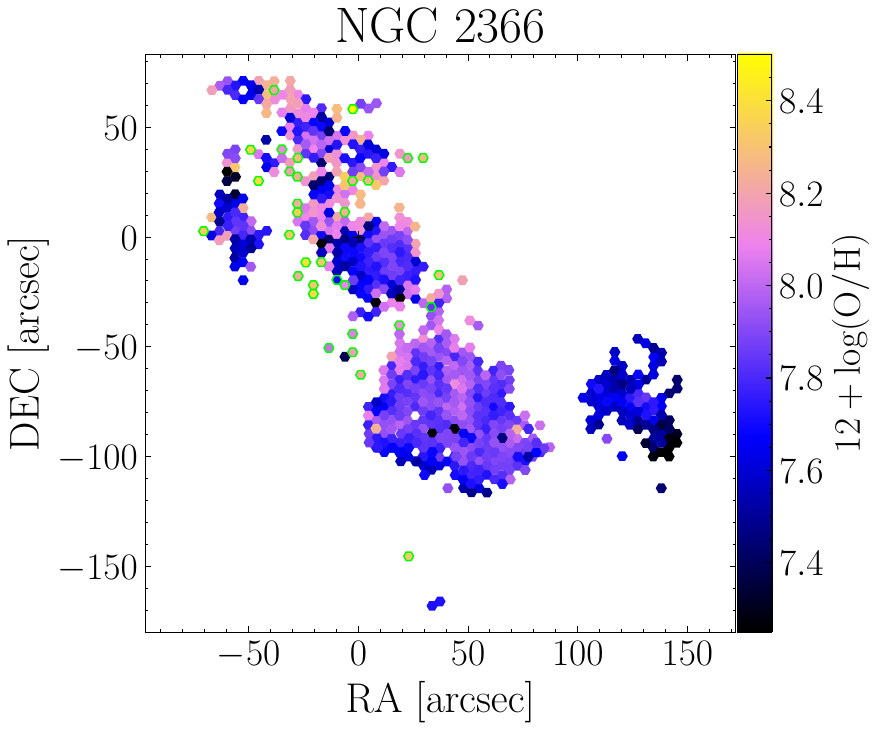}
                
         \caption{Metallicity maps for the observed galaxies, labelled accordingly. The fibres shown are all corrected for extinction, filtered such that $\rm SNR > 3$ and earlier categorized as SF in their respective [N II] diagnostic BPT diagram. Fibres identified as DIG have been flagged in green.}
         \label{fig:metmaps_1}
\end{figure}

\begin{figure*}[h]
\vspace{0.5cm}
\centering
        
        \includegraphics[scale=0.39]{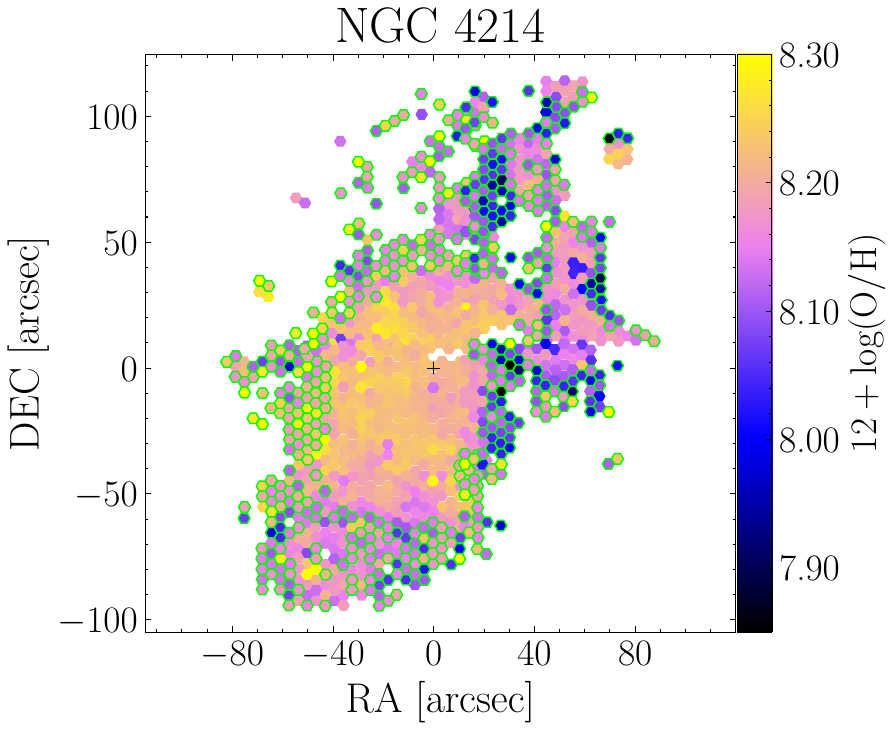}
        \hspace{0.1cm}
        \includegraphics[scale=0.39]{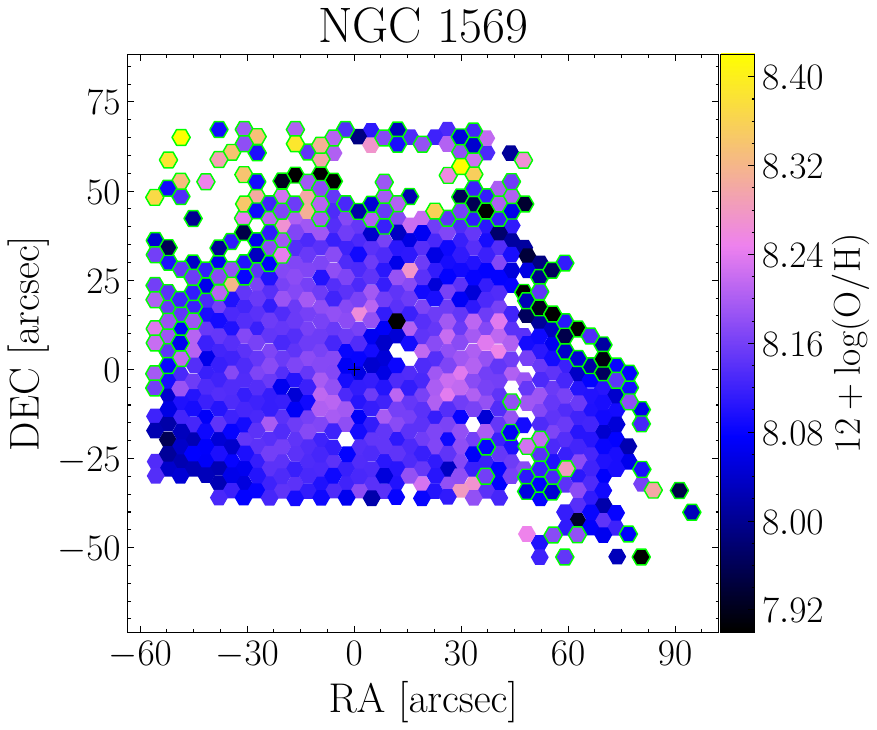}
        \hspace{0.1cm}
        \includegraphics[scale=0.39]{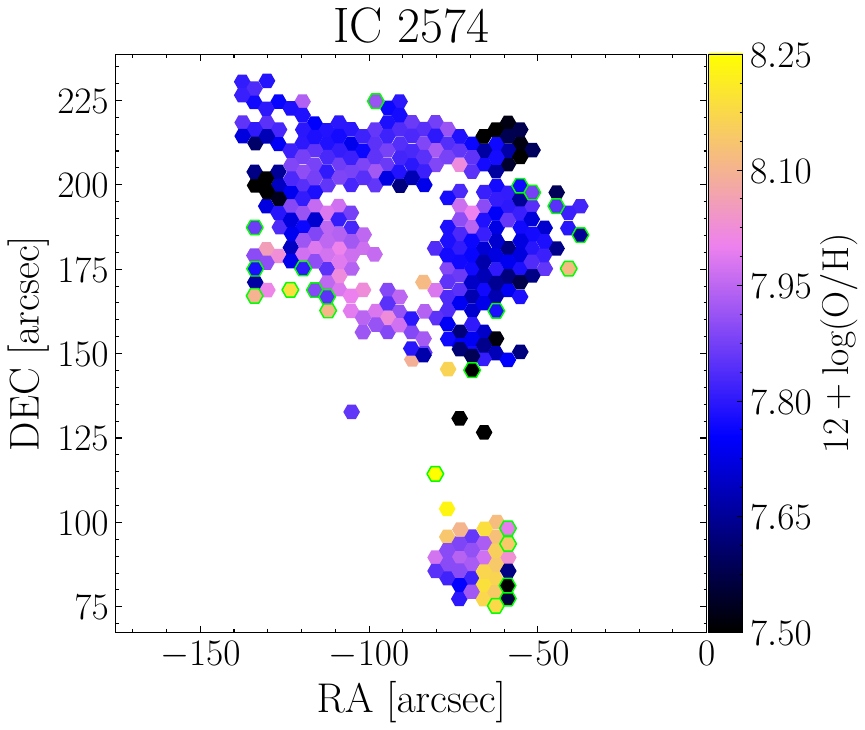}

        \vspace{0.75cm}

        \includegraphics[scale=0.39]{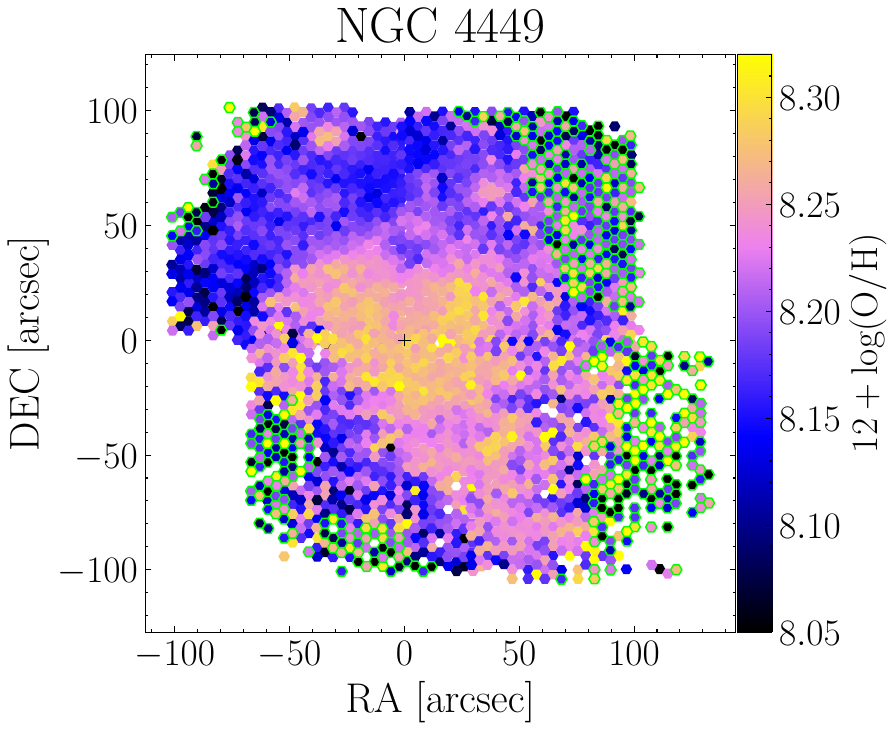}
        \hspace{0.1cm}
        \includegraphics[scale=0.39]{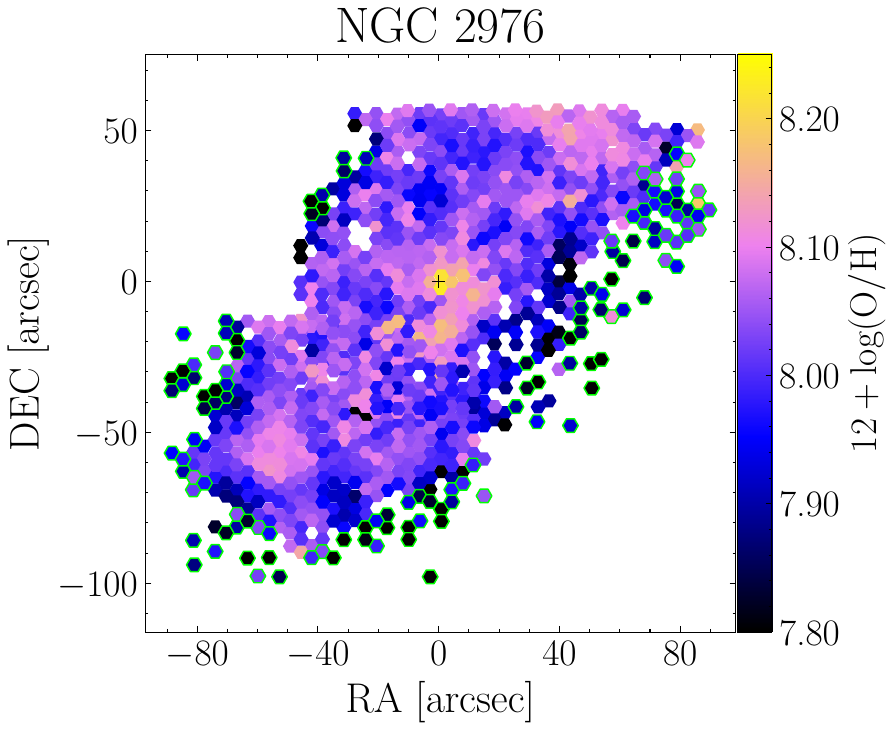}
        \hspace{0.1cm}
        \includegraphics[scale=0.39]{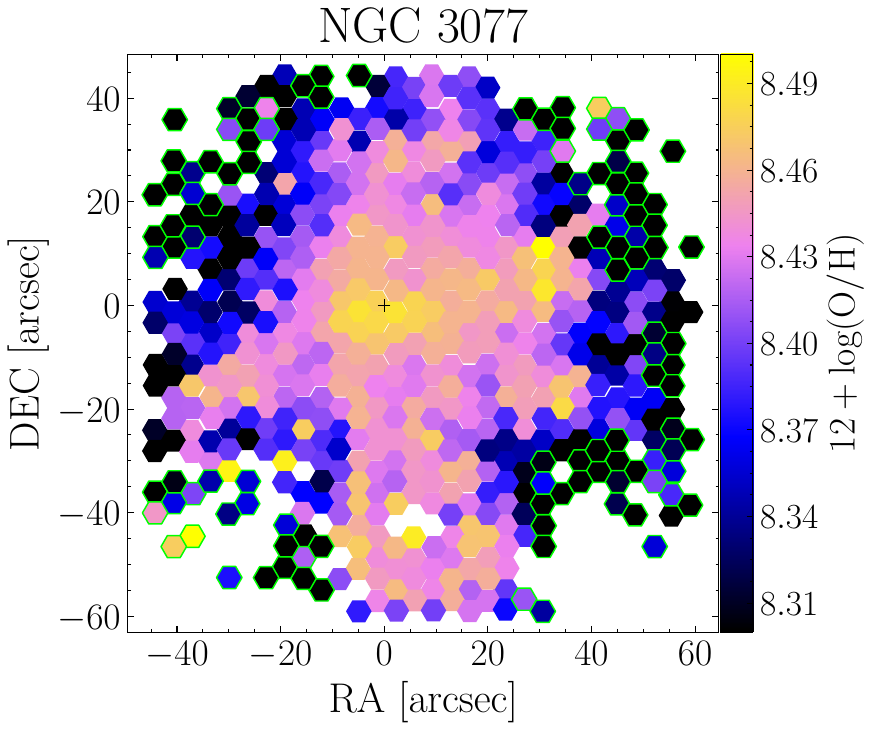}

        \vspace{0.75cm}
        
        \includegraphics[scale=0.39]{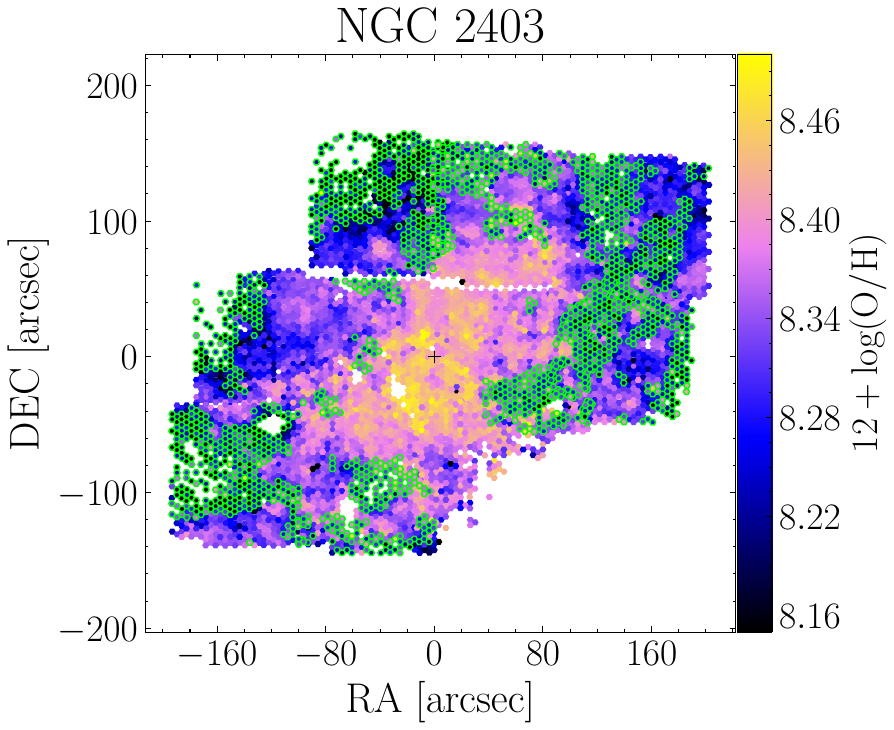}
        \hspace{0.1cm}
        \includegraphics[scale=0.39]{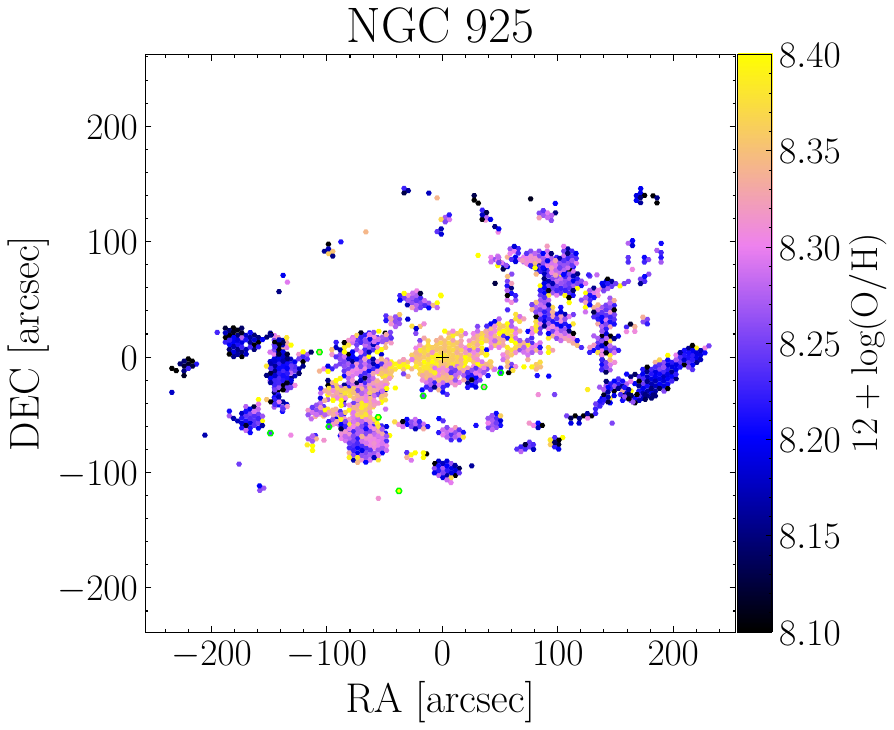}
        \hspace{0.1cm}
        \includegraphics[scale=0.39]{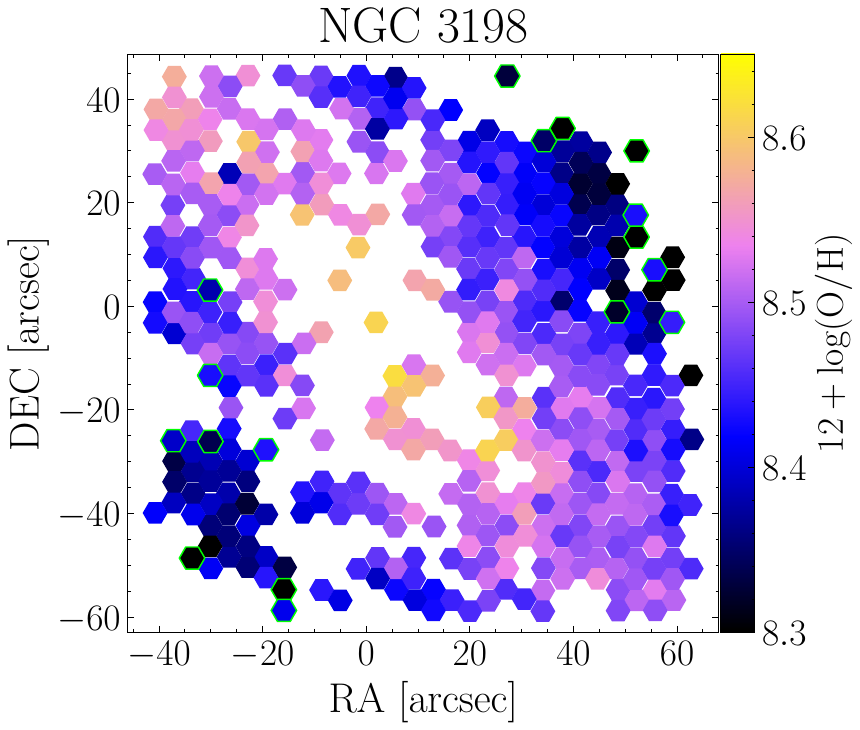}

        \vspace{0.25cm}

        \includegraphics[scale=0.39]{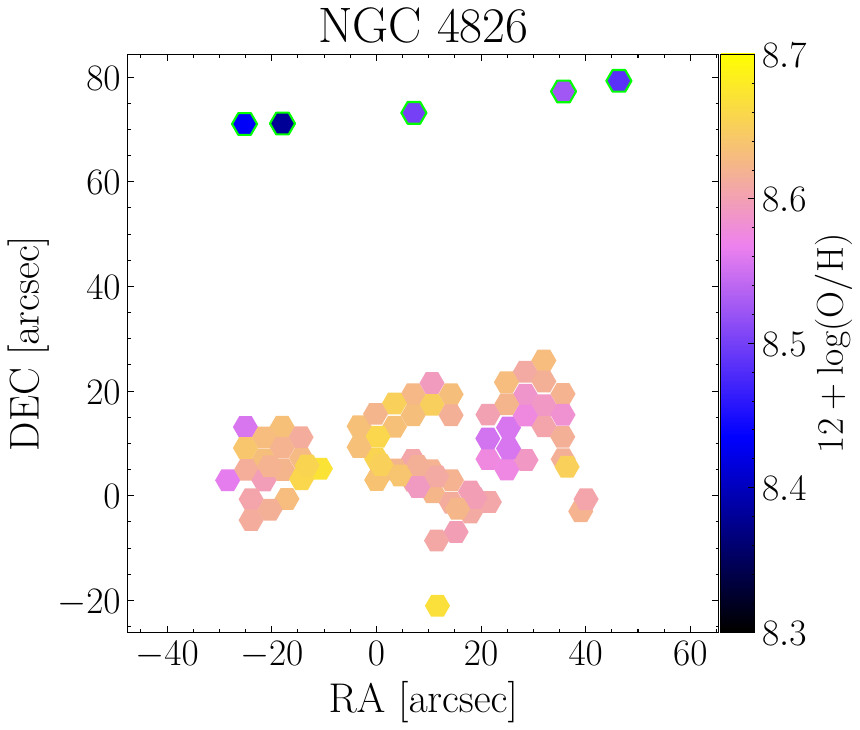}
        \hspace{0.1cm}
        \includegraphics[scale=0.39]{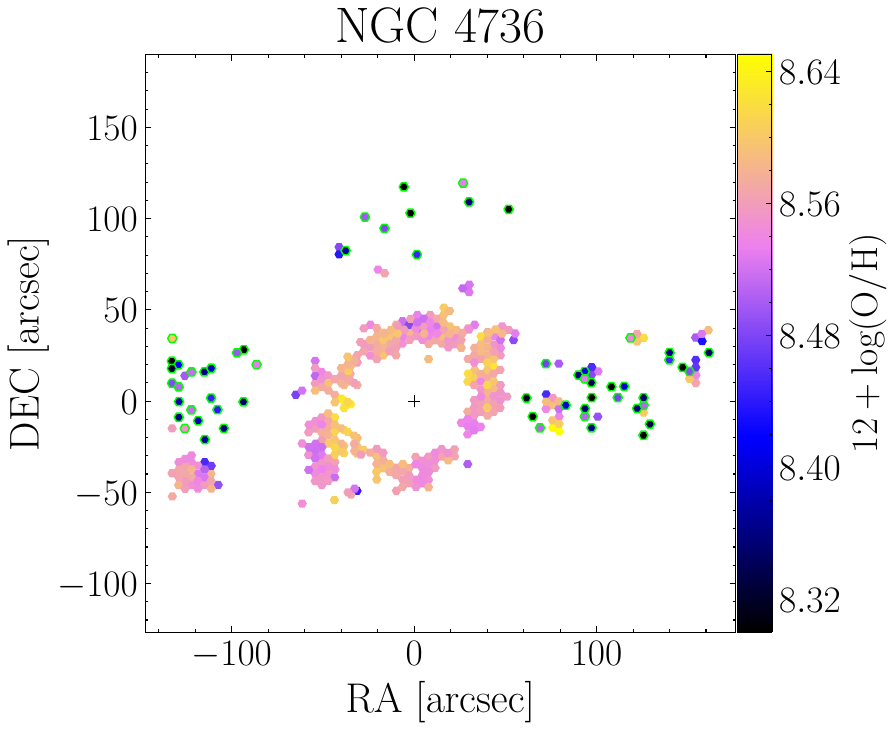}
        \hspace{0.1cm}
        \raisebox{1cm}{\includegraphics[scale=0.39]{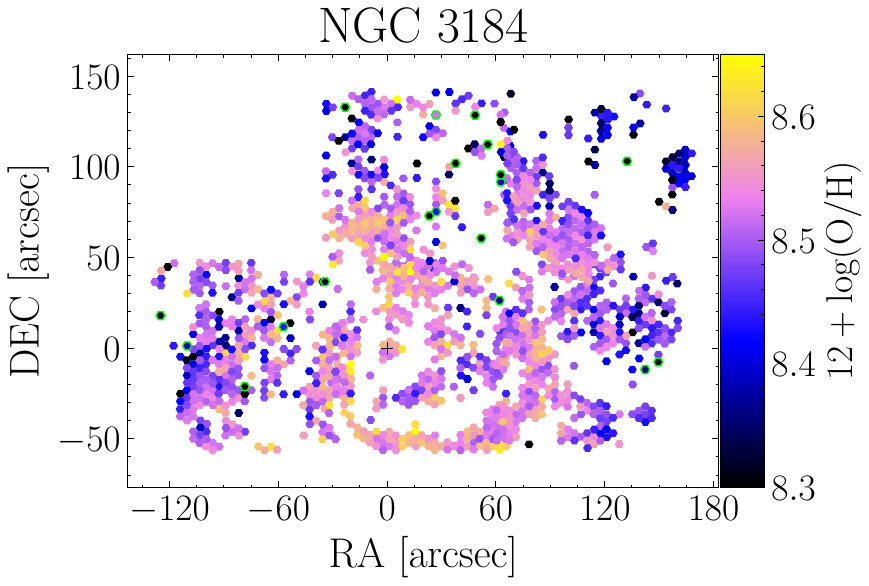}}
        
        \caption{Same as Figure \ref{fig:metmaps_1}, but showing the second part of the sample}
         \label{fig:metmaps_2}
\end{figure*}

\clearpage

\begin{figure}[H]
\vspace{0.5cm}
\centering

        \raisebox{1cm}{\includegraphics[scale=0.39]{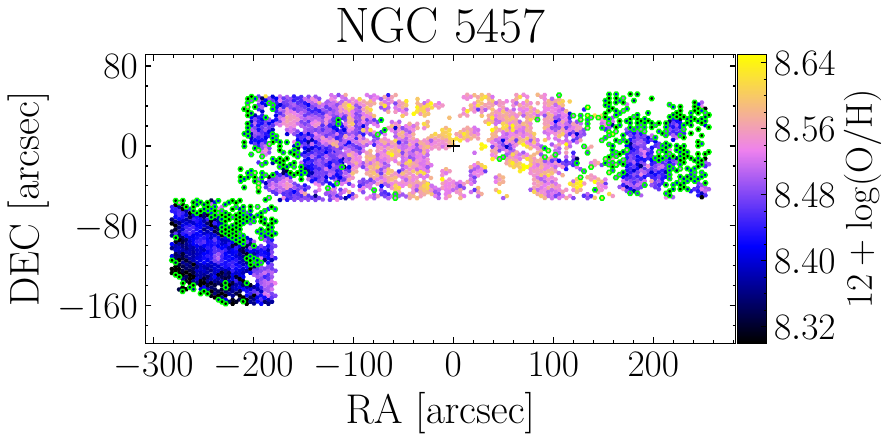}}
        \hspace{0.05cm}
        \includegraphics[scale=0.39]{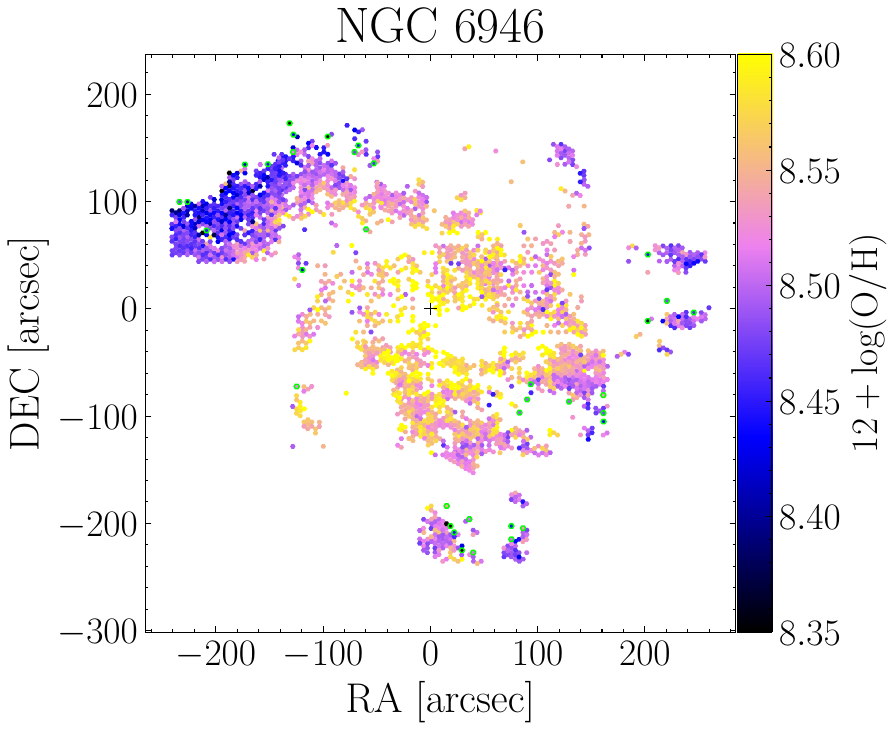}
        \hspace{0.05cm}
        \raisebox{1cm}{\includegraphics[scale=0.39]{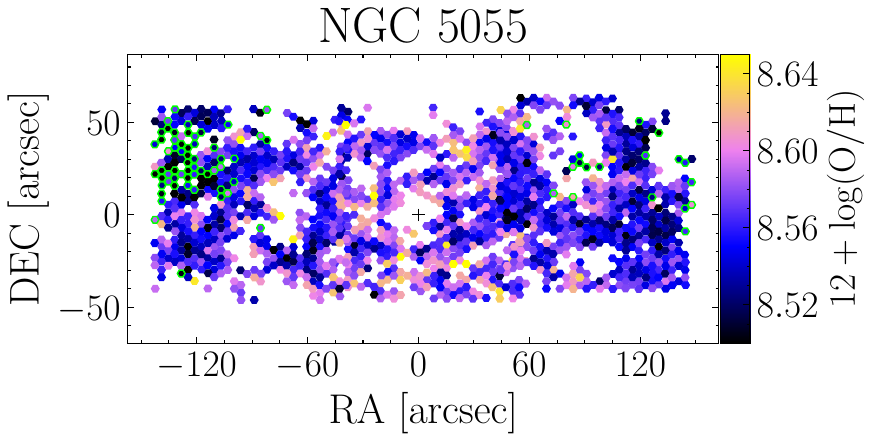}}

        \vspace{0.25cm}

        \includegraphics[scale=0.39]{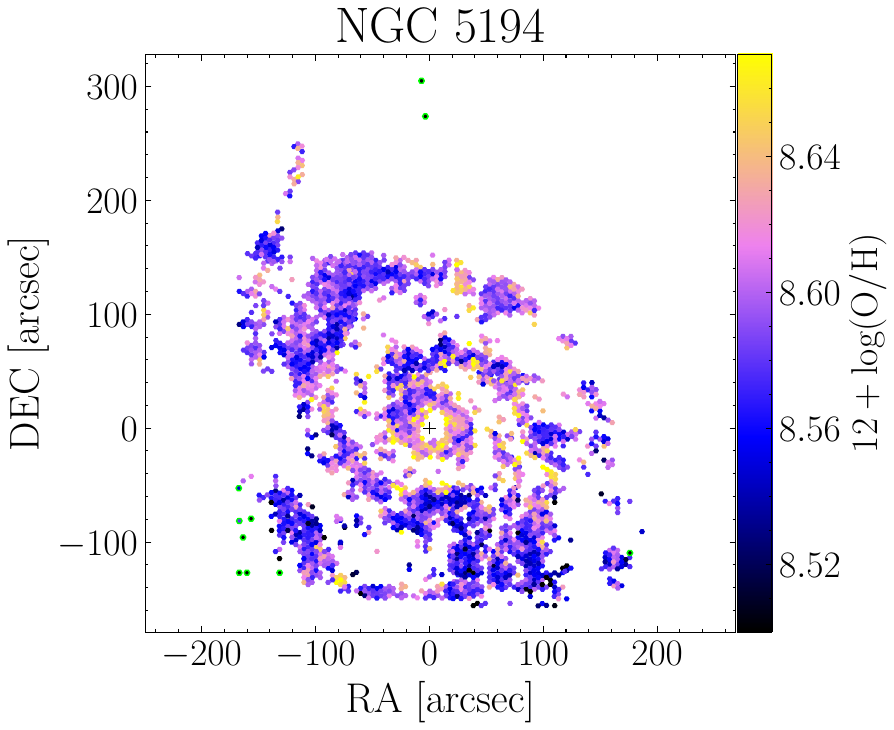}
        \hspace{0.8cm}
        \includegraphics[scale=0.39]{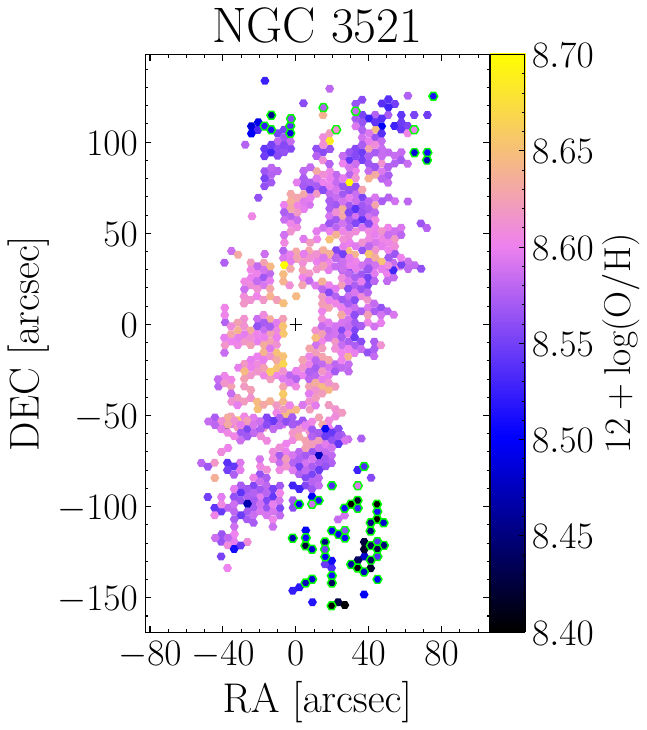}
        \hspace{0.8cm}
        \includegraphics[scale=0.39]{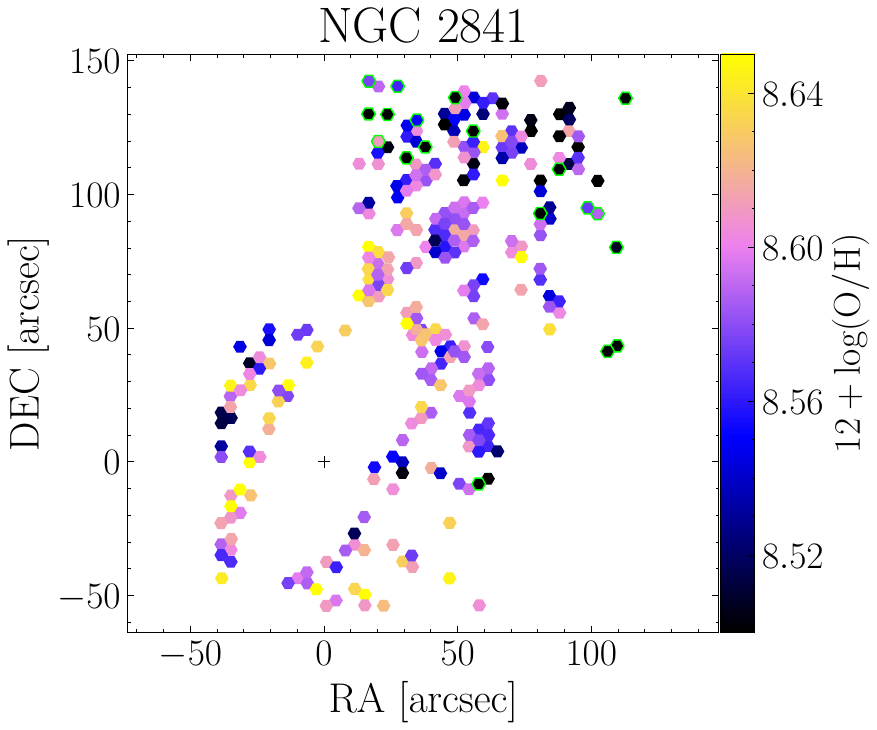}

        \vspace{0.25cm}
        
        \includegraphics[scale=0.39]{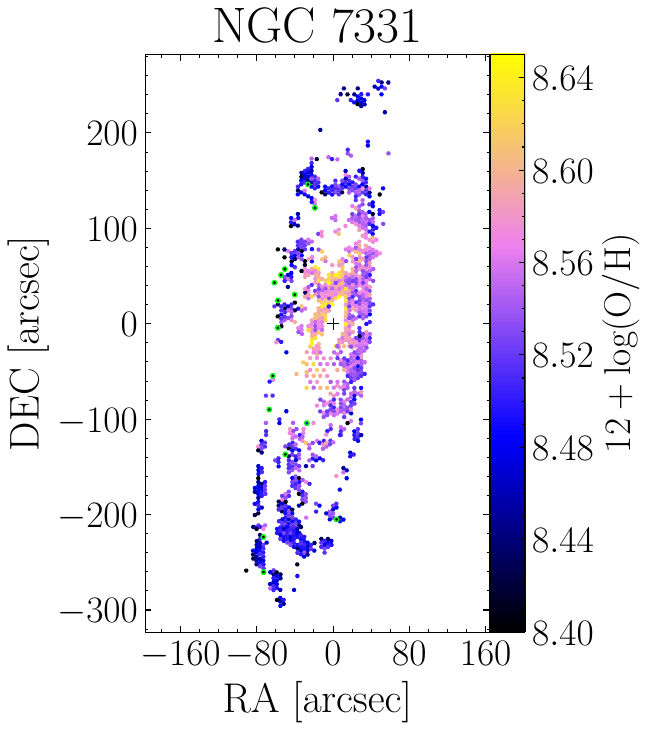}
        
        \caption{Same as Figures \ref{fig:metmaps_1} and \ref{fig:metmaps_2}, but showing the final part of the sample}
         \label{fig:metmaps_3}
\end{figure}

\onecolumn

\clearpage

\begin{sidewaystable*}[h]

\section{Metallicity gradients in all bands} \label{other_grads_table}

\centering
\renewcommand{\arraystretch}{1.25}
\begin{tabular}{>{\rowmac}c|>{\rowmac}c>{\rowmac}c>{\rowmac}c>{\rowmac}c>{\rowmac}c|>{\rowmac}c>{\rowmac}c>{\rowmac}c>{\rowmac}c>{\rowmac}c>{\rowmac}c<{\clearrow}}

\hline
\hline
1 & 2 & 3 & 4 & 5 & 6 & 7 & 8 & 9 & 10 & 11 & ... \\

\begin{tabular}[c]{@{}c@{}c@{}} Galaxy \\ \textbf{} \\ \textbf{} \end{tabular} &

\begin{tabular}[c]{@{}c@{}c@{}} $R_{\rm eff, \, r}$ \\ \textbf{} \\ $\rm [arcsec]$  \end{tabular} &
\begin{tabular}[c]{@{}c@{}c@{}}$ \rm \nabla(O/H) $ \\ \small no DIG \normalsize\\ {[}$ \rm dex/R_{eff} ${]} \end{tabular} &
\begin{tabular}[c]{@{}c@{}c@{}}$ \rm O/H_0$ \\ \small no DIG \normalsize \\ \textbf{} \end{tabular} &
\begin{tabular}[c]{@{}c@{}c@{}}$ \rm \nabla(O/H) $ \\ \small with DIG \normalsize\\ {[}$ \rm dex/R_{eff} ${]} \end{tabular} &
\begin{tabular}[c]{@{}c@{}c@{}}$ \rm O/H_0$ \\ \small with DIG \normalsize \\ \textbf{} \end{tabular} &

\begin{tabular}[c]{@{}c@{}c@{}} $R_{\rm eff, \, Ks}$ \\ \textbf{} \\ $\rm [arcsec]$ \end{tabular} &
\begin{tabular}[c]{@{}c@{}c@{}}$ \rm \nabla(O/H) $ \\ \small no DIG \normalsize \\ {[}$ \rm dex/R_{eff} ${]} \end{tabular} &
\begin{tabular}[c]{@{}c@{}c@{}}$ \rm O/H_0$ \\ \small no DIG \normalsize \\ \textbf{} \end{tabular} &
\begin{tabular}[c]{@{}c@{}c@{}}$ \rm \nabla(O/H) $ \\ \small with DIG \normalsize \\ {[}$ \rm dex/R_{eff} ${]} \end{tabular} &
\begin{tabular}[c]{@{}c@{}c@{}}$ \rm O/H_0$ \\ \small with DIG \normalsize \\ \textbf{} \end{tabular} &

\begin{tabular}[c]{@{}c@{}c@{}} \textbf{} \\ \textbf{} \\ \textbf{} \end{tabular}

\\ \hline

DDO 53      & 329.62  & -0.47$\pm$0.36   & 7.751$\pm$0.09  & -0.47$\pm$0.36   & 7.75$\pm$0.09   & $-$    & $-$              & $-$             & $-$              & $-$             \\
DDO 154     & 38.08   & -0.08$\pm$0.18   & 7.83$\pm$0.26   & -0.079$\pm$0.18  & 7.825$\pm$0.262 & $-$    & $-$              & $-$             & $-$              & $-$             \\
Holmberg I  & 80.33   & 0.50$\pm$0.21    & 7.40$\pm$0.17   &  0.23$\pm$0.20   & 7.60$\pm$0.17   & $-$    & $-$              & $-$             & $-$              & $-$             \\
M81 DwB     & 17.51   & 0.01$\pm$0.06    & 8.01$\pm$0.04   & 0.01$\pm$0.06    & 8.01$\pm$0.04   & $-$    & $-$              & $-$             & $-$              & $-$             \\
Holmberg II & 94.40   & -0.130$\pm$0.023 & 7.790$\pm$0.025 & -0.126$\pm$0.024 & 7.798$\pm$0.025 & $-$    & $-$              & $-$             & $-$              & $-$             \\
NGC 2366    & 106.32  & -0.19$\pm$0.02   & 7.99$\pm$0.01   & -0.19$\pm$0.02   & 8.01$\pm$0.02   & $-$    & $-$              & $-$             & $-$              & $-$             \\
NGC 4214    & 55.78   & -0.024$\pm$0.002 & 8.214$\pm$0.003 & -0.026$\pm$0.003 & 8.202$\pm$0.004 & 57.86  & -0.025$\pm$0.003 & 8.214$\pm$0.003 & -0.027$\pm$0.003 & 8.202$\pm$0.004 \\
NGC 1569    & 26.87   & -0.019$\pm$0.004 & 8.153$\pm$0.006 & -0.012$\pm$0.004 & 8.145$\pm$0.007 & 33.70  & -0.024$\pm$0.005 & 8.153$\pm$0.006 & -0.015$\pm$0.005 & 8.145$\pm$0.007 \\
IC 2574     & 192.35  & -0.21$\pm$0.03   & 8.08$\pm$0.04   & -0.20$\pm$0.03   & 8.07$\pm$0.04   & 37.64  & -0.041$\pm$0.006 & 8.08$\pm$0.04   & -0.039$\pm$0.006 & 8.07$\pm$0.04   \\
NGC 4449    & 55.41   & -0.029$\pm$0.001 & 8.260$\pm$0.003 & -0.025$\pm$0.002 & 8.254$\pm$0.003 & 74.2   & -0.038$\pm$0.002 & 8.260$\pm$0.003 & -0.034$\pm$0.002 & 8.254$\pm$0.003 \\
NGC 2976    & 68.47   & -0.041$\pm$0.005 & 8.366$\pm$0.006 & -0.069$\pm$0.006 & 8.384$\pm$0.007 & 72.65  & -0.043$\pm$0.006 & 8.366$\pm$0.006 & -0.073$\pm$0.006 & 8.384$\pm$0.007 \\
NGC 3077    & 38.26   & -0.055$\pm$0.006 & 8.464$\pm$0.007 & -0.091$\pm$0.007 & 8.488$\pm$0.009 & 50.64  & -0.073$\pm$0.008 & 8.464$\pm$0.007 & -0.121$\pm$0.009 & 8.488$\pm$0.009 \\
NGC 2403    & 91.89   & -0.058$\pm$0.002 & 8.442$\pm$0.003 & -0.066$\pm$0.002 & 8.442$\pm$0.003 & 125.49 & -0.079$\pm$0.002 & 8.442$\pm$0.003 & -0.091$\pm$0.002 & 8.442$\pm$0.003 \\
NGC 925     & 67.48   & -0.043$\pm$0.002 & 8.35$\pm$0.004  & -0.043$\pm$0.002 & 8.35$\pm$0.004  & 128.17 & -0.082$\pm$0.004 & 8.35$\pm$0.004  & -0.081$\pm$0.004 & 8.35$\pm$0.004  \\
NGC 3198    & 79.0    & -0.121$\pm$0.003 & 8.603$\pm$0.004 & -0.124$\pm$0.004 & 8.605$\pm$0.004 & 77.93  & -0.120$\pm$0.003 & 8.603$\pm$0.004 & -0.123$\pm$0.004 & 8.605$\pm$0.004 \\
NGC 4826    & 57.11   & -0.011$\pm$0.013 & 8.620$\pm$0.008 & -0.064$\pm$0.006 & 8.648$\pm$0.005 & 70.59  & -0.013$\pm$0.017 & 8.620$\pm$0.008 & -0.079$\pm$0.007 & 8.648$\pm$0.005 \\
NGC 4736    & 23.37   & -0.008$\pm$0.001 & 8.582$\pm$0.004 & -0.026$\pm$0.002 & 8.617$\pm$0.007 & 31.03  & -0.011$\pm$0.002 & 8.582$\pm$0.004 & -0.035$\pm$0.003 & 8.617$\pm$0.007 \\
NGC 3184    & 81.79   & -0.061$\pm$0.003 & 8.575$\pm$0.004 & -0.063$\pm$0.003 & 8.575$\pm$0.004 & 96.88  & -0.073$\pm$0.004 & 8.575$\pm$0.004 & -0.075$\pm$0.004 & 8.575$\pm$0.004 \\
NGC 5457    & 174.58  & -0.127$\pm$0.002 & 8.596$\pm$0.002 & -0.146$\pm$0.002 & 8.602$\pm$0.003 & 191.35 & -0.139$\pm$0.002 & 8.596$\pm$0.002 & -0.160$\pm$0.003 & 8.602$\pm$0.003 \\
NGC 6946    & 85.67   & -0.043$\pm$0.001 & 8.598$\pm$0.002 & -0.044$\pm$0.001 & 8.599$\pm$0.002 & 119.76 & -0.060$\pm$0.001 & 8.598$\pm$0.002 & -0.061$\pm$0.001 & 8.599$\pm$0.002 \\
NGC 5055    & 64.6    & -0.014$\pm$0.001 & 8.585$\pm$0.002 & -0.016$\pm$0.001 & 8.588$\pm$0.002 & 80.77  & -0.017$\pm$0.002 & 8.585$\pm$0.002 & -0.020$\pm$0.002 & 8.588$\pm$0.002 \\
NGC 5194    & 145.74  & -0.018$\pm$0.002 & 8.607$\pm$0.001 & -0.022$\pm$0.002 & 8.609$\pm$0.002 & 83.61  & -0.011$\pm$0.001 & 8.607$\pm$0.002 & -0.012$\pm$0.001 & 8.609$\pm$0.002 \\
NGC 3521    & 65.15   & -0.029$\pm$0.001 & 8.633$\pm$0.002 & -0.037$\pm$0.001 & 8.643$\pm$0.002 & 63.61  & -0.029$\pm$0.001 & 8.633$\pm$0.002 & -0.036$\pm$0.001 & 8.643$\pm$0.002 \\
NGC 2841    & 56.88   & -0.025$\pm$0.004 & 8.630$\pm$0.008 & -0.041$\pm$0.005 & 8.655$\pm$0.009 & 71.05  & -0.025$\pm$0.004 & 8.630$\pm$0.008 & -0.042$\pm$0.005 & 8.655$\pm$0.010 \\
NGC 7331    & 48.25   & -0.026$\pm$0.001 & 8.614$\pm$0.005 & -0.027$\pm$0.001 & 8.616$\pm$0.005 & 84.51  & -0.046$\pm$0.002 & 8.614$\pm$0.005 & -0.048$\pm$0.002 & 8.616$\pm$0.005 \\
\hline

\end{tabular}
\caption{Linear model parameters for each galaxy. Each column shows information as follows: 1) galaxy designation; 2) effective radius in the r-band; 3) and 4) linear model parameters and errors without DIG-dominated fibres with r-band effective radius normalisation; 5) and 6) linear model parameters and errors with all fibres with r-band effective radius normalisation. Columns 7), 8), 9) and 10) show the same parameters as the previous four columns but for $K_s$-band effective radius normalisation. Each linear model was fitted using a Bayesian MCMC routine.}
\label{tab:grads_data_1}
\end{sidewaystable*}

\clearpage

\begin{sidewaystable*}[h]

\centering
\renewcommand{\arraystretch}{1.25}
\begin{tabular}{>{\rowmac}c|>{\rowmac}c>{\rowmac}c>{\rowmac}c>{\rowmac}c>{\rowmac}c|>{\rowmac}c>{\rowmac}c>{\rowmac}c<{\clearrow}}

\hline
\hline
1 & 12 & 13 & 14 & 15 & 16 & 17 & 18 & 19 \\

\begin{tabular}[c]{@{}c@{}c@{}} Galaxy \\ \textbf{} \\ \textbf{} \end{tabular} &

\begin{tabular}[c]{@{}c@{}c@{}} $R_{\rm 25}$ \\ \textbf{} \\ $\rm [arcsec]$ \end{tabular} &

\begin{tabular}[c]{@{}c@{}c@{}}$ \rm \nabla(O/H) $ \\ \small no DIG \normalsize \\ {[}$ \rm dex/R_{eff} ${]} \end{tabular} &
\begin{tabular}[c]{@{}c@{}c@{}}$ \rm O/H_0$ \\ \small no DIG \normalsize \\ \textbf{} \end{tabular} &
\begin{tabular}[c]{@{}c@{}c@{}}$ \rm \nabla(O/H) $ \\ \small with DIG \normalsize \\ {[}$ \rm dex/R_{eff} ${]} \end{tabular} &
\begin{tabular}[c]{@{}c@{}c@{}}$ \rm O/H_0$ \\ \small with DIG \normalsize \\ \textbf{} \end{tabular} &

\begin{tabular}[c]{@{}c@{}c@{}}$ \rm \nabla(O/H)_{\, 1} $ \\ \small no DIG \normalsize \\ {[}$ \rm dex/R_{eff} ${]} \end{tabular} &
\begin{tabular}[c]{@{}c@{}c@{}}$ \rm \nabla(O/H)_{\, 2} $ \\ \small no DIG \normalsize \\ {[}$ \rm dex/R_{eff} ${]} \end{tabular} &
\begin{tabular}[c]{@{}c@{}c@{}}Breakpoint \\ \small no DIG \normalsize \\ {[}$R_{\rm eff, \; NUV}${]} \end{tabular}
\\ \hline

DDO 53      & $-$     & $-$              & $-$             & $-$              & $-$             & $-$ & $-$ & $-$ \\
DDO 154     & 57.56   & -0.12$\pm$0.27   & 7.83$\pm$0.26   & -0.12$\pm$0.27   & 7.83$\pm$0.26   & $-$ & $-$ & $-$ \\
Holmberg I  & 97.70   & 0.64$\pm$0.27    & 7.38$\pm$0.18   & 0.28$\pm$0.25    & 7.60$\pm$0.17   & $-$ & $-$ & $-$ \\
M81 DwB     & $-$     & $-$              & $-$             & $-$              & $-$             & $-$ & $-$ & $-$ \\
Holmberg II & 224.47  & -0.31$\pm$0.08   & 7.79$\pm$0.03   & -0.30$\pm$0.06   & 7.80$\pm$0.03   & $-$ & $-$ & $-$ \\
NGC 2366    & $-$     & $-$              & $-$             & $-$              & $-$             & -0.05$\pm$0.08   & -0.55$\pm$0.03   & 0.95$\pm$0.04 \\
NGC 4214    & 206.27  & -0.089$\pm$0.009 & 8.214$\pm$0.003 & -0.10$\pm$0.01   & 8.202$\pm$0.004 & $-$ & $-$ & $-$ \\
NGC 1569    & 119.43  & -0.08$\pm$0.02   & 8.153$\pm$0.006 & -0.05$\pm$0.02   & 8.145$\pm$0.008 & $-$ & $-$ & $-$ \\
IC 2574     & 386.75  & -0.42$\pm$0.06   & 8.08$\pm$0.04   & -0.40$\pm$0.06   & 8.07$\pm$0.04   & $-$ & $-$ & $-$ \\
NGC 4449    & 137.51  & -0.071$\pm$0.003 & 8.260$\pm$0.003 & -0.062$\pm$0.004 & 8.254$\pm$0.003 & $-$ & $-$ & $-$ \\
NGC 2976    & 217.72  & -0.13$\pm$0.02   & 8.366$\pm$0.006 & -0.22$\pm$0.02   & 8.384$\pm$0.008 & $-$ & $-$ & $-$ \\
NGC 3077    & 162.84  & -0.24$\pm$0.03   & 8.464$\pm$0.007 & -0.39$\pm$0.03   & 8.488$\pm$0.009 & $-$ & $-$ & $-$ \\
NGC 2403    & 470.54  & -0.295$\pm$0.009 & 8.442$\pm$0.003 & -0.339$\pm$0.009 & 8.442$\pm$0.003 & $-$ & $-$ & $-$ \\
NGC 925     & 318.37  & -0.203$\pm$0.009 & 8.35$\pm$0.004  & -0.202$\pm$0.009 & 8.35$\pm$0.004  & $-$ & $-$ & $-$ \\
NGC 3198    & 194.31  & -0.298$\pm$0.009 & 8.603$\pm$0.004 & -0.306$\pm$0.009 & 8.605$\pm$0.004 & $-$ & $-$ & $-$  \\
NGC 4826    & $-$     & $-$              & $-$             & $-$              & $-$             & -0.03$\pm$0.09 & -0.14$\pm$0.36 & 1.63$\pm$0.09  \\
NGC 4736    & 232.60  & -0.082$\pm$0.013 & 8.582$\pm$0.004 & -0.262$\pm$0.020 & 8.617$\pm$0.007 & $-$ & $-$ & $-$ \\
NGC 3184    & 221.13  & -0.165$\pm$0.009 & 8.575$\pm$0.004 & -0.171$\pm$0.009 & 8.575$\pm$0.004 & $-$ & $-$ & $-$ \\
NGC 5457    & 719.65  & -0.523$\pm$0.009 & 8.596$\pm$0.002 & -0.602$\pm$0.01  & 8.602$\pm$0.002 & $-$ & $-$ & $-$ \\
NGC 6946    & 342.61  & -0.172$\pm$0.004 & 8.598$\pm$0.002 & -0.175$\pm$0.004 & 8.599$\pm$0.002 & -0.115$\pm$0.002 & 0.12$\pm$0.03    & 1.13$\pm$0.02 \\
NGC 5055    & 355.35  & -0.074$\pm$0.008 & 8.585$\pm$0.002 & -0.088$\pm$0.007 & 8.588$\pm$0.002 & -0.010$\pm$0.008 & -0.055$\pm$0.008 & 0.65$\pm$0.06 \\
NGC 5194    & 232.05  & -0.029$\pm$0.002 & 8.607$\pm$0.002 & -0.035$\pm$0.003 & 8.609$\pm$0.002 & $-$ & $-$ & $-$ \\
NGC 3521    & 249.6   & -0.112$\pm$0.005 & 8.633$\pm$0.002 & -0.14$\pm$0.005  & 8.643$\pm$0.002 & -0.043$\pm$0.003 & -0.14$\pm$0.02   & 1.70$\pm$0.10 \\
NGC 2841    & 207.73  & -0.074$\pm$0.013 & 8.630$\pm$0.008 & -0.124$\pm$0.014 & 8.655$\pm$0.009 & $-$ & $-$ & $-$  \\
NGC 7331    & 275.02  & -0.149$\pm$0.007 & 8.614$\pm$0.005 & -0.156$\pm$0.008 & 8.616$\pm$0.005 & $-$ & $-$ & $-$ \\
\hline

\end{tabular}
\caption{Linear model parameters for each galaxy. Each column shows information as follows: 1) galaxy designation; 12) isophotal radius; 13) and 14) linear model parameters and errors without DIG-dominated fibres with isophotal radius normalisation; 15) and 16) linear model parameters and errors with all fibres with isophotal radius normalisation. Each linear model was fitted using a Bayesian MCMC routine. Columns 17), 18), 19) respectively show the slopes and breakpoint for the piecewise linear fits for the NUV-band normalisation for those galaxies where a change in the tendency of their metallicity distribution was spotted.}
\label{tab:grads_data_2}
\end{sidewaystable*}

\clearpage

\section{Fitting parameters of presented relations} \label{fit_parameters_rel}

\begin{table}[H]
\centering \normalsize
\renewcommand{\arraystretch}{1.25}
\begin{tabular}{>{\rowmac}c>{\rowmac}c>{\rowmac}c>{\rowmac}c>{\rowmac}c>{\rowmac}c<{\clearrow}}

\hline
\hline
1 & 2 & 3 & 4 & 5 & 6 \\

Normalisation&
\multicolumn{5}{c}{M-T + SAMI} \\

radius &
breakpoint&
$a_{1}$ &
$b_{1}$ &
$a_{2}$ &
$b_{2}$
\\

&
{[}$\log(M_{\rm star}/M_\odot)${]} &
{[}$ \rm dex/R_{eff} ${]} &
{[}$ \rm dex/R_{eff} ${]} &
{[}$ \rm dex/R_{eff} ${]} &
{[}$ \rm dex/R_{eff} ${]}
\\

\hline

$R_{\rm eff,\, NUV}$ & 9.6$\pm$0.2 & -0.04$\pm$0.03 & 0.35$\pm$0.10 & 0.04$\pm$0.01 & -0.43$\pm$0.14 \\
$R_{\rm eff,\, r}$   & 9.4$\pm$0.2 & -0.07$\pm$0.05 & 0.54$\pm$0.12 & 0.03$\pm$0.01 & -0.39$\pm$0.13 \\
$R_{\rm eff,\, Ks}$  & 9.6$\pm$0.1 & -0.04$\pm$0.01 & 0.30$\pm$0.11 & 0.04$\pm$0.04 & -0.45$\pm$0.13 \\

\hline

\end{tabular}
\caption{Fitting parameters of all plots in Fig. \ref{fig:grad_mass_fit_HI_r}, where each column is given as follows: 1) is the radius for the normalisation of the metallicity gradients, 2) is the break in the stellar mass for the piecewise function, 3), 4), 5) and 6) are the parameters for the linear fit to the data, $f(x)_i = a_i\cdot x + b_i$ for $i = 1,2$, where $1$ represents the function to the left to the breakpoint, and $2$ for the right.}
\label{tab:fig9_table}
\end{table}

\vspace{-0.5cm}

\begin{table*}[h]
\centering
\renewcommand{\arraystretch}{1.2}
\begin{tabular}{>{\rowmac}c>{\rowmac}c>{\rowmac}c>{\rowmac}c>{\rowmac}c<{\clearrow}}

\hline
\hline
1 & 2 & 3 & 4 & 5\\


breakpoint & $a_1$ & $b_1$ & $a_2$ & $b_2$ \\

[$\log (M_{\rm star} / M_\odot)$] & [$\mathrm{dex / R_{eff}}$] & [$\mathrm{dex / R_{eff}}$] & [$\mathrm{dex / R_{eff}}$] & [$\mathrm{dex / R_{eff}}$] \\

\hline

9.6$\pm$0.2 & -0.04$\pm$0.01 & 0.29$\pm$0.10 & 0.04$\pm$0.01 & $-$0.47$\pm$0.12 \\

\hline

\end{tabular}
\caption{Fitting parameters for the SAMI sample in Fig. \ref{fig:grad_mass_fit_HI_r}, found in \citet{Poetrodjojo_2021}. 1) is the break in the stellar mass for the piecewise function, 3), 4) and 5) are the parameters for the linear fit to the data, $f(x)_i = a_i\cdot x + b_i$, where $1$ represents the function to the left to the breakpoint, and $2$ for the right.}
\label{tab:SAMI_pwlf}
\end{table*}
\vspace{-0.5cm}

\begin{table}[H]
\centering \normalsize
\renewcommand{\arraystretch}{1.25}
\begin{tabular}{>{\rowmac}c>{\rowmac}c>{\rowmac}c>{\rowmac}c>{\rowmac}c>{\rowmac}c<{\clearrow}}

\hline
\hline
1 & 2 & 3 & 4 & 5 & 6 \\

\begin{tabular}[c]{@{}c@{}c@{}} Normalisation \\ radius \end{tabular} &
\begin{tabular}[c]{@{}c@{}c@{}} $f_{\rm g, \, H \, I }$ \\ breakpoint \end{tabular} &

\begin{tabular}[c]{@{}c@{}c@{}}$ a_{1} $ \\ {[}$ \rm dex/R_{eff} ${]} \end{tabular} &
\begin{tabular}[c]{@{}c@{}c@{}}$ b_{1} $ \\ {[}$ \rm dex/R_{eff} ${]} \end{tabular} &

\begin{tabular}[c]{@{}c@{}c@{}}$ a_{2} $ \\ {[}$ \rm dex/R_{eff} ${]} \end{tabular} &
\begin{tabular}[c]{@{}c@{}c@{}}$ b_{2} $ \\ {[}$ \rm dex/R_{eff} ${]} \end{tabular}

\\
\hline

$R_{\rm eff,\, NUV}$ & 0.74$\pm$0.10 & -0.19$\pm$0.13 & -0.03$\pm$0.05 & 1.23$\pm$0.83 & -1.08$\pm$0.09 \\
$R_{\rm eff,\, r}$   & 0.75$\pm$0.10 & -0.14$\pm$0.13 & -0.01$\pm$0.05 & 1.18$\pm$0.83 & -1.00$\pm$0.10 \\
$R_{\rm eff,\, Ks}$  & 0.53$\pm$0.09 & -0.13$\pm$0.06 & -0.02$\pm$0.01 & 0.21$\pm$0.10 & -0.20$\pm$0.07 \\
$R_{25}$             & 0.51$\pm$0.13 & -0.50$\pm$0.24 & -0.03$\pm$0.09 & 0.78$\pm$0.38 & -0.69$\pm$0.16 \\

\hline

\end{tabular}
\caption{Fitting parameters of all plots in Fig. \ref{fig:metgrad-gasfrac}. 1) is the radius for the normalisation of the metallicity gradients, 2) is the break in the atomic gas fraction for the piecewise function, 3), 4), 5) and 6) are the parameters for the linear fit to the data, $f(x)_i = a_i\cdot x + b_i$,  where $1$ represents the function to the left to the breakpoint, and $2$ for the right.}
\label{tab:fig10_table}
\end{table}
\vspace{-0.5cm}

\label{LastPage}

\begin{multicols}{2}

\section{Results for the metallicity recalibration} \label{other}

\begin{figure}[H]
    \centering
    \includegraphics[width = 0.8\columnwidth]{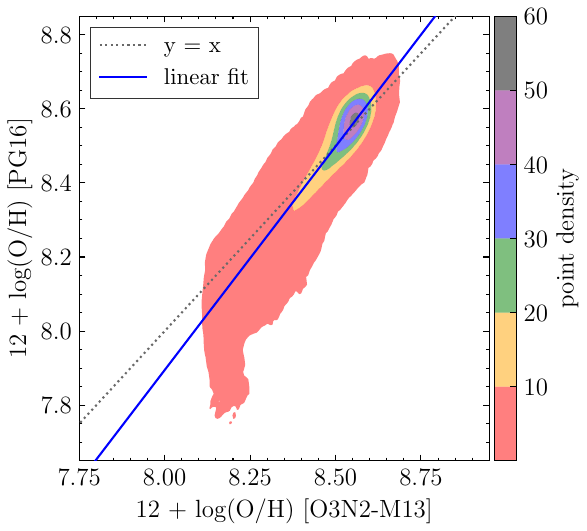}
    \caption{Correlation between the gas-phase metallicities of SDSS galaxies using the PG16 and the O3N2-M13 calibrations.}
    \label{fig:met_recal}
\end{figure}

The empirical formulation for the O3N2-M13 calibration, as given in \citet{O3N2_M13_paper}, is
\begin{equation*}
    12 + \log(\mathrm{O/H})_{\textrm{O3N2$-$M13}} = 8.533\pm0.012 - (0.214\pm0.012) \times \mathrm{O3N2}
\end{equation*}
where
\begin{equation*}
    \mathrm{O3N2} = \log \left( \frac{[\mathrm{O \; III}] \lambda5007}{\mathrm{H\beta}} \times \frac{\mathrm{H\alpha}}{\mathrm{[N \; II]\lambda6583}} \right)
\end{equation*}

Given this indicator, PG16 and the SDSS data, the correlation between these two calibrations is as follows:
\begin{equation*}
    12+\log(\mathrm{O/H})_{\textrm{PG16}} = -1.76 + 1.21 \times ( 12+\log(\mathrm{O/H})_{\textrm{O3N2$-$M13}} ) .
\end{equation*}

\end{multicols}

\end{document}